\documentclass[fleqn,usenatbib]{mnras}

\usepackage{newtxtext,newtxmath}
\usepackage{placeins}

\usepackage[T1]{fontenc}

\DeclareRobustCommand{\VAN}[3]{#2}
\let\VANthebibliography\thebibliography
\def\thebibliography{\DeclareRobustCommand{\VAN}[3]{##3}\VANthebibliography}

\usepackage{graphicx}	
\usepackage{amsmath}	
\usepackage{multicol}        
\usepackage{bm}		
\usepackage{pdflscape}	
\usepackage{float}
\usepackage{placeins}
\usepackage{enumitem}
\usepackage{caption}
\usepackage{minted}
\usepackage{xcolor}
\usepackage{rotating}
\usepackage{hyperref}
\usepackage{orcidlink}
\usepackage{subfigure}

\numberwithin{equation}{section}
\let\originalleft\left
\let\originalright\right
\renewcommand{\left}{\mathopen{}\mathclose\bgroup\originalleft}
\renewcommand{\right}{\aftergroup\egroup\originalright}
\hypersetup{
urlcolor=blue}
\defcitealias{lilo17}{L17}

\title[LILO with UNIONS]{Revisiting the `Lensing is Low' Problem with UNIONS}

\author[M. Campbell et al.]{
Martine C. T. Campbell\orcidlink{0009-0006-9109-8729}$^{1,2,3}$\thanks{E-mail: mctcampb@uwaterloo.ca},
Jack Elvin-Poole\orcidlink{0000-0001-5148-9203}$^{1,2}$,
Michael J. Hudson\orcidlink{0000-0002-1437-3786}$^{1,2,4}$,
Thomas de Boer\orcidlink{0000-0001-5486-2747}$^{5}$,
\newauthor
Sacha Guerrini\orcidlink{0009-0004-3655-4870}$^{6}$,
Fabian Hervas-Peters\orcidlink{0009-0008-1839-2969}$^{6}$,
Hendrik Hildebrandt\orcidlink{0000-0002-9814-3338}$^{7}$,
Martin Kilbinger\orcidlink{0000-0001-9513-7138}$^{6}$,
Eugene A. Magnier\orcidlink{0000-0002-7965-2815}$^{5}$,
\newauthor
Alan W. McConnachie\orcidlink{0000-0003-4666-6564}$^{8}$,
Charlie T. Mpetha\orcidlink{0000-0002-7805-2500}$^{9}$,
Romain Paviot\orcidlink{0009-0002-8108-3460}$^{6}$,
Ludovic Van Waerbeke\orcidlink{0000-0002-2637-8728}$^{10}$, 
Anna Wittje\orcidlink{0000-0002-8173-3438}$^{7}$
\\
$^{1}$Department of Physics and Astronomy, University of Waterloo, 200 University Ave. W, Waterloo, ON N2L 3G1, Canada\\
$^{2}$Waterloo Centre for Astrophysics, University of Waterloo, 200 University Ave. W, Waterloo, ON N2L 3G1, Canada\\
$^{3}$Institut de Física d’Altes Energies (IFAE), The Barcelona Institute of Science and Technology, Campus UAB, Bellaterra (Barcelona), 08193, Spain\\
$^{4}$Perimeter Institute for Theoretical Physics, 31 Caroline St. N, Waterloo, ON N2L 2Y5, Canada\\
$^{5}$Institute for Astronomy, University of Hawaii, 2680 Woodlawn Dr., Honolulu, HI 96822, USA\\
$^{6}$Université Paris-Saclay, Université Paris Cité, CEA, CNRS, AIM, Gif-sur-Yvette, 91191, France\\
$^{7}$Ruhr University Bochum, Faculty of Physics and Astronomy, Astronomical Institute (AIRUB), German Centre for Cosmological Lensing, Bochum, 44780, Germany\\
$^{8}$National Research Council Herzberg Astronomy and Astrophysics, 5071 W. Saanich Rd., Victoria, BC V8Z6M7, Canada\\
$^{9}$NASA Goddard Space Flight Center, 8800 Greenbelt Rd., Greenbelt, MD 20771, USA\\
$^{10}$Department of Physics and Astronomy, University of British Columbia, 6224 Agricultural Rd., Vancouver, BC V6T 1Z1, Canada
}

\date{Accepted XXX. Received YYY; in original form ZZZ}

\pubyear{\the\year{}}

\begin{document}
\label{firstpage}
\pagerange{\pageref{firstpage}--\pageref{lastpage}}
\maketitle

\begin{abstract}
We present new measurements of the galaxy--galaxy lensing (GGL) signal around Baryon Oscillation Spectroscopic Survey (BOSS) CMASS galaxies using background sources from the Ultraviolet Near-Infrared Optical Northern Survey (UNIONS). With high-quality imaging of background sources and a survey overlap of approximately 2650 square degrees, we obtain precise large-scale GGL measurements. Building on these new measurements, we revisit the so-called `lensing is low' problem, wherein galaxy--halo connection models calibrated on galaxy clustering (GC) data over-predict the GGL signal by 20--40$\%$ assuming CMB-based cosmological parameters. We model the galaxy--halo connection using a halo occupation distribution (HOD), and perform joint fits to both GGL and GC signals across a wide range of scales, as well as a GC-only fit. In contrast to previous work, we do not find a significant `lensing is low' effect in the CMASS sample, although the best joint fits are achieved by decreasing the amplitude of the matter power spectrum slightly relative to the Planck cosmological parameters. Overall, we find that two models describe our observables similarly well: one where HOD and cosmological parameters are free, and one where HOD, cosmological, and feedback parameters are free. Importantly, we emphasise the role of large scales in constraining the lensing is low effect, shifting the narrative away from an exclusively small-scale issue.
\end{abstract}




\section{Introduction}
Over the past three decades, the $\Lambda$CDM concordance model has emerged as a cornerstone of modern cosmology, providing a successful framework for describing the large-scale structure (LSS) of the Universe. Within this paradigm, the halo model has proven effective at capturing non-linear structure formation, serving as a powerful tool for extracting cosmological information \citep{2000MNRAS.318..203S,COORAY_2002,Zehavi_2004}. Crucially, it enables modelling of the galaxy--halo connection---the underlying relationship between galaxies and their host dark matter halos. 

A complementary combination of observables is galaxy--galaxy lensing (GGL) and galaxy clustering (GC) \citep{Cacciato_2009}. GGL measures the average projected mass distribution around foreground lens galaxies via the coherent distortion of light from background source galaxies. On the other hand, GC quantifies the `clumpiness' of galaxies across different projected scales. Together, these observables provide constraints on the galaxy--halo connection and the underlying matter distribution \citep{Miyatake_2022}.

Several studies have simultaneously analysed GGL and GC. Notably, \citeauthor{lilo17} (\citeyear{lilo17}; hereafter, \citetalias{lilo17}) first identified a tension between the two signals in the Baryon Oscillation Spectroscopic Survey\footnote{\url{https://www.sdss3.org/surveys/boss.php}} (BOSS; \citealp{2011AJ....142...72E,2013AJ....145...10D}) CMASS Luminous Red Galaxy (LRG) sample. They used standard galaxy--halo connection models (e.g., the halo occupation distribution, subhalo abundance matching) calibrated on GC data to predict the GGL signal on small--intermediate scales (up to $12\,h^{-1}\,\mathrm{Mpc}$), adopting a Planck-like cosmology \citep{planck18}. The resulting GGL predictions exceeded their measurements by 20--40$\%$. They referred to this as the `lensing is low' problem. Although first identified in CMASS galaxies, the lensing is low problem is also evident in other galaxy samples (e.g., LOWZ; \citealp{Lange_2019,2020MNRAS.492.2872W,2024A&A...690A.221P}). 

Since the work of \citetalias{lilo17}, there have been further investigations into the origin(s) of the lensing is low problem, with much of the work focusing on the limitations of the standard halo occupation distribution (HOD; \citealp{2000MNRAS.318.1144P,2002ApJ...575..587B,zheng05}). Astrophysical effects such as assembly bias (AB), feedback, and satellite galaxy segregation have all been proposed as potential explanations (\citetalias{lilo17}; \citealp{Lange_2019,2021MNRAS.502.3582Y,Amon23,ChavesMontero23}). The standard HOD, which assigns galaxies to halos based solely on halo mass, does not capture these astrophysical effects. Instead, it bypasses galaxy formation physics by distributing galaxies in a statistical manner. 

It is possible that the galaxy--halo connection depends on halo properties beyond mass, an effect known as `assembly bias' \citep{2005MNRAS.363L..66G,2006ApJ...652...71W,10.1111/j.1365-2966.2006.11230.x}. Candidate secondary properties include halo formation time, environment (e.g., local over-density), and halo concentration \citep{Xu_2021,2021MNRAS.502.3582Y,2025ApJ...988..280W}. \cite{2021MNRAS.502.3582Y} and \cite{ChavesMontero23} show that neglecting AB results in the standard HOD placing too few galaxies in low mass halos, causing an over-predicted GGL signal. The standard HOD can be extended to include AB---such models are referred to as \textit{decorated} HODs \citep{Hearin_2016}. \cite{2024A&A...690A.221P} demonstrate that a decorated HOD provides a better fit to the eBOSS \citep{2016AJ....151...44D} LRG GGL signal than a standard HOD. While this modification shifts GGL predictions in the right direction, it does not fully resolve the lensing is low tension. Note that currently, there is no broad consensus on whether AB is present in observations. 

Many HOD analyses use halo catalogues from dark-matter-only N-body simulations (e.g., \citealp{Reid_2014}, \citealp{2017MNRAS.465.4853A}, \citealp{2024MNRAS.535.2469Z}). These simulations do not include baryonic feedback, which describes the complex interplay between baryons and dark matter. Hydrodynamical cosmological simulations reveal that feedback processes can significantly redistribute the matter within halos \citep{2011MNRAS.417.2020S,2018MNRAS.475..676S,medlock2025constrainingbaryonicfeedbackeffects}. For example, active galactic nuclei (AGN) and supernovae tend to push gas towards the outskirts of halos, whereas baryonic cooling can pull material towards the centre \citep{Blumenthal_Faber_1986,2010MNRAS.407..435A,medlock2025constrainingbaryonicfeedbackeffects}. These competing effects can increase or decrease the GGL signal at different scales. In the context of the lensing is low problem, feedback has been invoked to reduce the amplitude of the predicted GGL signal, but this effect alone cannot account for the tension between GGL and GC \citep{Lange_2019,Amon23,ChavesMontero23}. On the other hand, feedback calibrated via the kinetic Sunyaev-Zeldovich (kSZ; \citealp{1972CoASP...4..173S,1980MNRAS.190..413S}) effect has been shown to have a significant effect on cosmic shear \citep{2024MNRAS.534..655B}, so the matter is not settled.

In the standard HOD implementation, satellite galaxies are assigned positions based on the mass distribution within their host halo \citep{Zheng_2007}. For example, if satellite galaxies are distributed according to a Navarro--Frenk--White (NFW; \citealp{Navarro_1996}) profile, their concentration is set by the halo concentration. Alternatively, satellite galaxies can be placed at the locations of dark matter particles within the halo \citep{Miyatake_2022}. However, satellite galaxies may not strictly follow the underlying mass distribution, particularly for samples selected by properties such as colour or star formation rate \citep{ChavesMontero23}. This describes satellite galaxy segregation. \cite{ChavesMontero23} found that, for IllustrisTNG \citep{2018MNRAS.475..676S} galaxies representative of the BOSS sample, the concentrations of the satellite galaxy populations are significantly lower than those of their host halos. If satellite galaxies are placed farther from the halo centre, thereby probing a less dense environment, the predicted GGL signal would decrease. The same paper demonstrates that simultaneously incorporating multiple astrophysical effects, including AB and feedback, largely explains the lensing is low tension. It is important to note that these effects primarily modify GGL on small scales, with AB extending to intermediate scales. However, GGL predictions converge around the 2-halo term (see Fig. 10 in \citealp{ChavesMontero23}), and so the larger scales are insensitive to these effects.

An alternative solution to the lensing is low problem is to reduce the amplitude of the matter power spectrum relative to Planck (\citetalias{lilo17}; \citealp{Lange_2019,Amon23}). This suppression is often quantified by the parameter $S_8\equiv\sigma_8(\Omega_{\mathrm{m}}/0.3)^{0.5}$, where $\sigma_8$ is the root-mean-square amplitude of matter density fluctuations on $8\,h^{-1}\,\mathrm{Mpc}$ scales. Reconciling GGL and GC through cosmological changes typically requires lowering $S_8$ by $2\text{--}3\sigma$ relative to the Planck 2018 constraint from primary cosmic microwave background (CMB) anisotropies (\citetalias{lilo17}; $S_8=0.834\pm0.016$). This explanation remains debated, given that astrophysical effects can account for much of the small-scale tension. Furthermore, some joint analyses of BOSS GGL and GC obtain $S_8$ values consistent with Planck (within $1\sigma$). For example, \cite{zhang2025modellinggalaxyclusteringtomographic} report $S_8=0.804\pm0.051$ using Hyper Suprime-Cam\footnote{\url{https://hsc.mtk.nao.ac.jp/ssp}} Year 3 (HSC Y3; \citealp{2022PASJ...74..247A,2022PASJ...74..421L}) data and a minimum-bias model with a point-mass correction term for GGL. They model the GGL signal from $2\text{--}70\,h^{-1}\,\mathrm{Mpc}$ and the GC signal from $8\text{--}80\,h^{-1}\,\mathrm{Mpc}$. \cite{More_2015} find $S_8\sim0.817\pm0.035$ for various CMASS subsamples using Canada--France--Hawaii Telescope Lensing Survey (CFHTLenS; \citealp{2012MNRAS.427..146H,2013MNRAS.429.2858M}) data and an HOD approach. Their scales of interest span $0.1\text{--}20\,h^{-1}\,\mathrm{Mpc}$ for GGL and $0.85\text{--}80\,h^{-1}\,\mathrm{Mpc}$ for GC. Historically, $S_8$ values inferred from low-$z$ probes have tended to fall below CMB-based constraints, giving rise to the so-called `$S_8$ tension' \citep{2013MNRAS.432.2433H,2020MNRAS.492.2872W,kids1000_3x2,2022PhRvD.105b3514A,2023MNRAS.520.5373L,2025arXiv251215962L}. However, several low-$z$ analyses, based on cosmic shear or 3 x 2-point analyses, now find consistency with the Planck $S_8$ (e.g., \citealp{2024MNRAS.534..655B}, \citealp{wright2025kidslegacycosmologicalconstraintscosmic}). 

To differentiate astrophysical and cosmological effects, it is essential to precisely measure and model GGL and GC signals across a wide range of scales, from the non-linear to the linear regime. This is made possible with weak-lensing data from the Ultraviolet Near-Infrared Optical Northern Survey\footnote{\url{https://www.skysurvey.cc}} (UNIONS; \citealp{gwyn25}). UNIONS delivers exquisite galaxy imaging and a substantial overlap with the CMASS sample on the sky (over approximately 2650 square degrees), enabling precise large-scale GGL measurements. Using these new measurements, we revisit the lensing is low problem and jointly model GGL and GC with an HOD implemented by the \textsc{Dark Emulator}\footnote{\url{https://github.com/DarkQuestCosmology/dark_emulator_public}} \citep{Nishimichi_2019,Miyatake_2022}. We model the GGL signal over $0.15\text{--}62\,h^{-1}\,\mathrm{Mpc}$ and the GC signal over $0.15\text{--}80\,h^{-1}\,\mathrm{Mpc}$. We also explore various extensions to the standard HOD, including feedback with an analytic prescription. 

In this paper, distances are expressed in comoving coordinates, and halo masses are reported in units of $h^{-1}M_{\odot}$. The structure of this paper is as follows. In Section \ref{sec:data} we detail the galaxy catalogues used in this work. In Section \ref{sec:measurements} we describe our GGL and GC measurements. In Section \ref{sec:theory} we explain our theoretical GGL and GC models. In section \ref{sec:ia} we estimate the potential impact of intrinsic alignments (IAs) on our analysis. In Section \ref{sec:param_est} we describe our likelihood procedure for determining best-fit models. In Section \ref{sec:results} we present the results of this work, and interpret them in the context of the lensing is low problem. We finally summarise in Section \ref{sec:conclusions}, and provide additional methodological details and plots in the appendices. Measurements were performed assuming a flat $\Lambda$CDM cosmology with $\Omega_{\mathrm{m,0}}=0.3111$, where $\Omega_{\mathrm{m}}$ includes $\Omega_\nu$. Predictions are rescaled to ensure consistency with the measurement cosmology, following the equations in Appendix \ref{sec:cosmo_dependence}.

\section{Data}
\label{sec:data}
This section describes the survey details of our lens and source galaxy samples. Both samples are used for GGL measurements, while the lens galaxy sample is used for GC measurements. We use UNIONS galaxies as our sources, and BOSS CMASS galaxies as our lenses. 

\subsection{UNIONS Background Source Galaxy Sample}
\label{subsec:unions}
We use an early version of the UNIONS \textsc{ShapePipe} v1.4.5 catalogue \citep{hervaspeters2026unions3500weaklensingi} as our background source galaxy sample. This catalogue is internal to the UNIONS collaboration and has not been publicly released. UNIONS is an ongoing deep, wide-field imaging survey that exploits three telescopes in Hawaii: the Canada--France--Hawaii Telescope\footnote{\url{https://www.cfht.hawaii.edu}} (CFHT; \citealp{1981JRASC..75..305R}), the Panoramic Survey Telescope and Rapid Response System\footnote{\url{https://about.ifa.hawaii.edu/research/surveys-2}} (Pan-STARRS; \citealp{2004AN....325..636H}), and the Subaru Telescope\footnote{\url{https://subarutelescope.org}} \citep{2021PJAB...97..337I}. It is expected to deliver ground-based $ugriz$ photometry for the Euclid space mission\footnote{\url{https://www.euclid-ec.org}} \citep{2011arXiv1110.3193L,2018LRR....21....2A}, enabling photometric redshift estimation. Data in the $u$- and $r$-bands is obtained with CFHT and provided by the Canada--France Imaging Survey (CFIS), whereas Pan-STARRS delivers $i$- and $z$-band data. The Subaru Telescope also contributes to the $z$-band imaging through the Wide Imaging with Subaru Hyper Suprime-Cam of the Euclid Sky (WISHES) programme, and to the $g$-band imaging through the Waterloo--Hawaii Institute for Astronomy $g$-band Survey (WHIGS). The final survey footprint will cover 5900 square degrees (5614 in the North Galactic Cap/NGC and 286 in the South Galactic Cap/SGC), reaching magnitudes of 24.3, 25.2, 24.9, 24.3, and 24.1 in the $u$-, $r$-, $g$-, $i$-, and $z$-bands, respectively. Euclid Data Release 3 (DR3) is expected to overlap with UNIONS across 5815 square degrees. Aside from its synergy with Euclid, UNIONS is also strategically designed to overlap significantly with BOSS, providing background source galaxies for BOSS GGL measurements. For more detailed information, see \cite{gwyn25}. 

Our shape catalogue was generated with the \textsc{ShapePipe} software package\footnote{\url{https://github.com/CosmoStat/shapepipe}} \citep{Farrens_2022,guinot22}, which provides ellipticity components $\epsilon_1$ and $\epsilon_2$ (derived from $r$-band imaging) for over 85 million galaxies. Here, we summarise key aspects of the shape measurement pipeline and catalogue-specific choices. The point-spread function (PSF) is modelled using \textsc{PSFEx}\footnote{\url{https://github.com/astromatic/psfex}} \citep{2011ASPC..442..435B}. A loose PSF size cut of $\mathrm{HLR}_{\mathrm{gal}}/\mathrm{HLR}_{\mathrm{PSF}}>0.5$ is applied to galaxies, where HLR denotes the half-light radius. This removes galaxies whose sizes are too small compared to the PSF from the final sample. We also require $n_{\mathrm{epoch}}\geq2$ and $n_{\mathrm{point}}\geq3$, where $n_{\mathrm{epoch}}$ and $n_{\mathrm{point}}$ are the numbers of observed epochs (i.e., distinct observation times) and pointings for a given object. Galaxy shapes are measured using \textsc{NGMIX}\footnote{\url{https://github.com/esheldon/ngmix}} \citep{2017ApJ...841...24S}, which models galaxy profiles as sums of Gaussians and performs a $\chi^2$ minimisation. The \textsc{Metacalibration} \citep{huff2017metacalibrationdirectselfcalibrationbiases,2017ApJ...841...24S} framework is used to calibrate galaxy shapes. Unlike simulation-based approaches, this technique relies solely on the imaging data. Essentially, a small artificial shear is applied to each galaxy in the $\pm\epsilon_1$ and $\pm\epsilon_2$ directions, and the shape measurement algorithm's response is recorded in a $2\times2$ matrix $\bm{R}$. Each element $R_{ij}$ corresponds to the $i$th ellipticity component measured after shearing the image along the $\pm\epsilon_j$ directions. The weighted mean response matrix across all galaxies is used to calibrate ellipticities as follows: 
\begin{equation}
\bm{\epsilon}_{\mathrm{cal}}=\langle\bm{R}\rangle^{-1}\bm{\epsilon}_{\mathrm{meas}},
\end{equation}
where $\bm{\epsilon}_{\mathrm{cal}}$ is the final catalogue ellipticity, containing both real and imaginary components. We use the mean $\bm{R}$ instead of individual matrices, as the inverse of a noisy quantity is biased. 

Despite calibration, residual multiplicative biases often remain in ellipticity estimates. Additive biases may also be present, although these are generally for GGL since they are uncorrelated with lens galaxy positions. One can incorporate both these biases and write 
\begin{equation}
    \bm{\epsilon}_{\rm cal} \simeq(1+m) \bm{\epsilon}_{\rm true} + \bm{c} \,.
\end{equation}
When the work for this paper was done, it was expected that the $m$-bias would be small, of order of a percent or two.  However, after the work here was completed,  \cite{hervaspeters2026unions3500weaklensingi} estimated $m = -0.0572 \pm 0.0141$ for an updated version of the catalogue. In this work, we do not incorporate this correction into the measurements, but comment on its impact where appropriate.   

The \textsc{ShapePipe} catalogue also contains individual weights for each galaxy. We intended to follow the Dark Energy Survey\footnote{\url{https://www.darkenergysurvey.org}} (DES; \citealp{thedarkenergysurveycollaboration2005darkenergysurvey}) prescription from \cite{2021MNRAS.504.4312G}, which defines the weights as:
\begin{equation}
w_s=(1/\sigma^2_{\epsilon})\langle R\rangle^2, 
\end{equation}
where $R=(R_{11}+R_{22})/2$. The weights above are computed for and uniformly assigned to an ensemble of galaxies in a given $\mathrm{HLR}_{\mathrm{gal}}/\mathrm{HLR}_{\mathrm{PSF}}$, signal-to-noise (S/N) bin. The variance in $\epsilon$ is given by:
\begin{equation}
\sigma^2_{\epsilon}\,=\,\frac{1}{2}\left(\frac{\sum{\epsilon_1}^2}{N_{\mathrm{gal}}}\,+\,\frac{\sum{\epsilon_2}^2}{N_{\mathrm{gal}}}\right),
\end{equation}
where the sum runs over all galaxies in the $\mathrm{HLR}_{\mathrm{gal}}/\mathrm{HLR}_{\mathrm{PSF}}$, S/N bin, and $N_{\mathrm{gal}}$ is the total number of galaxies in that bin. While assigning weights, the axes in the bottom right panel of Fig. 4 (S/N vs. $T/T_{\mathrm{PSF}}$) in \cite{2021MNRAS.504.4312G} were flipped, resulting in a suboptimal (but consistently applied) weighting scheme. We verified that using the proper DES weights produces minimal changes in our GGL signal.

\begin{figure}
\includegraphics[width = \columnwidth]{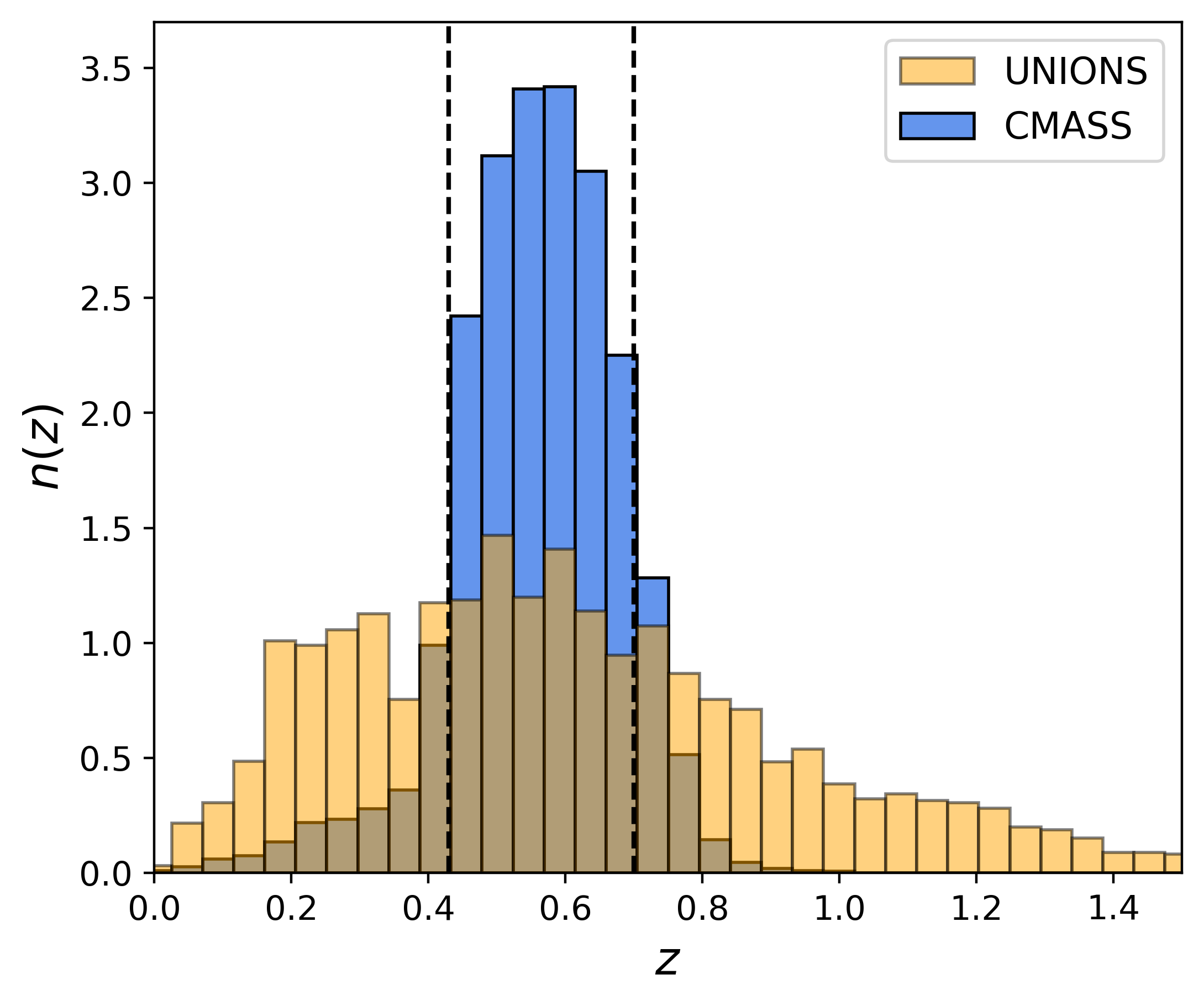}
\caption{Effective redshift distribution (shifted by the mean SOM bias, see Section \ref{subsec:esd}) of UNIONS \textsc{ShapePipe} v1.4.5 galaxies (yellow) compared to the CMASS redshift distribution (blue) without any redshift cuts. Vertical black lines indicate the CMASS redshift cuts of $z=0.43$ and $z=0.7$.}
\label{fig:nz}
\end{figure}

Photometric redshifts are not yet available for UNIONS galaxies. As a workaround, we estimate their effective redshift distribution, $n(z_{\mathrm{s}})$. Here, we summarise the $n(z_{\mathrm{s}})$ construction, which is based on the procedure in \cite{2021A&A...647A.124H}. A detailed description of the implementation for UNIONS can be found in \cite{li2024blackholetohalomassrelationunions}. First, $ugriz$ photometry is assigned to UNIONS $r$-band galaxies within the CFHTLenS W3 field. A self-organizing map (SOM; \citealp{Kohonen,2015ApJ...813...53M}) is then trained on a spectroscopic calibration sample with CFHTLenS $ugriz$ photometry \citep{2012MNRAS.421.2355H}. After the SOM has produced a 2D representation of the high-dimensional magnitude space, it is populated with the UNIONS galaxies. Each SOM cell therefore contains both UNIONS objects and galaxies from the calibration sample, and their ratio defines a weight $w_{\mathrm{SOM}}$. Note that this weight incorporates the individual source galaxy weights $w_{\mathrm{s}}$. Then, the spectroscopic redshift distributions in each cell are re-weighted using $w_{\mathrm{SOM}}$, and the effective redshift distribution is constructed as follows:
\begin{equation}
\begin{split}
n(z_{\mathrm{s}})\,&=\,\int{w_{\mathrm{SOM}}(z_{\mathrm{s}})\langle R(z_{\mathrm{s}})\rangle n_{\mathrm{spec}}(z_{\mathrm{s}})\mathrm{d}z_{\mathrm{s}}}, \\
&=\,\sum_i w^{(i)}_{\mathrm{SOM}}\langle R\rangle^{(i)} n^{(i)}_{\mathrm{spec}}(z_{\mathrm{s}}),
\end{split}
\end{equation}
where $i$ denotes the $i$th SOM cell. The resulting redshift distribution is shown in Fig. \ref{fig:nz}. Note that the procedure above assumes that UNIONS galaxies in the W3 field are representative of the entire sample. 

\subsection{BOSS Foreground Lens Galaxy Sample}
\label{subsec:cmass}
We use BOSS DR12 CMASS galaxies as our foreground spectroscopic lens sample. The corresponding catalogues are publicly available through the Science Archive Server\footnote{\url{https://data.sdss.org/sas}} (SAS), within the LSS directories. BOSS is a spectroscopic redshift survey within the Sloan Digital Sky Survey (SDSS; \citealp{2000AJ....120.1579Y})-III programme\footnote{\url{https://sdss3.org}}, designed to measure the Baryon Acoustic Oscillation (BAO) scale though the clustering of matter \citep{2013AJ....145...10D}. Data is collected with the 2.5-meter Sloan Foundation Telescope \citep{2006AJ....131.2332G} at Apache Point Observatory in New Mexico, and spectroscopy is performed with the `BOSS' spectrographs. The CMASS sample primarily targets LRGs with approximately Constant stellar MASS ($\mathrm{log}_{10}\,(M_*/M_{\odot})\approx{11.2\text{--}11.4}$; \citealp{Sonnenfeld_2019}), hence the term `CMASS'. Readers are referred to \cite{10.1093/mnras/stv2382} for a more detailed description of the CMASS sample and targeting strategies. 

\begin{figure}
\includegraphics[width = \columnwidth]{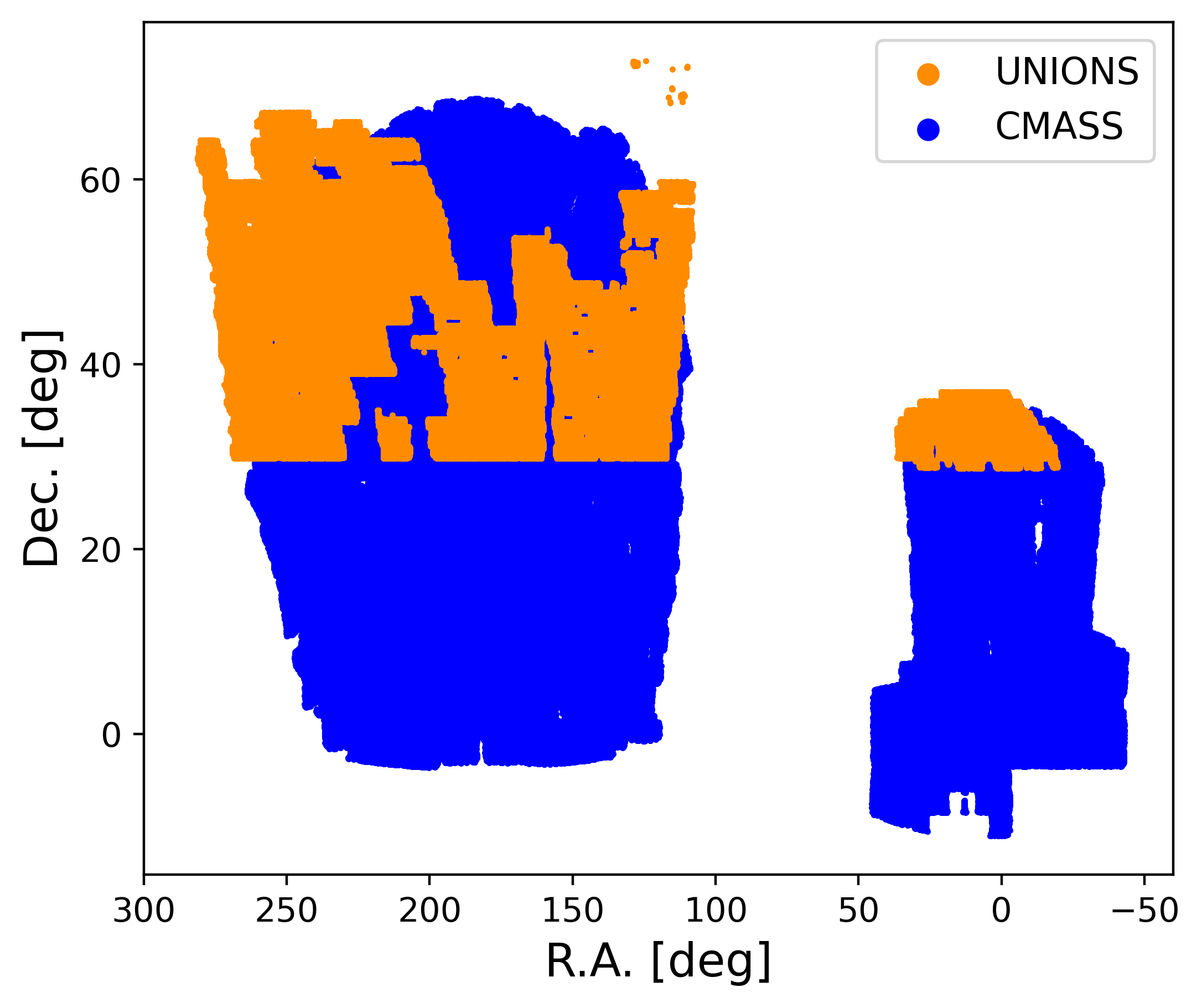}
\caption{On-sky overlap between UNIONS \textsc{ShapePipe} v1.4.5 galaxies (orange) and CMASS galaxies in the redshift range $0.43<z<0.7$ (blue). The overlap area is approximately 2650 square degrees.}
\label{fig:sky}
\end{figure}

For our analysis, we merge the CMASS catalogues from the NGC and SGC into a single sample. Following previous work, we apply a redshift cut of $0.43<z<0.7$ to the full catalogue, without further subdivision into tomographic bins. The final catalogue contains 777,202 galaxies over a total area of 10,252 square degrees \citep{10.1093/mnras/stv2382}. 273,662 of these galaxies enter our GGL calculation, overlapping with UNIONS over approximately 2650 square degrees. The angular overlap between surveys is shown in Fig. \ref{fig:sky}. We also use the NGC and SGC random catalogues provided by SDSS, merging them into a single sample and applying the same redshift cut. The final random catalogue is downsampled to contain 30 times more objects than the galaxy catalogue, enabling a faster computation of correlation functions. 

\section{Measurements}
\label{sec:measurements}
This section details our GGL and GC measurements. Although the CMASS GC signal has been extensively measured, we perform our own measurements to estimate its covariance matrix via jackknife resampling \citep{5685f092-0b53-33cc-a858-e3a8f62b8477,Quenouille_1956,Tukey_1958}. 

\subsection{Excess Surface Density, $\bf{\Delta\Sigma}$}
\label{subsec:esd}
We measure the GGL signal around CMASS galaxies using \textsc{dsigma}\footnote{\url{https://github.com/johannesulf/dsigma}} \citep{dsigma}. The GGL signal is formally defined as the mean excess surface mass density (ESD) around a set of foreground lens galaxies:
\begin{equation}
\Delta\Sigma(r_{\mathrm{p}})\,=\,\overline{\Sigma({<}r_{\mathrm{p}})}\,-\,\langle \Sigma(r_{\mathrm{p}})\rangle.
\label{eqn:dsigma_def}
\end{equation}
The first term in the expression above is the mean surface density around lens galaxies inside a circle of projected radius $r_{\mathrm{p}}$. The second term is the expected surface density at $r_{\mathrm{p}}$. 

In practice, we cannot measure the ESD around real galaxies using Eq. \ref{eqn:dsigma_def} since dark matter is not directly observable. We can, however, infer the ESD by measuring the shapes of background source galaxies. The mass distribution of a lens system induces a tangential shear, $\gamma_{\mathrm{t}}$, in the images of background objects. This shear is directly related to the ESD via:
\begin{equation}
\gamma_{\mathrm{t}}(r_{\mathrm{p}})\,=\,\Sigma^{-1}_{\mathrm{crit}}(z_{\mathrm{l}},\,z_{\mathrm{s}})\Delta\Sigma(r_{\mathrm{p}}),
\label{eqn:dsigma_shear}
\end{equation} 
where $z_{\mathrm{l}}$ is the lens galaxy redshift, and $\Sigma_{\mathrm{crit}}$---the critical surface density---encodes the geometry of a lens--source pair. This factor ensures consistent GGL measurements regardless of the chosen background source galaxies, whose distances affect the observed shear. The critical surface density is given by:
\begin{equation}
\Sigma_{\mathrm{crit}}(z_{\mathrm{l}},\,z_{\mathrm{s}})\,=\,\frac{c^2}{4 \pi G}\frac{D_{\mathrm{A}}(z_{\mathrm{s}})}{D_{\mathrm{A}}(z_{\mathrm{l}})D_{\mathrm{A}}(z_{\mathrm{l}},\,z_{\mathrm{s}})},
\end{equation}
where $D_{\mathrm{A}}$ is an angular diameter distance. 

Since our source galaxies lack photometric redshifts, we compute an effective $\Sigma^{-1}_{\mathrm{crit}}$ for each lens galaxy by averaging over the $n(z_{\mathrm{s}})$:
\begin{equation}
\Sigma^{-1}_{\mathrm{crit,\,eff}}(z_{\mathrm{l}})\,=\,\int{\Sigma^{-1}_{\mathrm{crit}}}(z_{\mathrm{l}},\,z_{\mathrm{s}})n(z_{\mathrm{s}})\mathrm{d}z_{\mathrm{s}}.
\end{equation}
The procedure above has been used in many analyses where photometric redshifts were either unreliable or unavailable \citep{2004AJ....127.2544S,2005MNRAS.361.1287M,li2024blackholetohalomassrelationunions,2025ApJ...992..171C}. Thus, our estimator of the mean ESD around many lens galaxies is:
\begin{equation}
\Delta\Sigma(r_{\mathrm{p}})\,=\,\frac{\sum_{l,s}w_lw_{l,s}\Sigma_{\mathrm{crit,\,eff}}(z_{\mathrm{l}})\epsilon_{\mathrm{t}}}{\sum_{l,s}w_lw_{l,s}},
\label{eqn:estimator}
\end{equation}
where the sum runs over all lens--source galaxy pairs in a given $r_{\mathrm{p}}$ bin, $w_l$ are the lens galaxy systematic weights ($w_{\mathrm{tot}}$ in Section \ref{subsec:proj_corr}), $w_{l,s}$ are lens--source pair weights, and $\epsilon_{\mathrm{t}}$ is the tangential ellipticity. The tangential ellipticity is a good approximation of $\gamma_{\mathrm{t}}$ when many source galaxies are stacked \citep{Prat_2026}. The lens--source pair weights are computed as follows:
\begin{equation}
w_{l,s}\,=\,w_{s}\Sigma^{-2}_{\mathrm{crit,\,eff}}(z_{\mathrm{l}}).
\label{eqn:pair}
\end{equation}

\begin{figure}
\includegraphics[width = \columnwidth]{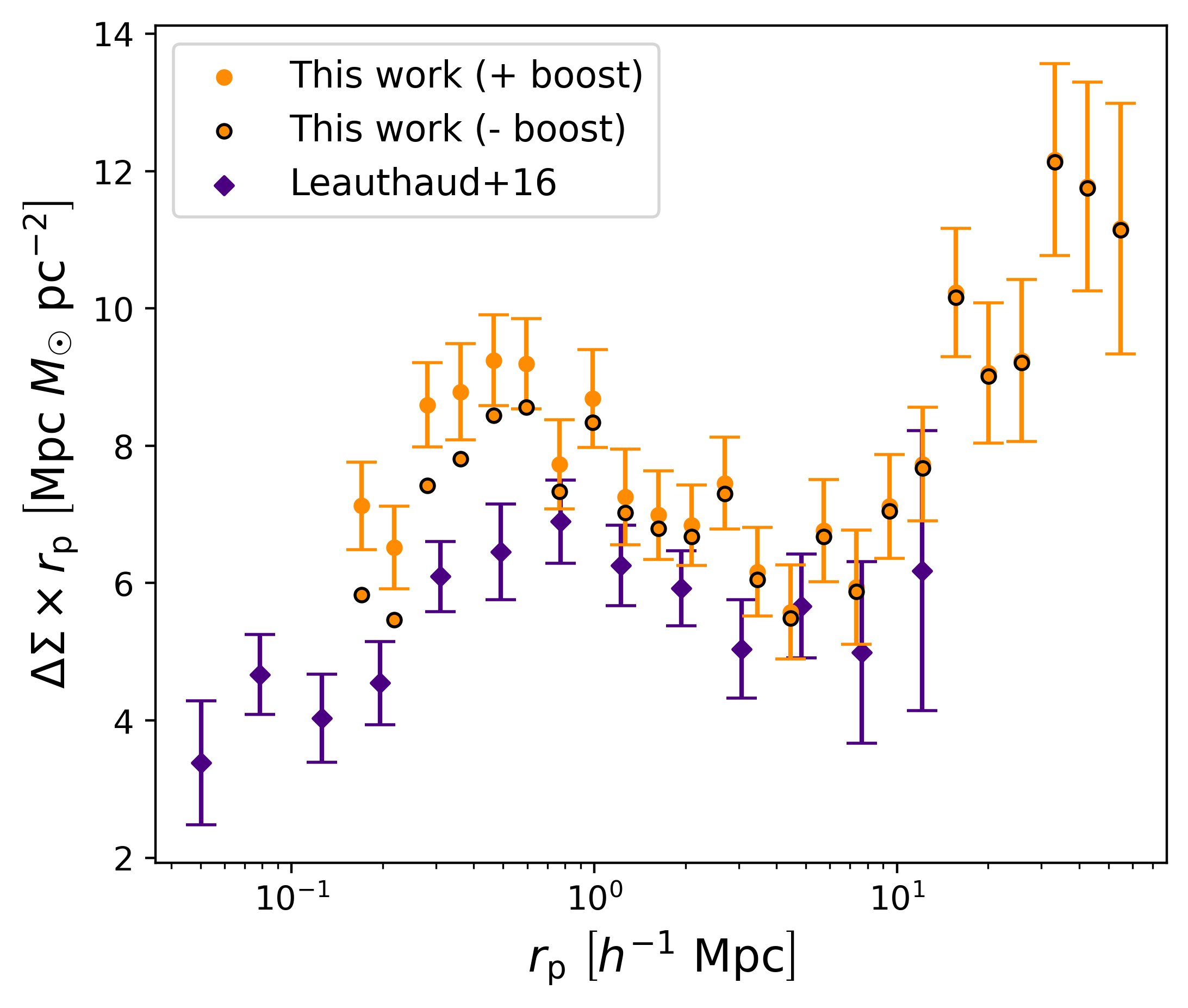}
\caption{CMASS GGL signal. Orange circles show our measurements using UNIONS \textsc{ShapePipe} v1.4.5 source galaxies with no correction for $m$-bias. Purple diamonds show results from 
\protect\citetalias{lilo17} using sources from the CFHT Stripe 82 Survey (CS82; 
\protect{\citealp{2014RMxAC..44..202M,2018MNRAS.479.1170C}}) and CFHTLenS. Orange circles with black borders indicate measurements without the boost factor. Only statistical $1\sigma$ errors are shown in this plot.}
\label{fig:esd}
\end{figure}

The estimator in Eq. \ref{eqn:estimator} is further refined through two additional corrections: random subtraction and the boost factor, $b$ \citep{dsigma}. In theory, the GGL signal around random objects should be consistent with zero. However, the survey geometry and residual systematics can result in a non-zero signal. Subtracting the GGL signal around randoms from that around `true' lenses can mitigate such systematics. On the other hand, the boost factor accounts for contamination from physically associated lens--source pairs, which dilutes the GGL signal. This correction is particularly important in our case due to the significant overlap in redshift between lens and source galaxies, as shown in Fig. \ref{fig:nz}. With these corrections, the GGL signal becomes: 
\begin{equation}
\Delta\Sigma(r_{\mathrm{p}})\,=\,b(r_{\mathrm{p}})\Delta\Sigma_{\mathrm{true}}(r_{\mathrm{p}})\,-\,\Delta\Sigma_{\mathrm{rand}}(r_{\mathrm{p}}),
\end{equation} 
where $b$ is given by: 
\begin{equation}
b(r_{\mathrm{p}})\,=\,\frac{\sum_{l,s}w_lw_{l,s}}{\sum_{r,s}w_rw_{r,s}}.
\end{equation}
In the expression above, $r$ denotes randoms, and $w_{r,s}$ are computed following Eq. \ref{eqn:pair}. The boost factor quantifies the degree of physical associations by computing the excess of sources near lenses relative to randoms. In the absence of physical associations, we do not expect correlations between lens and source positions. Our measurement (with these corrections) of the GGL signal around CMASS galaxies is shown in Fig. \ref{fig:esd}.

A non-negligible systematic in this analysis is the uncertainty in $n(z_{\mathrm{s}})$. Internal tests using image simulations show that for our source catalogue, the SOM method in Section \ref{subsec:unions} overestimates the mean of $n(z_{\mathrm{s}})$ by $0.029\pm0.016$. Following the procedure in \cite{Bridle_2002}, we analytically marginalise over the $n(z_{\mathrm{s}})$ uncertainty, thereby introducing an additive component to our covariance matrix. Specifically, this procedure adds positive correlations, as biases in the mean of $n(z_{\mathrm{s}})$ systematically shift the GGL signal up or down. This approach avoids adding a nuisance parameter to our theoretical model for $\Delta\Sigma$, keeping the parameter space minimal. We summarize the methodology below. 

We first compute $\Delta\Sigma$ using the original $n(z_{\mathrm{s}})$ shifted by $\delta z=-0.029$. This defines our fiducial $\Delta\Sigma$ measurement data vector, $\bm{d}_{\mathrm{fid}}$, which is used in parameter inference. Then, we recompute $\Delta\Sigma$ using $n(z_{\mathrm{s}})$ shifted by $\pm1\sigma$ of the SOM bias, yielding data vectors $\bm{d}_+$ and $\bm{d}_-$, respectively. We now define:
\begin{equation}
\bm{\Delta}\,=\,\frac{\bm{d}_--\bm{d}_+}{2},
\end{equation}
which represents the $1\sigma$ shift in $\bm{d}_{\mathrm{fid}}$ due to residual $n(z_{\mathrm{s}})$ uncertainties. Following Eq. 8 in \cite{Bridle_2002}, we compute the final covariance matrix as follows:
\begin{equation}
\bm{C}_{\mathrm{final}}\,=\,\bm{C}(\bm{d}_{\mathrm{fid}})\,+\,\bm{\Delta}\bm{\Delta}^{\mathrm{T}},
\end{equation}
where $\bm{C}(\bm{d}_{\mathrm{fid}})$ is the statistical covariance matrix described below. Uncertainties from this systematic are at the $5\%$ level. 

\begin{figure}
\includegraphics[width = \columnwidth]{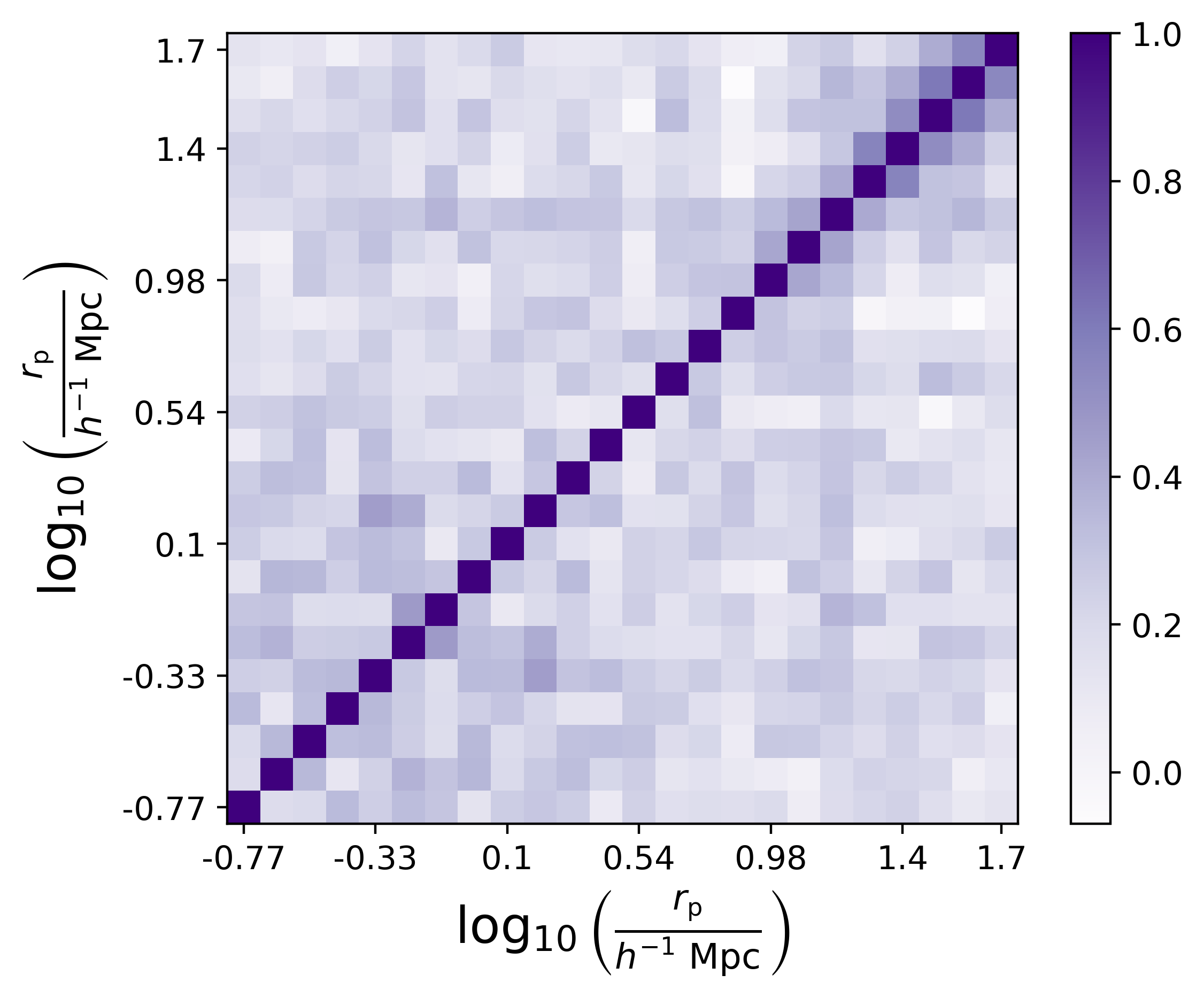}
\caption{Correlation matrix computed from the $\Delta\Sigma$ covariance matrix. By construction, every matrix element must have a value between -1 and 1. Small-scale $r_{\mathrm{p}}$ bins are uncorrelated with other bins, as shape noise dominates on these scales. Large-scale bins are mildly correlated due to LSS.}
\label{fig:esd_corr}
\end{figure}

We estimate the GGL signal's statistical covariance matrix using 150 jackknife resamplings. Jackknifing is built into \textsc{dsigma}, and the underlying code constructs patches such that they cover roughly the same on-sky area. The jackknife covariance matrix is given by:
\begin{equation}
{C}_{\mathrm{JK}}(\hat{f}_i,\,\hat{f}_j)\,=\,\frac{N_{\mathrm{JK}}\,-\,1}{N_{\mathrm{JK}}}\sum^{N_{\mathrm{JK}}}_{k\,=\,1}\left(\hat{f}^{(k)}_i\,-\,\overline{\hat{f}}_{i}\right)\left(\hat{f}^{(k)}_j\,-\,\overline{\hat{f}}_{j}\right),
\label{eqn:jk}
\end{equation}
where $\hat{f}$ is an estimator (here, our $\Delta\Sigma$ estimator), $(i,j)$ represents an element of the covariance matrix, $N_{\mathrm{JK}}$ is the number of jackknife patches, $\hat{f}^{(k)}_{i/j}$ is the estimator's evaluation when omitting the $k$th jackknife region, and $\overline{\hat{f}}_{i/j}$ is the mean evaluation across all realisations. We choose $N_{\mathrm{JK}}=150$ to obtain reliable covariance estimates at large scales, while preserving error estimates and correlation structure observed with larger $N_{\mathrm{JK}}$. The $\Delta\Sigma$ jackknife covariance matrix can be seen in Fig. \ref{fig:block_cov}. The corresponding correlation matrix is shown in Fig. \ref{fig:esd_corr}. 

Since $\bm{C}_{\mathrm{JK}}$ is noisy, its inverse is biased. To correct for this, we rescale $\bm{C}_{\mathrm{JK}}$ by the inverse Hartlap factor \citep{hartlap07}, $\mathcal{H}^{-1}$:
\begin{equation}
\bm{C}_{\mathrm{JK,\,final}}\,=\,\bm{C}_{\mathrm{JK}}\mathcal{H}^{-1}\,=\,\bm{C}_{\mathrm{JK}}\times\frac{N_{\mathrm{JK}}\,-\,1}{N_{\mathrm{JK}}\,-\,m\,-\,2},
\label{eqn:hartlap}
\end{equation}
where $m$ is the rank of the covariance matrix (here, the number of $r_{\mathrm{p}}$ points). We verified that applying the inverse Hartlap factor to a numerical covariance matrix, subsequently combined with additional covariances, yields an unbiased inverse covariance matrix.

We measure $\Delta\Sigma$ in 25 logarithmically spaced bins from $r_{\mathrm{p}}=0.15\text{--}62\,h^{-1}\,\mathrm{Mpc}$. The lower bound of our $r_{\mathrm{p}}$ bins does not extend below $0.1\,h^{-1}\,\mathrm{Mpc}$, as the \textsc{Dark Emulator} predictions start to become unreliable at this scale \citep{Nishimichi_2019}.

\subsection{Projected Correlation Function, $\bf{w_{\mathrm{p}}}$}
\label{subsec:proj_corr}
We measure the GC signal of CMASS galaxies using \textsc{pycorr}\footnote{\url{https://github.com/cosmodesi/pycorr}} \citep{10.1007/978-981-13-7729-7_1,2020MNRAS.491.3022S}. The GC signal, also known as the `projected correlation function', is defined as: 
\begin{equation}
w_{\mathrm{p}}(r_{\mathrm{p}})\,=\,\int_{-\pi_{\mathrm{max}}}^{\pi_{\mathrm{max}}}{\xi(r_{\mathrm{p}},\,r_{\pi})}\mathrm{d}r_{\pi},
\end{equation}
where $r_{\pi}$ are separation bins parallel to the line-of-sight (LOS), and $\xi(r_{\mathrm{p}},\,r_{\pi})$ is the 2D correlation function (2DCF), which measures the excess probability (relative to randoms) of finding a galaxy pair in a given $r_{\mathrm{p}}$, $r_{\pi}$ bin. Typically, the integration limit ($\pi_{\mathrm{max}}$) is selected to be large to minimize redshift space distortions (RSDs). We adopt $\pi_{\mathrm{max}}=100\,h^{-1}\,\mathrm{Mpc}$, which is greater than our largest $r_{\mathrm{p}}$ bin. To compute $\xi(r_{\mathrm{p}},\,r_{\pi})$, we use the Landy-Szalay minimum-variance estimator \citep{Landy_Szalay_1993}:
\begin{equation}
\xi(r_{\mathrm{p}},\,r_{\pi})\,=\,\frac{DD(r_{\mathrm{p}},\,r_{\pi})\,-\,2DR(r_{\mathrm{p}},\,r_{\pi})\,+\,RR(r_{\mathrm{p}},\,r_{\pi})}{RR(r_{\mathrm{p}},\,r_{\pi})}.
\label{eqn:2d_corr}
\end{equation} 
$DD$, $DR$, and $RR$ denote data--data (galaxy--galaxy), data--random, and random--random (normalised) pair counts, respectively. 

When performing the pair counts in Eq. \ref{eqn:2d_corr}, we apply the following BOSS weights to each galaxy: 
\begin{equation}
w_{\mathrm{tot}}\,=\,w_{\mathrm{FKP}}w_{\mathrm{sys}}(w_{\mathrm{zf}}\,+\,w_{\mathrm{cp}}\,-\,1).
\end{equation}
The Feldman-Kaiser-Peacock (FKP; \citealp{1994ApJ...426...23F}) weights account for spatial variations in galaxy number density and improve the clustering S/N, while the systematic weights correct for seeing and stellar contamination \citep{10.1093/mnras/stv2382}. The weights $w_{\mathrm{zf}}$ and $w_{\mathrm{cp}}$ correct for redshift failures and close-pairs, respectively. In BOSS, spectroscopy is performed using fibres plugged into plates and centred on target objects \citep{2013AJ....145...10D}. However, these fibres subtend an angle of 62", so any two galaxies separated by less than this `fibre collision' scale cannot be observed simultaneously. Such galaxies are said to form a `close pair'. If the missed galaxies remain unobserved in subsequent passes, $w_{\mathrm{p}}(r_{\mathrm{p}})$ will be underestimated. Approximately $5.8\%$ of DR12 CMASS targets are affected by fibre collisions \citep{10.1093/mnras/stv2382}. The redshift-failure and close-pair weights are constructed using a `nearest-neighbour' up-weighting scheme (i.e., if a galaxy is missed, its nearest neighbour on the sky receives a weight of +1), and their default values are unity. Note that randoms are only assigned FKP weights. 

The nearest-neighbour close-pair weights successfully recover the underlying clustering statistics on intermediate to large scales, but are insufficient near or below the fibre collision scale \citep{2012ApJ...756..127G,2017MNRAS.472.1106B}. So, we use `angular up-weights' \citep{Hawkins_2003} in place of the BOSS close-pair weights on small scales. The angular up-weights are given by: 
\begin{equation}
w_{\mathrm{pair}}(\theta)\,=\,\frac{1\,+\,w_{\mathrm{t}}(\theta)}{1\,+\,w_{\mathrm{s}}(\theta)},
\end{equation}
where $w_{\mathrm{t}}(\theta)$ and $w_{\mathrm{s}}(\theta)$ are the angular correlation functions (which do not require spectroscopic information) of target and spectroscopic samples, respectively. The spectroscopic sample is the final galaxy catalogue (i.e., DR12 CMASS catalogues on the SAS), while the target sample contains the spectroscopic sample plus galaxies with missing redshifts due to fibre collisions. Below, we detail our methodology for constructing the target sample. 

We begin with the BOSS DR12 target list, and identify CMASS targets using a bit-field mask\footnote{See \url{https://www.sdss4.org/dr17/tutorials/lss_galaxy}.}. Next, we generate `TARGETOBJIDs\footnote{TARGETOBJIDs are unique object IDs, combining the following fields: RUN, CAMCOL, FIELD, ID, and RERUN. Conversion between these fields and TARGETOBJID is handled by \textsc{sdsspy} \citep{sdsspy}.}' for each object in the spectroscopic sample and in the CMASS target list. We then determine which CMASS targets did not receive a fibre due to membership in a collision group using the `ICOLLIDED' flag\footnote{See \url{https://data.sdss.org/datamodel/files/BOSS_LSS_REDUX/bosstile-final-collated-boss2-bossN.html}.}. After, we iterate through all CMASS targets and keep those whose IDs (a) appear in the spectroscopic sample, and (b) are missing from the `specObj\footnote{The specObj file contains spectroscopic information. See \url{https://data.sdss.org/datamodel/files/SPECTRO_REDUX/specObj.html}.}' file and have ICOLLIDED = 1. By keeping the targets that are absent from this file and have ICOLLIDED = 1, we recover the galaxies missed to fibre collisions rather than redshift failures in the spectroscopic pipeline. Finally, we apply the masks used on galaxies in the spectroscopic sample to our final CMASS targets. Applying these masks ensures consistency in the selection function between our two samples. These masks account for the survey geometry, regions around bright stars, tiling completeness, etc.

A caveat of the angular up-weighting method is its assumption that missed galaxy pairs have the same radial properties as spectroscopically observed pairs, which is not strictly valid \citep{2017MNRAS.472.1106B}. Moreover, $w_{\mathrm{pair}}(\theta)$ approaches unity on large scales, necessitating an alternative correction scheme. When angular up-weights are used, the galaxy weights become:
\begin{equation}
\tilde{w}_{\mathrm{tot}}\,=\,w_{\mathrm{FKP}}w_{\mathrm{sys}}w_{\mathrm{zf}}.
\end{equation}

The pair counts in Eq. \ref{eqn:2d_corr} satisfy the following relationships: 
\begin{equation}
DD(r_{\mathrm{p}},\,r_{\pi})\,\propto\,\sum_{i \neq j}w^{(i)}_Dw^{(j)}_D,
\label{eq:dd}
\end{equation}
\vspace{-1mm}
\begin{equation}
DR(r_{\mathrm{p}},\,r_{\pi})\,\propto\,\sum_{i,j}w^{(i)}_Dw^{(j)}_R,
\label{eq:dr}
\end{equation}
\vspace{-1mm}
\begin{equation}
RR(r_{\mathrm{p}},\,r_{\pi})\,\propto\,\sum_{i \neq j}w^{(i)}_Rw^{(j)}_R,
\label{eq:rr}
\end{equation}
where $w$ denotes the weight of a galaxy or a random in a pair. Note that the expressions above are not normalised. With our weighting scheme, the summands in Eqs. \ref{eq:dd}, \ref{eq:dr}, and \ref{eq:rr} are: 
\begin{equation}
w^{(i)}_Dw^{(j)}_D\,=\,
\begin{cases}
\tilde{w}^{(i)}_{\mathrm{tot}}\tilde{w}^{(j)}_{\mathrm{tot}}w_{\mathrm{pair}} & (r_{\mathrm{p}}\,\leq\,r_{\mathrm{switch}})
\\
w^{(i)}_{\mathrm{tot}}w^{(j)}_{\mathrm{tot}} & (r_{\mathrm{p}}\,>\,r_{\mathrm{switch}})
\end{cases},
\end{equation}
\vspace{-1mm}
\begin{equation}
w^{(i)}_Dw^{(j)}_R\,=\,
\begin{cases}
\tilde{w}^{(i)}_{\mathrm{tot}}w^{(j)}_{\mathrm{FKP,\,rand}} & (r_{\mathrm{p}}\,\leq\,r_{\mathrm{switch}})
\\
w^{(i)}_{\mathrm{tot}}w^{(j)}_{\mathrm{FKP,\,rand}} & (r_{\mathrm{p}}\,>\,r_{\mathrm{switch}})
\end{cases},
\end{equation}
\vspace{-1mm}
\begin{equation}
w^{(i)}_Rw^{(j)}_R\,=\,w^{(i)}_{\mathrm{FKP,\,rand}}w^{(j)}_{\mathrm{FKP,\,rand}}.
\end{equation}
The parameter $r_{\mathrm{switch}}$ is the $r_{\mathrm{p}}$ value at which $w_{\mathrm{p}}$ measurements with and without angular up-weights intersect, and we find it to be $\sim0.7\,h^{-1}\,\mathrm{Mpc}$. 

\begin{figure}
\includegraphics[width = \columnwidth]{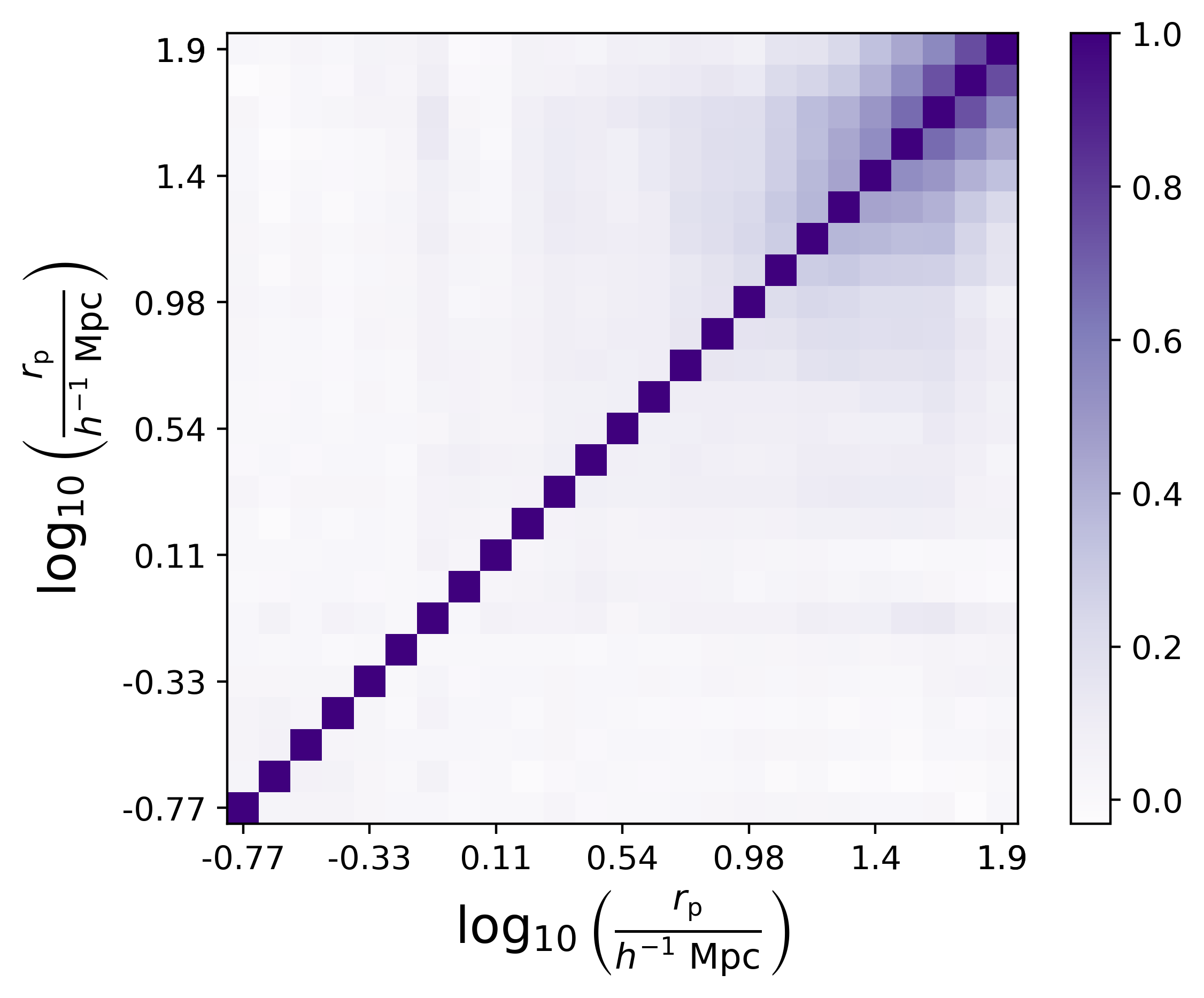}
\caption{Correlation matrix computed from the $w_{\mathrm{p}}$ covariance matrix. Small-scale $r_{\mathrm{p}}$ bins are dominated by shot noise. Large-scale bins are strongly correlated due to LSS.}
\label{fig:wp_corr}
\end{figure}

To estimate the GC signal's statistical covariance matrix, we again use the jackknife resampling technique (which is implemented in \textsc{PYCORR}) with 150 patches. The resulting covariance matrix is scaled by $\mathcal{H}^{-1}$, following the previous section. The $w_{\mathrm{p}}$ jackknife covariance matrix can be seen in Fig. \ref{fig:block_cov}, with its correlation matrix shown in Fig. \ref{fig:wp_corr}. 

The final $w_{\mathrm{p}}$ covariance matrix used for parameter estimation includes additional sources of uncertainty. First, we associate a $10\%$ uncertainty with our measurement of $w_{\mathrm{t}}$, following \cite{Reid_2014}. To propagate this uncertainty, we recompute $w_{\mathrm{p}}$ with $\pm10\%$ variations in $w_{\mathrm{t}}$, and use the differences in $w_{\mathrm{p}}$ from the original measurement to derive error bars from the angular up-weighting method. Second, we assign a 3$\%$ uncertainty to the \textsc{Dark Emulator} predictions of $w_{\mathrm{p}}$ (estimated from \citealp{Nishimichi_2019}). We do not include a model uncertainty for $\Delta\Sigma$, as it is small compared to the statistical uncertainties. The squares of these additional errors are added to the diagonal elements of the rescaled jackknife covariance matrix. We measure $w_{\mathrm{p}}$ in 26 logarithmically spaced bins from $r_{\mathrm{p}}=0.15\text{--}80\,h^{-1}\,\mathrm{Mpc}$.

\section{Theory}
\label{sec:theory}
This section outlines our HOD framework and the theoretical expressions for GGL and GC. We also describe our extensions to the standard HOD framework, including central galaxy off-centring and feedback. 

\subsection{The Halo Occupation Distribution}
\label{subsec:hod}
HODs are empirical models that connect galaxies and dark matter halos through statistical prescriptions, bypassing the complexities of galaxy formation physics. They are designed to model the distribution of luminosity/flux/mass-limited galaxy samples, with each halo treated as distinct (i.e., no subhalos; \citealp{Zu_2015}). In this analysis, we adopt the vanilla HOD introduced by \cite{zheng05}. Within the HOD framework, each halo can host a central galaxy and a distribution of satellite galaxies. Central galaxies reside at the centre of the dark matter halo (although this condition can be relaxed, as we show later on), while satellite galaxies occupy the rest of the halo. Moreover, central galaxies tend to be more massive, luminous, and older compared to satellite galaxies in hydrodynamical simulations \citep{zheng05}.

According to the HOD, the mean number of galaxies in halos of mass $M$ is given by the sum of the expected numbers of central galaxies and satellite galaxies: 
\begin{equation}
\langle N_{\mathrm{g}}(M)\rangle\,=\,\langle N_{\mathrm{c}}(M)\rangle\,+\,\langle N_{\mathrm{s}}(M)\rangle.
\end{equation}
The expected number of central galaxies is parametrised as follows: 
\begin{equation}
\langle N_{\mathrm{c}}(M)\rangle\,=\,\frac{1}{2}\left(1\,+\,\mathrm{erf}\left[\frac{\mathrm{log}\,(M/M_{\mathrm{min}})}{\sigma_{\mathrm{log}\,(M)}}\right]\right),
\label{eqn:Nc}
\end{equation}
where $M_{\mathrm{min}}$ is the minimum mass of halos that can host a central galaxy above some luminosity threshold, assuming no scatter in the galaxy luminosity--halo mass relationship ($\sigma_{\mathrm{log}\,(M)}\to0$). It is also the value where $\langle N_{\mathrm{c}}(M)\rangle=0.5$. At the low-mass end, $\langle N_{\mathrm{c}}(M)\rangle$ goes to 0. Conversely, $\langle N_{\mathrm{c}}(M)\rangle$ asymptotes towards 1 at the high-mass end. The form of $\langle N_{\mathrm{c}}(M)\rangle$ originates from the assumption that central galaxy luminosity, $L_{\mathrm{c}}$, at fixed halo mass is Gaussian-distributed:
\begin{equation}
\begin{split}
P(\mathrm{log}\,(L_{\mathrm{c}})|M)\,=\,&\frac{1}{\sqrt{2\pi}\sigma_{\mathrm{log}\,(L)}}\times \\
&\mathrm{exp}\left(\frac{\left[\mathrm{log}\,(L_{\mathrm{c}})\,-\,\mathrm{log}\,(\langle L_{\mathrm{c}}(M)\rangle)\right]^2}{2\sigma^2_{\mathrm{log}\,(L)}}\right). 
\end{split}
\label{eqn:luminosity_dist}
\end{equation}

The expected number of satellite galaxies assumes the following form:
\begin{equation}
\langle N_{\mathrm{s}}(M)\rangle\,=\,\langle N_{\mathrm{c}}(M) \rangle\,\lambda_{\mathrm{s}}(M)\,=\,\langle N_{\mathrm{c}}(M) \rangle\left(\frac{M\,-\,\kappa\,M_{\mathrm{min}}}{M_1}\right)^{\alpha}.
\label{eqn:Ns}
\end{equation}
The equation above is valid on the domain $M>\kappa M_{\mathrm{min}}$. If $M\leq\kappa M_{\mathrm{min}}$, then $\langle N_{\mathrm{s}}(M)\rangle=0$. Thus, $\kappa M_{\mathrm{min}}$ acts as a truncation mass. The parameter $\alpha$ controls how steep (shallow) the slope of $\langle N_{\mathrm{s}}(M)\rangle$ is at the high (low) mass end, while $M_1$ vertically shifts $\langle N_{\mathrm{s}}(M)\rangle$ in logspace. The above parametrisation was found to capture the behaviour of satellite galaxies in smoothed particle hydrodynamics (SPH) simulations \citep{Zheng_2007}. 

Eqs. \ref{eqn:Nc} and \ref{eqn:Ns} compute the mean numbers of galaxies in halos of mass $M$. They do not, however, determine the exact number of galaxies to place in a given halo. We must also know the second moments of the occupation numbers ($N_{\mathrm{c}}$ and $N_{\mathrm{s}}$) to model the two-point correlation function (2PCF) of galaxies \citep{Zheng_2007}. These points motivate defining probability distributions for the numbers of central and satellite galaxies. We assign the following distributions: central galaxies are drawn from a Bernoulli distribution with mean $\langle N_{\mathrm{c}}(M)\rangle$, and satellite galaxies are drawn from a Poisson distribution with mean $\lambda_{\mathrm{s}}(M)$:
\begin{equation}
P(N_{\mathrm{s}})\,=\,(\lambda_{\mathrm{s}})^{N_{\mathrm{s}}}\,\frac{\mathrm{exp}(-\lambda_{\mathrm{s}})}{N_{\mathrm{s}}!}.
\end{equation}
The assumption that $N_{\mathrm{s}}$ follows a Poisson distribution is consistent with theoretical predictions \citep{Zheng_2007}. As in \cite{Miyatake_2022}, we impose a `central condition' such that satellite galaxies cannot occupy a halo without a central galaxy\footnote{\cite{ChavesMontero23} show that the central condition can generate a slight lensing is low tension on scales below $1\,h^{-1}\,\mathrm{Mpc}$.}; i.e., $P(N_{\mathrm{s}}=0|N_{\mathrm{c}}=0)=1$. This condition is often imposed since central galaxies typically form before satellite galaxies in hydrodynamical simulations. 

\subsection{Theoretical Two-Point Correlation Functions}
\label{subsec:2pcf}
Using both the halo model and the HOD framework, we can formulate theoretical expressions for our observables of interest: $\Delta\Sigma$ and $w_{\mathrm{p}}$. We first define the 3D galaxy--matter power spectrum, $P_{\mathrm{gm}}$: 
\begin{equation}
\begin{split}
P_{\mathrm{gm}}(k,\,z)\,=\,&\frac{1}{\overline{n}_{\mathrm{g}}}\int[\langle N_{\mathrm{c}}(M) \rangle[1\,-\,p_{\mathrm{off}}\,+\,p_\mathrm{off}\tilde{u}_{\mathrm{off}}(k|M,\,z)]\,+\\
&\langle N_{\mathrm{s}}(M) \rangle\tilde{u}_{\mathrm{s}}(k|M,\,z)]\times P_{\mathrm{hm}}(k,\,M,\,z)\times \\
&\frac{\mathrm{d}n_{\mathrm{h}}}{\mathrm{d}M}\mathrm{d}M,
\end{split}
\end{equation}
where $k$ is a given wavenumber, $\overline{n}_{\mathrm{g}}$ is the mean number density of galaxies, $p_{\mathrm{off}}$ is the fraction of central galaxies that are offset from the halo centre, and $\tilde{u}_{\mathrm{off}}$ and $\tilde{u}_{\mathrm{s}}$ are the Fourier transforms of the normalised radial profiles of off-centred central galaxies and satellite galaxies, respectively. We select a Navarro--Frenk--White (NFW; \citealp{Navarro_1996}) profile for the radial distribution of satellite galaxies, and allow the concentration to be modified by a multiplicative factor $R_{\mathrm{c}}$. The default satellite galaxy concentration is given by the halo concentration, computed with \textsc{Colossus}\footnote{\url{https://github.com/Shiriny24/colossus}} \citep{2018ApJS..239...35D}. From here on, we refer to $R_{\mathrm{c}}$ as the `satellite concentration normalisation', and its fiducial value is 1. $P_{\mathrm{hm}}$ is the 3D halo--matter cross--power spectrum (including 1- and 2-halo terms), and $\mathrm{d}n_{\mathrm{h}}/\mathrm{d}M$ is the halo mass function (HMF). We compute $\overline{n}_{\mathrm{g}}$ via:
\begin{equation}
\overline{n}_{\mathrm{g}}\,=\,\int(\langle N_{\mathrm{c}}(M) \rangle\,+\,\langle 
N_{\mathrm{s}}(M)\rangle)\frac{\mathrm{d}n_{\mathrm{h}}}{\mathrm{d}M}\mathrm{d}M.
\end{equation}
Following \cite{More_2015}, \cite{2020PhRvD.102f3504K}, and \cite{Miyatake_2022}, $\tilde{u}_{\mathrm{off}}$ is given by:
\begin{equation}
\tilde{u}_{\mathrm{off}}(k|M,\,z)\,=\,\exp\left(-\frac{1}{2}k^2(R_{\mathrm{200m}}R_{\mathrm{off}})^2\right),
\label{eqn:off-centring}
\end{equation}
where $R_{\mathrm{200m}}$ follows the same definition as $R_{\mathrm{200c}}$ in Section \ref{subsec:feedback}, but is defined with respect to the mean matter density, $\overline{\rho}_{\mathrm{m}}$. The expression above is the Fourier transform of a Gaussian with width set by the dimensionless parameter $R_{\mathrm{off}}$. This translates to a mean offset radius of $2\sqrt{2/\pi}R_{\mathrm{200m}}R_{\mathrm{off}}$. Finally, we compute the GGL signal via: 
\begin{equation}
\Delta\Sigma(r_{\mathrm{p}},\,z)\,=\,\overline{\rho}_{\mathrm{m},0}\int_0^{\infty}\frac{k^2}{2\pi^2}P_{\mathrm{gm}}(k,\,z)J_2(kr_{\mathrm{p}})\mathrm{d}k,
\end{equation}
where $J_2$ is the 2nd-order Bessel Function of the first kind. 

We account for the impact of lens magnification on the GGL signal at the model level. Briefly, intervening LSS between an observer and a lens sample magnifies the lenses while inducing a shear in background sources \citep{Unruh_2020}. This magnification produces two competing effects. First, the local stretching of the sky reduces the observed lens number density. However, lenses whose fluxes previously fell just below the survey's detection threshold can now be observed due to the conservation of surface brightness \citep{Prat_2022}. This, in turn, increases the lens number density. The change in lens number density coupled with stronger shears along preferential directions biases GGL measurements, as GGL correlates the gravitationally lensed shapes of background galaxies with the lens galaxy density field \citep{Lamman_2024}. Whether the observed number density increases or decreases relative to the true number density depends on the particular lens sample---specifically, on the number of lenses just below the survey's flux limit. 

We model the lens magnification contribution following \cite{Unruh_2020}, where the observed $\gamma_{\mathrm{t}}$ is given by:
\begin{equation}
\begin{split}
\gamma_{\mathrm{t}}(\theta|z_{\mathrm{l}},\,z_{\mathrm{s}})\,=\,&\gamma^{\mathrm{nomagn}}_{\mathrm{t}}(\theta|z_{\mathrm{l}},\,z_{\mathrm{s}})\,+\,2(\alpha_{\mathrm{l}}(z_{\mathrm{l}})\,-\,1)\times \\
&\gamma^{\mathrm{LSS}}_{\mathrm{t}}(\theta|z_{\mathrm{l}},\,z_{\mathrm{s}}).
\label{eqn:gamma_t_tot}
\end{split}
\end{equation}
Here, $\theta$ is the angular version of $r_{\mathrm{p}}$, $\gamma^{\mathrm{nomagn}}_{\mathrm{t}}$ denotes the tangential shear without lens magnification, $\gamma^{\mathrm{LSS}}_{\mathrm{t}}$ is the tangential shear from LSS, and $\alpha_{\mathrm{l}}$ is the slope of the lens number counts at the flux limit. We set $\alpha_{\mathrm{l}}\approx2.98$, using a redshift-weighted average of the measurements in \cite{krolewski20}. The factor of $2(\alpha_{\mathrm{l}}(z_{\mathrm{l}})-1)$ arises from the weak-lensing approximation $\mu\approx1+2\kappa$. If $\alpha_{\mathrm{l}}>1$, the lens number counts are enhanced. Conversely, they are suppressed when $\alpha_{\mathrm{l}}<1$. There is no lens magnification bias when $\alpha_{\mathrm{l}}=1$. 

The LSS tangential shear is computed via: 
\begin{equation}
\begin{split}
\gamma^{\mathrm{LSS}}_{\mathrm{t}}(\theta|z_{\mathrm{l}},\,z_{\mathrm{s}})\,=\,&\frac{9H^3_0\Omega^2_{\mathrm{m,0}}}{8\pi c^3}\int_0^\infty\mathrm{J}_2(\ell\theta)\ell\mathrm{d}\ell\int_0^{z_{\mathrm{l}}}(1\,+\,z)^2\times \\
&\frac{H_0}{H(z)}\times\frac{D_{\mathrm{A}}(z,\,z_{\mathrm{l}})D_{\mathrm{A}}(z,\,z_{\mathrm{s}})}{D_{\mathrm{A}}(z_{\mathrm{l}})D_{\mathrm{A}}(z_{\mathrm{s}})}\times \\
&P_{\mathrm{mm}}\left(\frac{\ell\,+\,1/2}{D_{\mathrm{A}}(z)[1\,+\,z]};\,z\right)\mathrm{d}z,
\end{split}
\label{eqn:gamma_t_lss}
\end{equation}
where $P_{\mathrm{mm}}$ is the 3D matter power spectrum, which we compute using \textsc{CAMB}\footnote{\url{https://github.com/cmbant/CAMB}}$^,$\footnote{Includes the \textsc{HaloFit} non-linear corrections.} \citep{Lewis:1999bs,Howlett:2012mh}. The expression above is derived from the convergence ($\kappa$) power spectrum, related to $P_{\mathrm{mm}}$ via the Limber approximation \citep{1953ApJ...117..134L,2017MNRAS.472.2126K}. We evaluate Eq. \ref{eqn:gamma_t_lss} using the weighted mean redshifts of lens and source galaxies from our catalogues. After scaling $\gamma^{\mathrm{LSS}}_{\mathrm{t}}$ by the factor in Eq. \ref{eqn:gamma_t_tot}, we multiply it by the weighted mean $\Sigma_{\mathrm{crit}}$ obtained from \textsc{dsigma}. Lastly, we add this term to our $\Delta\Sigma$ model predictions. 

To compute $w_{\mathrm{p}}$, we first define the 3D galaxy power spectrum, $P_{\mathrm{gg}}$:
\begin{equation}
P_{\mathrm{gg}}(k,\,z)\,=\,P^{\mathrm{1h}}_{\mathrm{gg}}(k,\,z)\,+\,P^{\mathrm{2h}}_{\mathrm{gg}}(k,\,z).
\end{equation}
The 1- and 2-halo terms are calculated as follows: 
\begin{equation}
\begin{split}
P^{\mathrm{1h}}_{\mathrm{gg}}(k,\,z)\,=\,&\frac{1}{\overline{n}^2_{\mathrm{g}}}\int[2\lambda_{\mathrm{s}}(M)\tilde{u}_{\mathrm{s}}(k|M,\,z)[1\,-\,p_{\mathrm{off}}\,+ \\
&p_{\mathrm{off}}\tilde{u}_{\mathrm{off}}(k|M,\,z)]+\,\lambda_{\mathrm{s}}^2(M)\tilde{u}_{\mathrm{s}}^2(k|M,\,z)]\times \\
&\langle N_{\mathrm{c}}(M)\rangle\frac{\mathrm{d}n_{\mathrm{h}}}{\mathrm{d}M}\mathrm{d}M,
\end{split}
\end{equation}
\vspace{-1mm}
\begin{equation}
\begin{split}
P^{\mathrm{2h}}_{\mathrm{gg}}(k,\,z)\,=\,&\frac{1}{\overline{n}^2_{\mathrm{g}}}\int[1\,-\,p_{\mathrm{off}}\,+\,p_{\mathrm{off}}\tilde{u}_{\mathrm{off}}(k|M,\,z)\,+ \\
&\lambda_{\mathrm{s}}(M)\tilde{u}_{\mathrm{s}}(k|M,\,z)]\langle N_{\mathrm{c}}(M) \rangle\frac{\mathrm{d}n_{\mathrm{h}}}{\mathrm{d}M}\mathrm{d}M\times \\
&\int[1\,-\,p_{\mathrm{off}}\,+\,p_{\mathrm{off}}\tilde{u}_{\mathrm{off}}(k|M',\,z)\,+ \\
&\lambda_{\mathrm{s}}(M')\tilde{u}_{\mathrm{s}}(k|M',\,z)]\langle N_{\mathrm{c}}(M')\rangle\times \\
&\frac{\mathrm{d}n_{\mathrm{h}}}{\mathrm{d}M'}\mathrm{d}M'P_{\mathrm{hh}}(k|M,\,M',\,z).
\end{split}
\end{equation}
Now, we can convert $P_{\mathrm{gg}}$ into its configuration-space equivalent, $\xi_{\mathrm{gg}}$:
\begin{equation}
\xi_{\mathrm{gg}}(r,\,z)\,=\,\int_0^{\infty}\frac{k^2}{2\pi^2}P_{\mathrm{gg}}(k,\,z)j_0(kr)\mathrm{d}k,
\end{equation}
where $j_0$ is the 0th-order spherical Bessel Function. We finally compute the GC signal by integrating $\xi_{\mathrm{gg}}$ along the LOS:
\begin{equation}
w_{\mathrm{p}}(r_{\mathrm{p}},\,z)\,=\,2\int_0^{\pi_{\mathrm{max}}}{\xi_{\mathrm{gg}}\left(\sqrt{r^2_{\mathrm{p}}\,+\,r^2_{\pi}},\,z\right)\mathrm{d}r_{\pi}}.
\end{equation}
Note that all predictions are evaluated at the median CMASS redshift of 0.54. We also account for the cosmology dependence of measurements according to the equations in Appendix \ref{sec:cosmo_dependence}.

\subsection{The \textsc{Dark Emulator}}
\label{subsec:dark_emu}
We implement our HOD through the \textsc{Dark Emulator}---a fast and flexible emulation tool built on the Dark Quest\footnote{\url{https://darkquestcosmology.github.io}} suite of dark matter-only N-body simulations \citep{Nishimichi_2019,Miyatake_2022}. These simulations are specifically designed to capture halo clustering statistics. Each simulation evolves $2048^3$ particles within a comoving box size of 1 or 2 $h^{-1}\,\mathrm{Gpc}$. The emulator is trained on a set of 100 different flat $w$CDM cosmologies, and provides accurate predictions for $P_{\mathrm{hm}}$, $P_{\mathrm{hh}}$, and $\mathrm{d}n_{\mathrm{h}}/\mathrm{d}M$. All other quantities, including $\Delta\Sigma$ and $w_{\mathrm{p}}$, are computed internally using the analytic prescriptions previously described. The \textsc{Dark Emulator} also supports extensions to the standard HOD, including central galaxy off-centring and halo mass incompleteness. However, AB is not implemented in the \textsc{Dark Emulator}. Users can specify the following cosmological parameters when calling the \textsc{Dark Emulator}: $\omega_{\mathrm{b,0}}$, $\omega_{\mathrm{c,0}}$, $\Omega_{\mathrm{de,0}}$, $\mathrm{ln}\,(10^{10}A_{\mathrm{s}})$, $n_{\mathrm{s}}$, and $w$. The parameters below are derived internally:
\begin{equation}
\Omega_{\mathrm{m},0}\,=\,1\,-\,\Omega_{\mathrm{de,0}},
\label{eqn:omega_relation}
\end{equation}
\vspace{-1mm}
\begin{equation}
h\,=\,\sqrt{\frac{\omega_{\mathrm{b,0}}\,+\,\omega_{\mathrm{c,0}}\,+\,\omega_{\nu,0}}{\Omega_{\mathrm{m,0}}}},
\end{equation}
where the present-day physical neutrino density, $\omega_{\nu,0}$, is fixed at $6.4\times10^{-4}$ by default.

\subsection{Feedback}
\label{subsec:feedback}
We model baryonic feedback using the semi-analytic framework introduced in \cite{2015JCAP...12..049S} and implemented by \cite{shirasaki2024massessunyaevzeldovichgalaxyclusters} for use with the \textsc{Dark Emulator}, commonly known as the `baryonic correction' (BC) model. The advantage of this approach is twofold: (1) it circumvents running computationally expensive hydrodynamical simulations, and (2) it is highly flexible and allows marginalisation over feedback parameters. The procedure expands or contracts collisionless matter (CLM) shells from dark matter-only simulations in response to the presence of baryons. 

To implement the method, we first take the \textsc{Dark Emulator} output $\xi_{\mathrm{hm}}(r,\,M,\,z)$ and compute the average mass density profile, $\rho_{\mathrm{hm}}$:
\begin{equation}
\rho_{\mathrm{hm}}(r,\,M,\,z)\,=\,\overline{\rho}_{\mathrm{m},0}[1\,+\,\xi_{\mathrm{hm}}(r,\,M,\,z)].
\end{equation}
Then, we truncate $\rho_{\mathrm{hm}}$ around the transition scale between the 1- and 2-halo regimes, ensuring that the enclosed mass converges as $r\rightarrow\infty$. The \textsc{Dark Emulator} provides the total halo--matter cross--correlation, $\xi_{\mathrm{hm}}$, without a decomposition into 1- and 2-halo terms. Taking inspiration from \cite{2014ApJ...789....1D}, we define the truncation radius, $r_{\mathrm{out}}$, as the radius at which the logarithmic slope of $\rho_{\mathrm{hm}}$ is the most negative. Now, we can define a truncated average mass density profile: 
\begin{equation}
\rho_{\mathrm{hm+t}}(r,\,M,\,z)\,=\,
\begin{cases} 
\rho_{\mathrm{hm}}(r,\,M,\,z) & \left(r \leq r_{\mathrm{out}}\right) \\
\rho_{\mathrm{hm}}(r_{\mathrm{out}},\,M,\,z)\left(\frac{r}{r_{\mathrm{out}}}\right)^{\gamma}\left(\frac{r^2_{\mathrm{out}}\,+\,\tau^2}{r^2\,+\,\tau^2}\right) & \left(r > r_{\mathrm{out}}\right) 
\end{cases},
\label{eqn:rho_hmt}
\end{equation}
where $\gamma$ is the dimensionless slope of $\rho_{\mathrm{hm}}$ at $r=r_{\mathrm{out}}$, and we set $\tau=3r_{\mathrm{out}}$ as in \cite{shirasaki2024massessunyaevzeldovichgalaxyclusters}.

We next define density profiles for dark matter, stars, and gas. The dark matter profile is simply Eq. \ref{eqn:rho_hmt} rescaled by the global mass fraction of dark matter:
\begin{equation}
\rho_{\mathrm{DM}}(r,\,M,\,z)\,=\,\left(1\,-\,\frac{\Omega_{\mathrm{b}}}{\Omega_{\mathrm{m}}}\right)\rho_{\mathrm{hm+t}}(r,\,M,\,z).
\end{equation}

We divide the total stellar density profile into two components: one from the central galaxy and one from satellite galaxies. The contribution from stars in the central galaxy is:
\begin{equation}
\rho_{\mathrm{cga}}(r)\,=\,\frac{f_{\mathrm{cga}}M_{\mathrm{tot}}}{4\pi^{3/2}R_*r^2}\,\mathrm{exp}\left[-\left(\frac{r}{2R_*}\right)^2\right],
\end{equation}
where $f_{\mathrm{cga}}$ is the fraction of stars in the central galaxy, $M_{\mathrm{tot}}$ is the enclosed mass of $\rho_{\mathrm{hm+t}}$ at r = $\infty$, and $R_*$ is the stellar half-light radius. We compute $f_{\mathrm{cga}}$ via:
\begin{equation}
f_{\mathrm{cga}}(M_{\mathrm{200c}})\,=\,0.09\left(\frac{M_{\mathrm{200c}}}{M_{\mathrm{s}}}\right)^{-\eta_{\mathrm{cga}}},
\end{equation}
where $M_{\mathrm{200c}}$ is the mass enclosed within a radius $R_{\mathrm{200c}}$, inside which the mean density is $200\rho_{\mathrm{crit}}$. Parameters $M_{\mathrm{200c}}$ and $R_{\mathrm{200c}}$ are related as follows:
\begin{equation}
\begin{split}
M_{\mathrm{200c}}\,&=\,\frac{4\pi}{3}\,200\,\rho_{\mathrm{crit}}(z)\,R^3_{\mathrm{200c}}(1\,+\,z)^{-3}, \\
&=\,\int_0^{R_{\mathrm{200c}}}{4 \pi r^2\rho_{\mathrm{hm}}(r,\,M,\,z)\mathrm{d}r}.
\end{split}
\end{equation}
Note that $R_{\mathrm{200c}}$ is in comoving coordinates in the expression above. We set $R_*=0.015 R_{\mathrm{200c}}$, $M_{\mathrm{s}}=2.5\times10^{11}\,h^{-1}\,M_{\odot}$, and $\eta_{\mathrm{cga}}=0.6$ as in \cite{shirasaki2024massessunyaevzeldovichgalaxyclusters}. 

One can also compute $f_{\mathrm{sga}}$, the fraction of stars in satellite galaxies, with the following relationships: 
\begin{equation}
f_{\mathrm{star}}(M_{\mathrm{200c}})\,=\,0.09\left(\frac{M_{\mathrm{200c}}}{M_{\mathrm{s}}}\right)^{\eta_{\mathrm{star}}},
\end{equation}
\vspace{-1mm}
\begin{equation}f_{\mathrm{sga}}(M_{\mathrm{200c}})\,=\,f_{\mathrm{star}}(M_{\mathrm{200c}})\,-\,f_{\mathrm{cga}}(M_{\mathrm{200c}}),
\end{equation}
where $\eta_{\mathrm{star}}$ is set to 0.32 as in \cite{shirasaki2024massessunyaevzeldovichgalaxyclusters}. With $f_{\mathrm{sga}}$, we can define the density profile of stars in satellite galaxies. We assume that satellite galaxies are collisionless and model their stellar distribution after the dark matter density profile:
\begin{equation}
\rho_{\mathrm{sga}}(r,\,M,\,z)\,=\,f_{\mathrm{sga}}\rho_{\mathrm{hm+t}}(r,\,M,\,z).
\end{equation}
The total stellar density profile is given by:
\begin{equation}
\rho_{\mathrm{star}}(r,\,M,\,z)\,=\,\rho_{\mathrm{cga}}(r)\,+\,\rho_{\mathrm{sga}}(r,\,M,\,z).
\end{equation}

The gas density profile assumes the following form:
\begin{equation}
\begin{split}
\rho_{\mathrm{gas}}(r)\,\propto\,&\left(1\,+\,\frac{r}{0.1R_{\mathrm{200c}}}\right)^{-\beta_{\mathrm{g}}}\times\\
&\left(1\,+\,\left(\frac{r}{\theta_{\mathrm{ej}}R_{\mathrm{200c}}}\right)^{\gamma_{\mathrm{g}}}\right)^\frac{-(\delta_{\mathrm{g}}\,-\,\beta_{\mathrm{g}})}{\gamma_{\mathrm{g}}}.
\end{split}
\end{equation}
Following \cite{shirasaki2024massessunyaevzeldovichgalaxyclusters}, we set $\gamma_{\mathrm{g}}=2$ and $\delta_{\mathrm{g}}=7$. The parameter $\theta_{\mathrm{ej}}$ is the maximum radius of gas ejection defined in terms of $R_{\mathrm{200c}}$, while $\beta_{\mathrm{g}}$ controls the slope of the gas density profile and depends on halo mass as follows: 
\begin{equation}
\beta_{\mathrm{g}}(M_{\mathrm{200c}})\,=\,\frac{3(M_{\mathrm{200c}}/M_{\mathrm{c}})^\mu}{1\,+\,(M_{\mathrm{200c}}/M_{\mathrm{c}})^{\mu}}.
\end{equation}
The halo mass dependence of $\beta_{\mathrm{g}}$ implies that low mass halos lose a larger fraction of their gas compared to high mass halos, consistent with observational results \citep{2015JCAP...12..049S,2024MNRAS.534..655B}. The parameter $M_{\mathrm{c}}$ acts as a characteristic mass scale: adjusting it shifts the mass range over which halos efficiently lose gas. For example, increasing $M_{\mathrm{c}}$ allows more massive halos to experience significant gas loss. The parameter $\mu$ controls how rapidly $\beta_{\mathrm{g}}$ approaches 3 ($\beta_{\mathrm{g}}$ is bounded above by 3). We set $\mu=1.2$, following the parameter constraints in Fig. 16 from \cite{shirasaki2024massessunyaevzeldovichgalaxyclusters}. The normalisation of $\rho_{\mathrm{gas}}$ is determined by: 
\begin{equation}
\int_0^{\infty}4\pi r^2\rho_{\mathrm{gas}}(r)\mathrm{d}r\,=\,\left(\frac{\Omega_{\mathrm{b}}}{\Omega_{\mathrm{m}}}\,-\,f_{\mathrm{star}}\right)M_{\mathrm{tot}}.
\end{equation}

\begin{figure}
\centering
\includegraphics[width = \columnwidth]{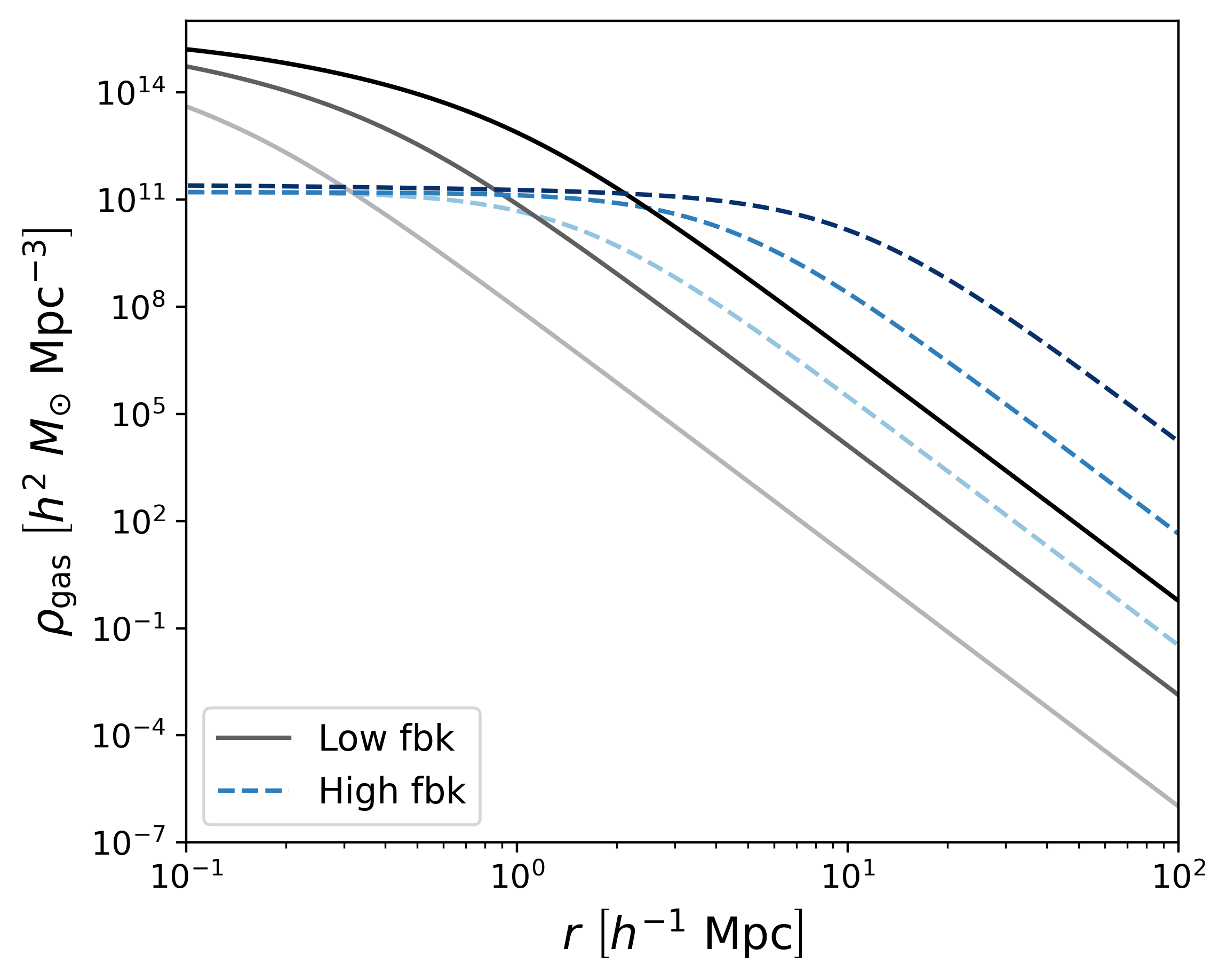} 
\caption{Gas density profiles for low (solid grey curves) and high (dashed blue curves) feedback. For the low feedback case, $\mathrm{log}\,M_{\mathrm{c}}=11.0$ and $\theta_{\mathrm{ej}}=0.5$. For the high feedback case, $\mathrm{log}\,M_{\mathrm{c}}=16.0$ and $\theta_{\mathrm{ej}}=5.0$. Larger halo masses are represented by darker curves. The following halo masses were used to generate these curves: $\mathrm{log}\,(M_{\mathrm{h}}/h^{-1}\,M_{\odot})=12.0,\,14.0,\,16.0$.}
\label{fig:fbk}
\end{figure}

In this analysis, we vary only $M_{\mathrm{c}}$ and $\theta_{\mathrm{ej}}$ since these parameters have the strongest impact on $P_{\mathrm{mm}}$\footnote{Note that marginalising over too few feedback parameters can introduce biases (see \citealp{2024MNRAS.534..655B}).} \citep{Schneider_2019}. As such, we show how they shape the gas density profile in Fig. \ref{fig:fbk} through two configurations: a `low' feedback case with low values for both parameters, and a `high' feedback case with high values. Low feedback produces more centrally peaked profiles, whereas high feedback produces more extended profiles. When low feedback is adopted, only low mass halos are significantly affected by gas-related feedback, and the gas is not ejected far (in our case, it is not driven beyond the halo's boundary). Conversely, halos across a broad mass range lose substantial amounts of gas beyond their boundaries when high feedback is adopted.

Now that we have density profiles for the different halo components, we can displace CLM spherical mass shells relative to the halo centre in response to the mass contributions from baryons. The initial and final radial positions of a CLM mass shell, $r_{\mathrm{i}}$ and $r_{\mathrm{f}}$, are related as follows:
\begin{equation}
\frac{r_{\mathrm{f}}}{r_{\mathrm{i}}}\,=\,\mathcal{A}\left[\left(\frac{M_{\mathrm{i}}}{M_{\mathrm{f}}}\right)^n\,-\,1\right]\,+\,1,
\label{eqn:displace}
\end{equation}
\vspace{-1mm}
\begin{equation}
M_{\mathrm{i}}\,=\,M_{\mathrm{hm+t}}(r_{\mathrm{i}}),
\end{equation}
\vspace{-1mm}
\begin{equation}
M_{\mathrm{f}}\,=\,M_{\mathrm{DM}}(r_{\mathrm{i}})\,+\,M_{\mathrm{sga}}(r_{\mathrm{i}})\,+\,M_{\mathrm{gas}}(r_{\mathrm{f}})\,+\,M_{\mathrm{cga}}(r_{\mathrm{f}}).
\end{equation}
The parameters $\mathcal{A}$ and $n$ govern the expansion/contraction of CLM shells. Setting both to unity corresponds to the adiabatic contraction case introduced by \cite{Blumenthal_Faber_1986}. We set $\mathcal{A}=0.3$ and $n=2$, as constrained by the gasdynamical simulations in \cite{2010MNRAS.407..435A}. \cite{Schneider_2019} investigate an alternative pair of values from \cite{2011MNRAS.414..195T}, and find that both choices yield nearly identical results. The enclosed mass, $M_{\mathrm{x}}$, of a given density profile (e.g., $\mathrm{x}$ = hm+t) is computed via: 
\begin{equation}
M_{\mathrm{x}}(r)\,=\,\int_0^r4\pi\tilde{r}^2\rho_{\mathrm{x}}(\tilde{r})\mathrm{d}\tilde{r}.
\end{equation}

One can solve for either $r_{\mathrm{i}}$ or $r_{\mathrm{f}}$ in Eq. \ref{eqn:displace}; we went with the former. In either case, one should identify $(r_{\mathrm{f}},\,r_{\mathrm{i}})$ pairs whose ratios satisfy the right-hand side of Eq. \ref{eqn:displace}. Once $r_{\mathrm{i}}$ or $r_{\mathrm{f}}$ is solved for, one should evaluate the CLM enclosed mass at $r_{\mathrm{i}}$, but associate it with $r_{\mathrm{f}}$: 
\begin{equation}
M_{\mathrm{CLM}}(r_{\mathrm{f}})\,=\,M_{\mathrm{DM}}(r_{\mathrm{i}})\,+\,M_{\mathrm{sga}}(r_{\mathrm{i}}).
\end{equation}
We then compute $\rho_{\mathrm{CLM}}$ by taking the radial derivative of $M_{\mathrm{CLM}}$:
\begin{equation}
\rho_{\mathrm{CLM}}(r)\,=\,\frac{1}{4\pi r^2}\frac{\mathrm{d}}{\mathrm{d}r}[M_{\mathrm{CLM}}(r)].
\end{equation}

Finally, we construct the halo--matter cross--correlation with feedback, $\xi_{\mathrm{hm+b}}$: 
\begin{equation}
\xi_{\mathrm{hm+b}}(r,\,M,\,z)\,=\,\xi^{\mathrm{1h}}_{\mathrm{hm+b}}(r,\,M,\,z)\,+\,\xi^{\mathrm{2h}}_{\mathrm{hm+b}}(r,\,M,\,z), 
\end{equation}
\vspace{-1mm}
\begin{equation}
\xi^{\mathrm{1h}}_{\mathrm{hm+b}}(r,\,M,\,z)\,=\,\frac{\rho_{\mathrm{hm,\,dmb}}(r,\,M,\,z)}{\overline{\rho}_{\mathrm{m,0}}},
\end{equation}
\vspace{-1mm}
\begin{equation}
\xi^{\mathrm{2h}}_{\mathrm{hm+b}}(r,\,M,\,z)\,=\,\frac{\rho_{\mathrm{hm}}(r,\,M,\,z)\,-\,\rho_{\mathrm{hm+t}}(r,\,M,\,z)}{\overline{\rho}_{\mathrm{m,0}}}\,-\,1,
\end{equation}
where $\rho_{\mathrm{hm,\,dmb}}=\rho_{\mathrm{CLM}}+\rho_{\mathrm{gas}}+\rho_{\mathrm{cga}}$, and `dmb' stands for dark--matter--baryon. We compute the Fourier transform of $\xi_{\mathrm{hm+b}}$ using the publicly available \textsc{FFTLog} code\footnote{\url{http://jila.colorado.edu/~ajsh/FFTLog}} \citep{TALMAN197835,10.1046/j.1365-8711.2000.03071.x}, and calculate the GGL signal as described in Section \ref{subsec:2pcf}. All of the computations mentioned here are performed outside of the \textsc{Dark Emulator}. 

Note that the modifications discussed in this section only affect the 1-halo term of $\Delta\Sigma$, and not $w_{\mathrm{p}}$. Since GGL directly probes the underlying matter distribution, which is altered by feedback mechanisms, we expect GGL to be more sensitive to these effects. Additionally, our feedback procedure modifies the $M_{\mathrm{200m}}$ of a given halo. When implementing the equations in Section \ref{subsec:2pcf}, the halo mass is taken to be the pre-baryonification $M_{\mathrm{200m}}$. Lastly, we verified that the lens magnification model in Section \ref{subsec:2pcf} remains valid even under extreme feedback. To test this, we modelled the suppression of $P_{\mathrm{mm}}$ due to feedback using \textsc{BCemu}\footnote{\url{https://github.com/sambit-giri/BCemu}} \citep{2021JCAP...12..046G}, which is based on the same BC model. The effect of baryons on the lens magnification signal was negligible since feedback is a small-scale effect, whereas lens magnification is primarily a large-scale effect \citep{Unruh_2020}. 

\section{Intrinsic Alignments}
\label{sec:ia}
In addition to gravitational lensing, galaxy alignment is also influenced by environmental effects and processes related to galaxy formation, known as `intrinsic alignments' (IAs; see \citealp{Lamman_2024} for an in-depth review). For GGL, this effect becomes more significant with increasing overlap between the lens and source redshift distributions, producing a higher fraction of physical associations amongst lens-source pairs. This leads to radial alignments that dilute the mean $\gamma_{\mathrm{t}}$. 

A commonly used IA model is the Non-linear Linear Alignment (NLA) model \citep{bridleking07,hirata07}, in which the intrinsic galaxy ellipticity is proportional to the tidal field by a factor $a_{\rm IA}(z)$. The model uses the non-linear matter power spectrum, and the redshift evolution of $a_{\rm IA}(z)$ is usually modelled with a power law:
\begin{equation}
a_{\rm IA}(z) = A_{\rm IA}\left(\frac{1+z}{1+z_0}\right)^{\eta},
\end{equation}
where $z_0$ is the pivot redshift at which $a_{\rm IA}(z) = A_{\rm IA}$. 

Studies find that red galaxies exhibit higher $A_{\rm IA}$, whereas blue galaxies are consistent with $A_{\rm IA}=0$ (e.g., \citealp{samuroff19,johnston19,2025A&A...699A.201H}). In fact, \cite{2025A&A...699A.201H} report a highly significant detection of IAs in the CMASS sample, with $A_{\rm IA}=4.02\pm0.31$. For a source sample with a broad colour selection---as in this UNIONS analysis---$A_{\rm IA}$ is typically expected to lie in the range 0--1 (\citealp{deskids}).

In this analysis, our fiducial choice is to fix cosmological parameters (see Table \ref{table:priors}) and to set $A_{\rm IA}=0$, allowing us to explore the lensing is low problem by separating small-scale effects (HOD properties, feedback) from large-scale effects (cosmology and IAs). In this section, we estimate the approximate maximum impact IAs are expected to have on our analysis. We leave marginalising over IA uncertainties to future work. 

To estimate the impact of IAs on $\gamma_{\mathrm{t}}$ (and hence $\Delta\Sigma$), we adopt the following modelling choices:
\begin{itemize}
\item[--]The redshift kernels are taken to be the CMASS and UNIONS redshift distributions,
\item[--]IAs are modelled using the NLA implementation of \cite{krause21} and \cite{secco22},
\item[--]$P_{\mathrm{mm}}$ is modelled using \textsc{CAMB}, including non-linear corrections (same as in Section \ref{subsec:2pcf}), 
\item[--]A linear galaxy bias model is assumed, with $b_{\mathrm{g}}=2.14$ for CMASS (taken from \citealp{krolewski20}),
\item[--]We assume no redshift evolution in $a_{\rm IA}$ (i.e., $\eta=0\implies a_{\rm IA}=A_{\rm IA}$),
\item[--]We include lens magnification, with the magnification bias fixed at $\alpha_{\mathrm{l}}\approx2.98$.
\end{itemize}
Note that this model does not contain small-scale effects for either $b_{\mathrm{g}}$ or the IA signal. Its purpose is to illustrate the potential large-scale impact of IAs and their degeneracy with cosmological parameters. 

Using our $\gamma_{\mathrm{t}}$ model, we find that an NLA amplitude of $A_{\rm IA}=1$ leads to a $\sim 20\%$ reduction in $\gamma_{\rm t}$ across all scales. The choice $A_{\rm IA} = 1$ lies near the upper limit allowed by the surveys in \cite{deskids}, and is close to the central value of the $A_{\mathrm{IA}}$ prior adopted in the UNIONS cosmic shear analysis ($0.83\pm0.78$; \citealp{goh2026unions3500weaklensingiii,guerrini2026unions3500weaklensingiv}). The substantial reduction in $\gamma_{\rm t}$ is primarily due to our use of a single, non-tomographic source sample, which overlaps significantly with the CMASS lens sample (see Fig. \ref{fig:nz}). Future work using tomography will substantially reduce this systematic.

In Section \ref{sec:results}, we test the impact of varying cosmological parameters on the lensing is low problem, but we urge caution in interpreting these results as a reliable constraint on $S_8$. To avoid misinterpretation, we only report the maximum a posteriori (MAP) values for cosmological parameters. Our IA estimate for $A_{\rm IA}=1$ suggests a potential bias of up to $\sim 20\%$ in the $\sigma_8$ posteriors from the combination of GGL and GC used in this work (a biased GGL signal is expected to propagate as $\sim \sigma_8$ on linear scales for the $2\times 2$-point combination). However, IA posteriors for broadly colour-selected source samples are generally consistent with $A_{\rm IA}=0$ within $2\sigma$ \citep{deskids}. Despite our test with $A_{\mathrm{IA}}=1$, we reiterate that $A_{\mathrm{IA}}=0$ is used by default in this paper.

\section{Parameter Estimation}
\label{sec:param_est}
We estimate the best-fit parameters using Bayesian inference. To sample the posterior distribution, $\mathcal{P}(\bm{\theta}|\bm{d})$, we perform a Markov Chain Monte Carlo (MCMC) analysis using \textsc{emcee}\footnote{\url{https://github.com/dfm/emcee}} \citep{emcee}. According to Bayes' Theorem, $\mathcal{P}(\bm{\theta}|\bm{d})$ is proportional to:
\begin{equation}
\mathcal{P}(\bm{\theta}|\bm{d})\,\propto\,\mathcal{L}(\bm{d}|\bm{\theta})\Pi(\bm{\theta}),
\end{equation}
where $\bm{\theta}$ is a set of free parameters, $\mathcal{L}(\bm{d}|\bm{\theta})$ is the likelihood of the data given a particular parameter configuration, and $\Pi(\bm{\theta})$ is the prior distribution. We assume that the likelihood follows a multivariate Gaussian distribution:
\begin{equation}
\mathrm{ln}\,\mathcal{L}(\bm{d}|\bm{\theta})\,=\,-\frac{1}{2}\chi^2,
\end{equation}
\vspace{-1mm}
\begin{equation}
\chi^2\,=\,\sum_{i,j}[d_i\,-\,t_i(\bm{\theta})](\bm{C}^{-1})_{ij}[d_j\,-\,t_j(\bm{\theta})],
\label{eqn:chi2}
\end{equation}
where $d$ represents an element of the data vector $\bm{d}$, $t$ is a model prediction at $\bm{\theta}$, and $\bm{C}^{-1}$ is the inverse covariance matrix. 

During parameter estimation, we perform joint fits to both $\Delta\Sigma$ and $w_{\mathrm{p}}$. We neglect any cross-covariance between the two and compute independent likelihoods for each (see Appendix \ref{sec:cross_cov_test} for a justification), such that the final log likelihood is:
\begin{equation}
\mathrm{ln}\,\mathcal{L}(\bm{d}|\bm{\theta})\,=\,-\frac{1}{2}(\chi_{\Delta\Sigma}^2\,+\,\chi^2_{w_{\mathrm{p}}}). 
\end{equation}

For all input model parameters, we impose uninformative, wide, flat priors. We also apply a `soft' Gaussian prior to $\overline{n}_{\mathrm{g}}$ with mean from Fig. 3 in \citetalias{lilo17}. Our models have up to 12 free parameters, with each including the 5 HOD parameters introduced in Section \ref{subsec:hod}. The degrees of freedom (dof) are calculated as: 24+25+1 - the number of free parameters. The numbers 24 and 25 come from the lengths of our $\Delta\Sigma$ and $w_{\mathrm{p}}$ data vectors, while the `+1' is due to the additional $\overline{n}_{\mathrm{g}}$ constraint, which is treated as an extra data point. Note that the effective dof may differ due to parameter degeneracies. 

Some models contain $\mathrm{ln}\,(10^{10}A_{\mathrm{s}})$ (amplitude of the primordial curvature power spectrum) and $\Omega_{\mathrm{m},0}$ as free parameters, as they directly influence $P_{\mathrm{mm}}$. The \textsc{Dark Emulator} takes $\Omega_{\mathrm{de,0}}$ as an input, so we vary $\Omega_{\mathrm{m,0}}$ using the relation in Eq. \ref{eqn:omega_relation}. When $\Omega_{\mathrm{m,0}}$ changes, either the baryon or cold dark matter (CDM) fraction must change---we allow the latter to vary. All other cosmological parameters are fixed at the Planck 2018 TT,TE,EE+lowE+lensing+BAO values \citep{planck18}. Other parameters that we free are listed in Table \ref{table:priors}, which also provides priors and fiducial values. We additionally show how each parameter impacts our observables in Appendix \ref{additional_plots}. To ensure the convergence of MCMC chains, we either require that (a) the length of a chain is at least 50 times the largest autocorrelation time estimated by \textsc{emcee}\footnote{See \url{https://emcee.readthedocs.io/en/stable/tutorials/autocorr}.}, or (b) the statistics from halves of a chain post burn-in are consistent. Lastly, we define the best-fit model as the parameter set drawn from the MAP. 
\begin{table*}
\centering
\begin{tabular}{lcccr}
\hline
Parameter & Type & Prior & Fiducial Value \\ \hline
$\mathrm{log}\,M_{\mathrm{min}}$ & HOD & $\mathcal{U}(10.0, 15.0)$ & -- \\
$\sigma^2_{\mathrm{log}\,M}$ & HOD & $\mathcal{U}(10^{-3}, 6.0)$ & -- \\
$\mathrm{log}\,M_1$ & HOD & $\mathcal{U}(12.0, 17.5)$ & -- \\
$\alpha$ & HOD & $\mathcal{U}(10^{-3}, 3.0)$ & -- \\
$\kappa$ & HOD & $\mathcal{U}(10^{-3}, 3.0)$ & -- \\
$R_{\mathrm{c}}$ & Extended HOD & $\mathcal{U}(0.1, 3.0)$ & 1.0 \\
$p_{\mathrm{off}}$ & Extended HOD & $\mathcal{U}(0.0, 1.0)$ & 0.0 \\
$R_{\mathrm{off}}$ & Extended HOD & $\mathcal{U}(10^{-2}, 5.0)$ & -- \\
$\mathrm{log}\,M_{\mathrm{c}}$ & Extended HOD & $\mathcal{U}(10.0, 17.5)$ & -- \\
$\theta_{\mathrm{ej}}$ & Extended HOD & $\mathcal{U}(0.1, 17.0)$ & -- \\
$\mathrm{ln}\,(10^{10}A_{\mathrm{s}})$ & Cosmological & $\mathcal{U}(2.4752, 3.7128)$ & 3.047 \\
$\Omega_{\mathrm{m,0}}$ & Cosmological & $\mathcal{U}(0.286, 0.338)$ & 0.3111 \\
$h$ & Cosmological & -- & 0.6766 \\
$\Omega_{\mathrm{b,0}}$ & Cosmological & -- & $4.897 \times 10^{-2}$ \\
$\Omega_{\mathrm{\nu,0}}$ & Cosmological & -- & $1.398 \times 10^{-3}$ \\
$n_{\mathrm{s}}$ & Cosmological & -- & 0.9665 \\
$w$ & Cosmological & -- & -1 \\
$\overline{n}_{\mathrm{g}}\,\left[h^3\,\mathrm{Mpc}^{-3}\right]$ & Derived & $\mathcal{G}(3.5,0.75) \times 10^{-4}$ & --
\end{tabular}
\caption{Priors and fiducial values used for parameter estimation. `Extended HOD' refers to parameters beyond the standard HOD framework that do not enter the expressions for $\langle N_{\mathrm{c}}(M) \rangle$ and $\langle N_{\mathrm{s}}(M) \rangle$. The symbol $\mathcal{U}(\mathrm{low, high})$ denotes a uniform prior with lower and upper bounds given in parentheses, and $\mathcal{G}(\mu,\sigma)$ denotes a Gaussian prior with mean $\mu$ and width $\sigma$. Priors on cosmological parameters are set by the \textsc{Dark Emulator}'s supporting range. Feedback is only incorporated when its corresponding parameters are allowed to vary.}
\label{table:priors}
\end{table*}

\section{Results and Discussion}
\label{sec:results}
This section presents the results of our analysis and is divided into 5 parts. The first subsection shows the results obtained with cosmological parameters fixed at the Planck 2018 values. The second subsection follows the first, but with free cosmological parameters---specifically, $\Omega_{\mathrm{m},0}$ and $\mathrm{ln}\,(10^{10}A_{\mathrm{s}})$. In the third subsection, we test a model with feedback parameters fixed at the \cite{2024MNRAS.534..655B} constraints. The final two subsections compare our findings with previous work and discuss them in the context of the lensing is low problem. 
\subsection{HOD with Planck Cosmology}
\label{subsec:hod_planck}
\begin{figure*}
\includegraphics[width = 0.75\textwidth]{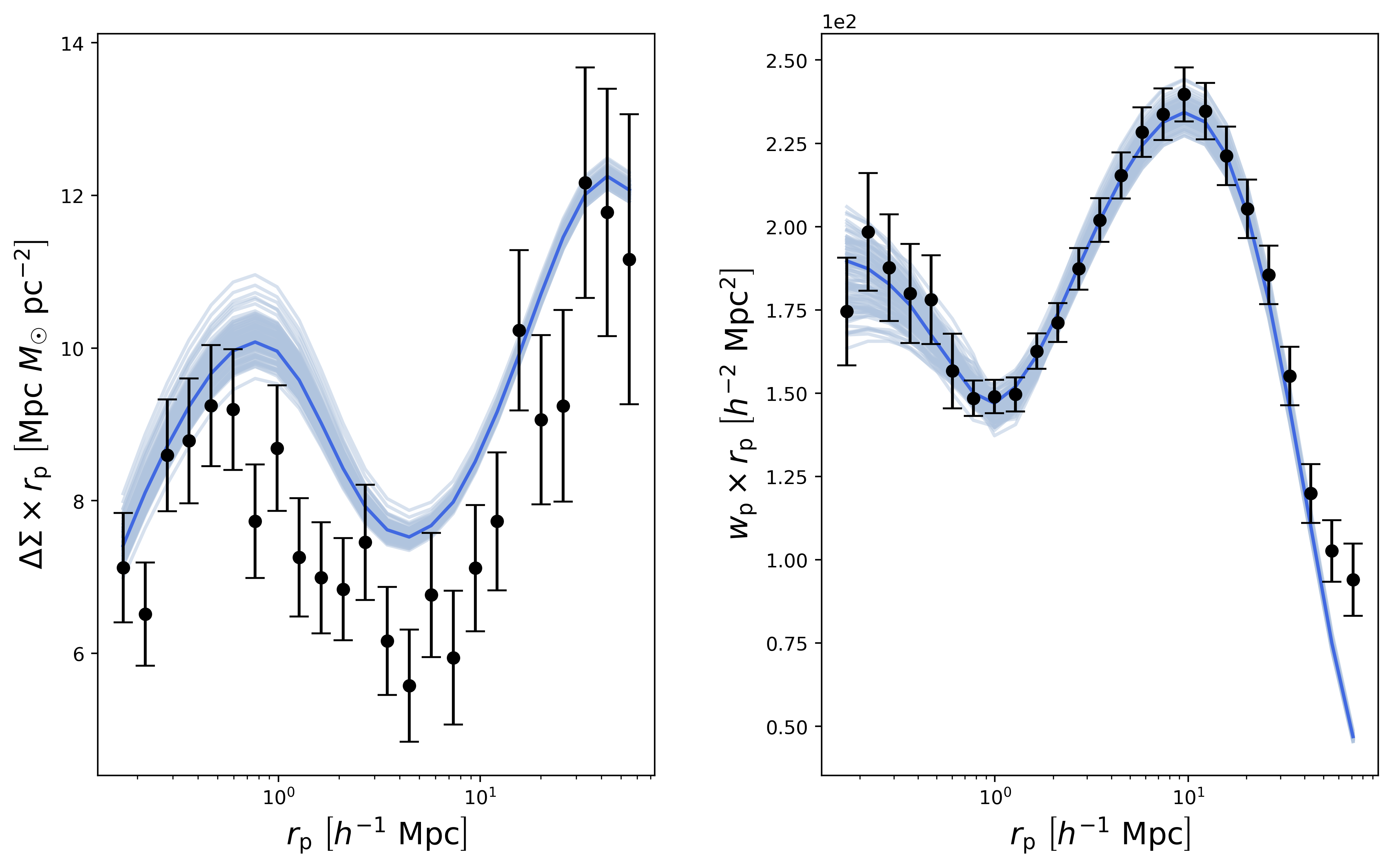}
\caption{Right panel: 5-parameter HOD fit (blue curve) to the CMASS GC signal, $w_{\mathrm{p}}$ (black points). Cosmological parameters are fixed at the Planck 2018 TT,TE,EE+lowE+lensing+BAO values, and extended HOD parameters are set to the fiducial values listed in Table \ref{table:priors}. Left panel: corresponding GGL prediction using the best-fit parameters from the $w_{\mathrm{p}}$ fit. Light blue curves show models drawn from the posterior distribution of the $w_{\mathrm{p}}$ fit. The $1\sigma$ error bars include both statistical and systematic uncertainties.}
\label{fig:lilo}
\end{figure*}

We now present the main results of this work, summarised in Table \ref{table:stats} which reports goodness of fit metrics. Although our primary goal is to test various models through joint fits to both $\Delta\Sigma$ and $w_{\mathrm{p}}$, we also perform a GC-only fit\footnote{This fit (and all others) still incorporates the galaxy abundance.} (Fig. \ref{fig:lilo}), where the 5 HOD parameters from Section \ref{subsec:hod} are varied and cosmological parameters are fixed at the Planck 2018 TT,TE,EE+lowE+lensing+BAO values. Following \citetalias{lilo17}, we use the best-fit parameters to predict the GGL signal. Consistent with previous studies, we observe a lensing is low effect in the CMASS sample; many data points, including their $1\sigma$ error bars, fall below the predicted curve. On average, the predicted GGL signal is about 13$\%$ larger than our $\bm{R}$-calibrated measurements. However, this does not account for the $m$-bias (see Section \ref{subsec:unions}). Correcting for this bias would increase the measurement by approximately 6\%, leaving only a 7\% discrepancy. The fit to $w_{\mathrm{p}}$ is quite good, although joint fits with other models achieve better agreement. Notably, the model struggles to fit the last two $w_{\mathrm{p}}$ data points---a trend also observed by \cite{More_2015}. This may be explained by the covariance matrix's off-diagonality at large scales. 

Before we compare different models, we introduce the Corrected Akaike Information Criterion (AICc; \citealp{10.1093/biomet/76.2.297,Akaike1998}) and the Bayesian Information Criterion (BIC; \citealp{c4048c8f-6ca9-3965-96a3-653ab8996955}). These two criteria facilitate model comparison/selection, balancing model simplicity and goodness of fit. For both, the preferred model amongst a set of candidates is the one with the lowest score. The AICc is defined as:
\begin{equation}
\mathrm{AICc}\,=\,2k\,-\,2\mathrm{ln}\,(\mathcal{L}_{\mathrm{max}})\,+\,\frac{2k^2\,+\,2k}{n\,-\,k\,-\,1}\,=\,2k\,+\,\chi^2_{\mathrm{min}}\,+\,\frac{2k^2\,+\,2k}{n\,-\,k\,-\,1},
\end{equation}
where $k$ is the number of free parameters, $n$ is the sample size (here, 50), $\mathcal{L}_{\mathrm{max}}$ is the maximum likelihood, and $\chi^2_{\mathrm{min}}$ is the corresponding $\chi^2$ value. We set $\chi^2_{\mathrm{min}}=\chi^2_{\mathrm{tot}}$, where $\chi^2_{\mathrm{tot}}$ is the total $\chi^2$ evaluated at the MAP across all observables: $\chi^2_{\mathrm{tot}}=\chi^2_{\Delta\Sigma}+\chi^2_{w_{\mathrm{p}}}+\chi^2_{\overline{n}_{\mathrm{g}}}$. In the limit of large sample sizes, the AICc converges to the general Akaike Information Criterion (AIC), which is equivalent to the AICc but without the 3rd term. The BIC is computed via: 
\begin{equation}
\mathrm{BIC}\,=\,k\mathrm{ln}\,(n)\,-\,2\mathrm{ln}\,(\mathcal{L}_{\mathrm{max}})\,=\,k\mathrm{ln}\,(n)\,+\,\chi^2_{\mathrm{min}}.
\end{equation}

\begin{figure*}
\includegraphics[width = 0.75\textwidth]{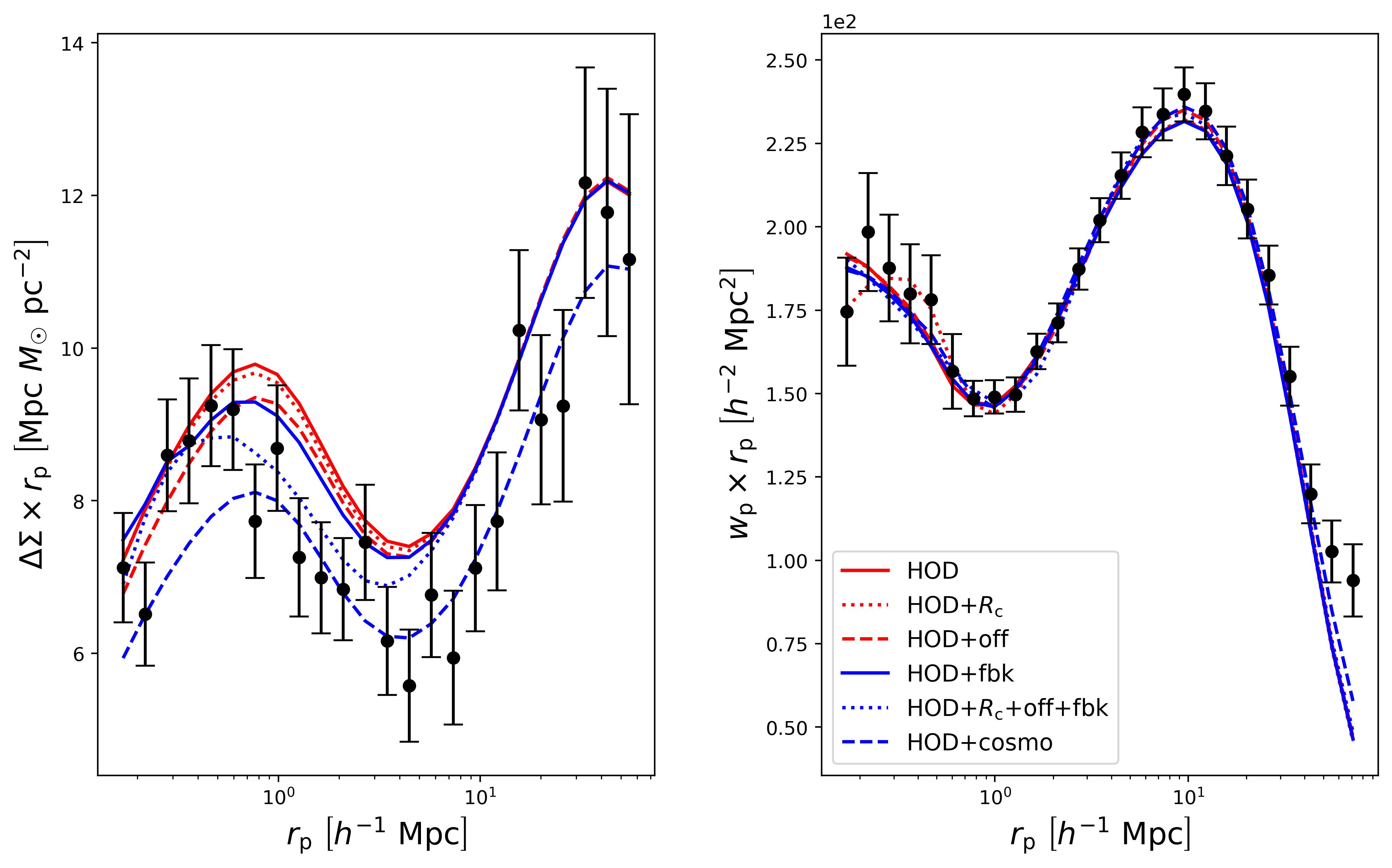}
\caption{Joint fits to both the GGL and GC signals of CMASS galaxies. The baseline model (solid red) varies only the 5 HOD parameters, with all other parameters fixed at their fiducial values. A `+' indicates that additional parameters are varied alongside the HOD parameters. Parameters characterizing a single extension to the standard HOD are grouped together. For example, `off' indicates that both off-centring parameters ($p_{\mathrm{off}}$ and $R_{\mathrm{off}}$) are varied. Feedback parameters are represented by `fbk', and cosmological parameters by `cosmo'.}
\label{fig:res_1}
\end{figure*}

Fig. \ref{fig:res_1} shows the results of a joint fit to $\Delta\Sigma$ and $w_{\mathrm{p}}$ using the same setup as the GC-only fit (HOD curves). For the next few paragraphs, we refer to this configuration as the `baseline' model. Interestingly, the joint fit and the fit to $w_{\mathrm{p}}$ alone yield similar GGL predictions; the maximum percent difference between the two curves is $\sim3\%$. Consequently, $\Delta\Sigma$ remains slightly over-predicted, indicating that the standard 5-parameter HOD at a Planck cosmology slighty over-predicts the amplitude of $\Delta\Sigma$ even when GGL information is included in the fitting routine.  

We also explore individual extensions to the standard HOD, with results shown in the same figure. We find that the model varying both HOD parameters and the satellite concentration normalisation (HOD+$R_{\mathrm{c}}$ curves) predicts a GGL signal quite similar to that of the baseline model. The baseline fit does not improve when introducing $p_{\mathrm{off}}$ and $R_{\mathrm{off}}$ as free parameters (HOD+off curves). The MAP estimates for $p_{\mathrm{off}}$ and $R_{\mathrm{off}}$ are 0.063 and 4.517, respectively. These values imply that $\sim6\%$ of central galaxies are offset from their halo centres according to the spatial distribution in Eq. \ref{eqn:off-centring}, where $R_{\mathrm{200m}}R_{\mathrm{off}}=4.517R_{\mathrm{200m}}$. Despite the large value of $R_{\mathrm{off}}$, very few central galaxies are impacted by this effect, making the overall mean offset radius (including `true' central galaxies) small.  

Introducing feedback and its associated parameters ($\mathrm{log}\,M_{\mathrm{c}}$ and $\theta_{\mathrm{ej}}$) to the baseline model increases the $p$-value and decreases the AICc and BIC. Notably, the MAP estimate for $\theta_{\mathrm{ej}}$ is 16.918---an extremely large maximum gas ejection radius. Although this is somewhat counteracted by a low value of $\mathrm{log}\,M_{\mathrm{c}}$ (11.432), the resulting parameter combination remains unphysical. Evidently, this model cannot provide a plausible description of the data. 

Building on all of these models, we constructed a model that simultaneously varies all parameters except cosmological ones (HOD+$R_{\mathrm{c}}$+off+fbk curves). The goal was to test whether expanding the parameter space and incorporating different physical effects could produce a sufficiently flexible model, capable of capturing the amplitudes of both observables without deviating from the fiducial Planck cosmology. While the $p$-value increases, the AICc and BIC also increase. Interestingly, the MAP estimate for $R_{\mathrm{c}}$ shifts from 0.377 to 2.486, implying a more centrally concentrated satellite galaxy distribution. This is partially offset by the off-centring effect, with MAP estimates of $p_{\mathrm{off}}=0.404$ and $R_{\mathrm{off}}=2.471$. Here, a substantial fraction of central galaxies are offset from their halo centres, reducing the central concentration of galaxies. Feedback is also more physically realistic, with $\mathrm{log}\,M_{\mathrm{c}}=14.028$ and $\theta_{\mathrm{ej}}=0.380$. 

\subsection{HOD with Trial Cosmologies}
\label{hod_trial}
When we allow freedom in our cosmology through the parameters $\mathrm{ln}\,(10^{10}A_{\mathrm{s}})$ and $\Omega_{\mathrm{m,0}}$ (HOD+cosmo curves), we observe reductions in both the AICc and BIC. Most notably, $\Delta\Sigma$ is no longer over-predicted on intermediate--large scales, although the small-scale signal is now under-predicted. The MAP estimates for cosmological parameters are $\mathrm{ln}\,(10^{10}A_{\mathrm{s}})=3.057$ and $\Omega_{\mathrm{m},0}=0.286$, corresponding to $\sigma_8=0.761$ and $S_8=0.743$. This, however, does not include any correction for $m$-bias. Assuming that $\sigma_8 \propto \Delta \Sigma$, correcting for $m$-bias would increase to $S_8 \sim 0.79$.

Moreover, due to our fiducial choice of $A_{\mathrm{IA}}=0$, and the expectation that $A_{\mathrm{IA}} > 0$, this $S_8$ value should be regarded as a lower limit. Although we do not report the errors on cosmological parameters, we note that, excluding the uncertainty in $A_{\mathrm{IA}}$, the uncertainty on $S_8$ is at the few-percent level. While our estimate for $\mathrm{ln}\,(10^{10}A_{\mathrm{s}})$ closely matches the fiducial value of 3.047, the estimate for $\Omega_{\mathrm{m},0}$ is substantially lower and hits the lower bound of our prior range. This low value of $\Omega_{\mathrm{m},0}$ drives the reduced $S_8$ and $\sigma_8$ values.

If $\Omega_{\mathrm{m},0}$ is reduced relative to some fiducial value, then the GGL signal is suppressed, while the GC signal is enhanced (see panel (d) of Fig. \ref{fig:rainbow_2}). This behaviour arises since $\Delta\Sigma\propto \Omega_{\mathrm{m,0}}b_{\mathrm{g}}\xi_{\mathrm{mm}}$, whereas $w_{\mathrm{p}}\propto b^2_{\mathrm{g}}\xi_{\mathrm{mm}}$. These proportionalities, which are strictly valid on linear scales, explain why combining GGL and GC breaks degeneracies between HOD and cosmological parameters. Lowering $\Omega_{\mathrm{m},0}$ decreases the amplitude of $\xi_{\mathrm{mm}}$, which, in turn, increases $b_{\mathrm{g}}$ for a fixed HOD since galaxies become more biased tracers of the underlying matter distribution. Due to the different powers of $b_{\mathrm{g}}$, changes in $\Omega_{\mathrm{m},0}$ affect $\Delta\Sigma$ and $w_{\mathrm{p}}$ in opposite directions. 

\begin{figure*}
\includegraphics[width = 0.75\textwidth]{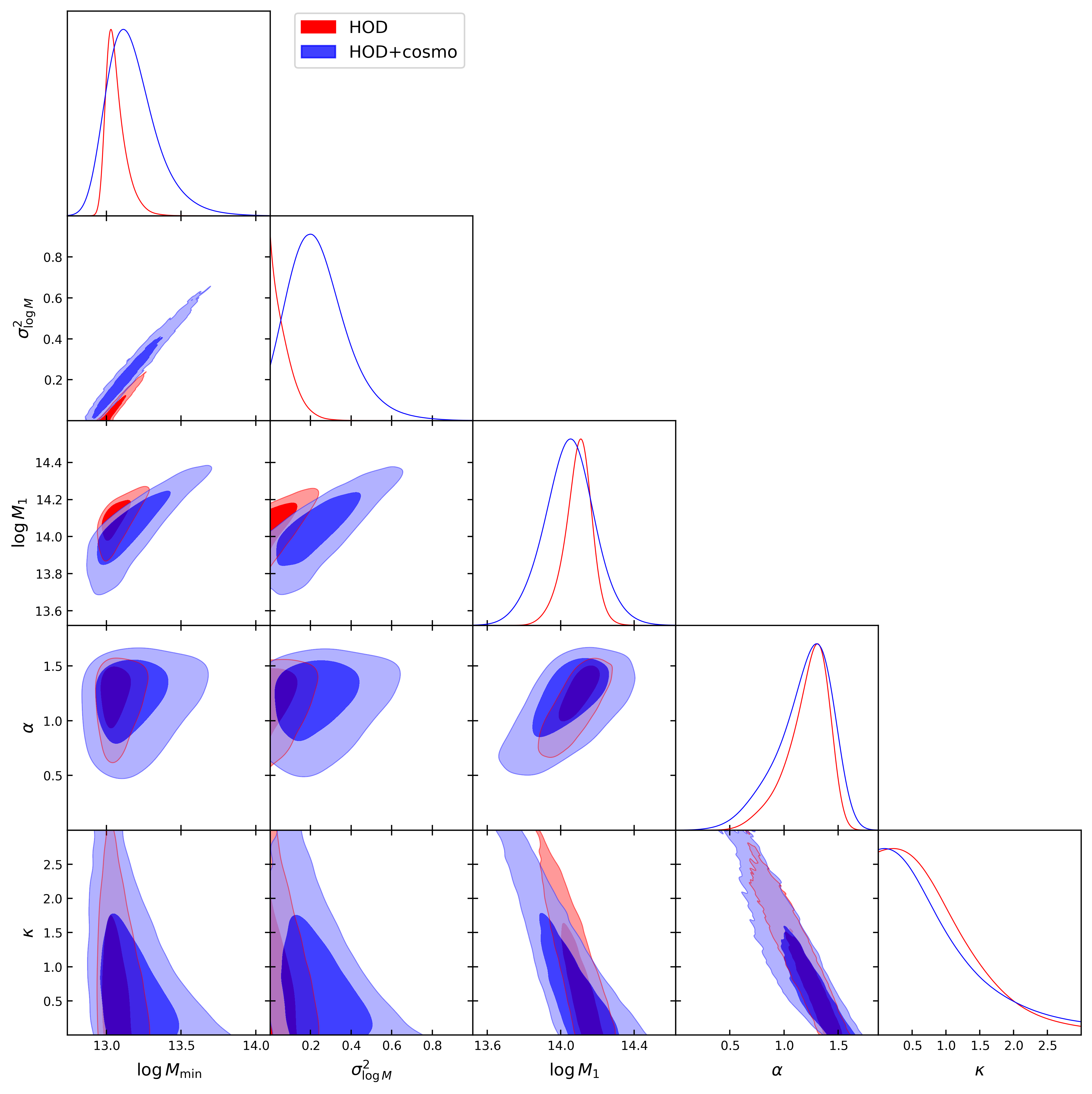}
\caption{HOD Posterior distributions for the HOD and HOD+cosmo models. Median values and their uncertainties are listed in Table \ref{table:median}. Note that $\mathrm{log}\,M_{\mathrm{min}}$ and $\sigma^2_{\mathrm{log}\,M}$ are highly degenerate (see panels (a) and (b) of Fig. \ref{fig:rainbow_1}). A lower value of $\mathrm{log}\,M_{\mathrm{min}}$ allows lower mass halos to host central galaxies, reducing $b_{\mathrm{g}}$. In contrast, a low value of $\sigma^2_{\mathrm{log}\,M}$ implies a sharper transition from $\langle N_{\mathrm{c}}\rangle=0$ to $\langle N_{\mathrm{c}}\rangle=1$, placing more central galaxies in high mass halos and increasing $b_{\mathrm{g}}$.}
\label{fig:res_2}
\end{figure*}

Adjusting $\Omega_{\mathrm{m},0}$ from its fiducial value of 0.3111 to 0.286 results in a roughly $16\%$ increase in $w_{\mathrm{p}}$ around its peak (at $\sim10\,h^{-1}\,\mathrm{Mpc}$). The GC signal is slightly enhanced relative to the baseline prediction, but not at the $16\%$ level, indicating that the HOD parameters must have compensated to maintain consistency with $w_{\mathrm{p}}$. Indeed, we observe non-negligible shifts in the MAP estimates and posterior distributions of HOD parameters (see Fig. \ref{fig:res_2}). The most notable change is a considerable increase in the MAP estimate for $\sigma^2_{\mathrm{log}\,M}$ from 0.001 to 0.304. This shift reflects a much more softened transition from 0 to 1 central galaxy in the mean occupation function ($\langle N_{\mathrm{c}}\rangle$), enabling more low mass halos to host central galaxies and thereby reducing $b_{\mathrm{g}}$ (see panel (b) of Fig. \ref{fig:rainbow_1}). Interestingly, the HOD+$R_{\mathrm{c}}$+off+fbk model yields similar MAP estimates for $\mathrm{log}\,M_{\mathrm{min}}$, $\sigma^2_{\mathrm{log}\,M}$, and $\mathrm{log}\,M_1$. 

\begin{figure*}
\includegraphics[width = 0.75\textwidth]{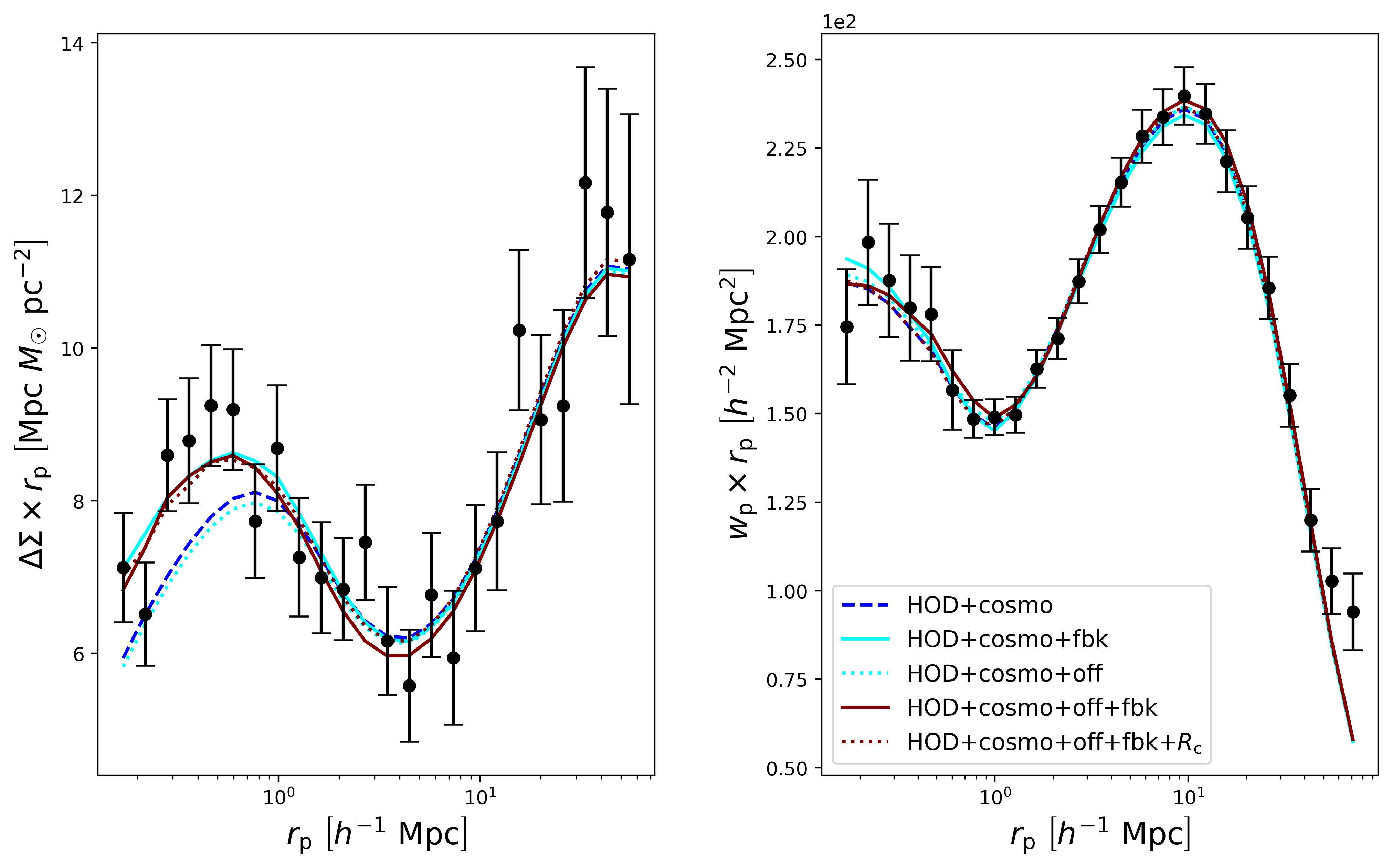}
\caption{Similar to Fig. \ref{fig:res_1}, except the baseline model is now `HOD+cosmo'.}
\label{fig:res_3}
\end{figure*}

\begin{figure}
\includegraphics[width = \columnwidth]{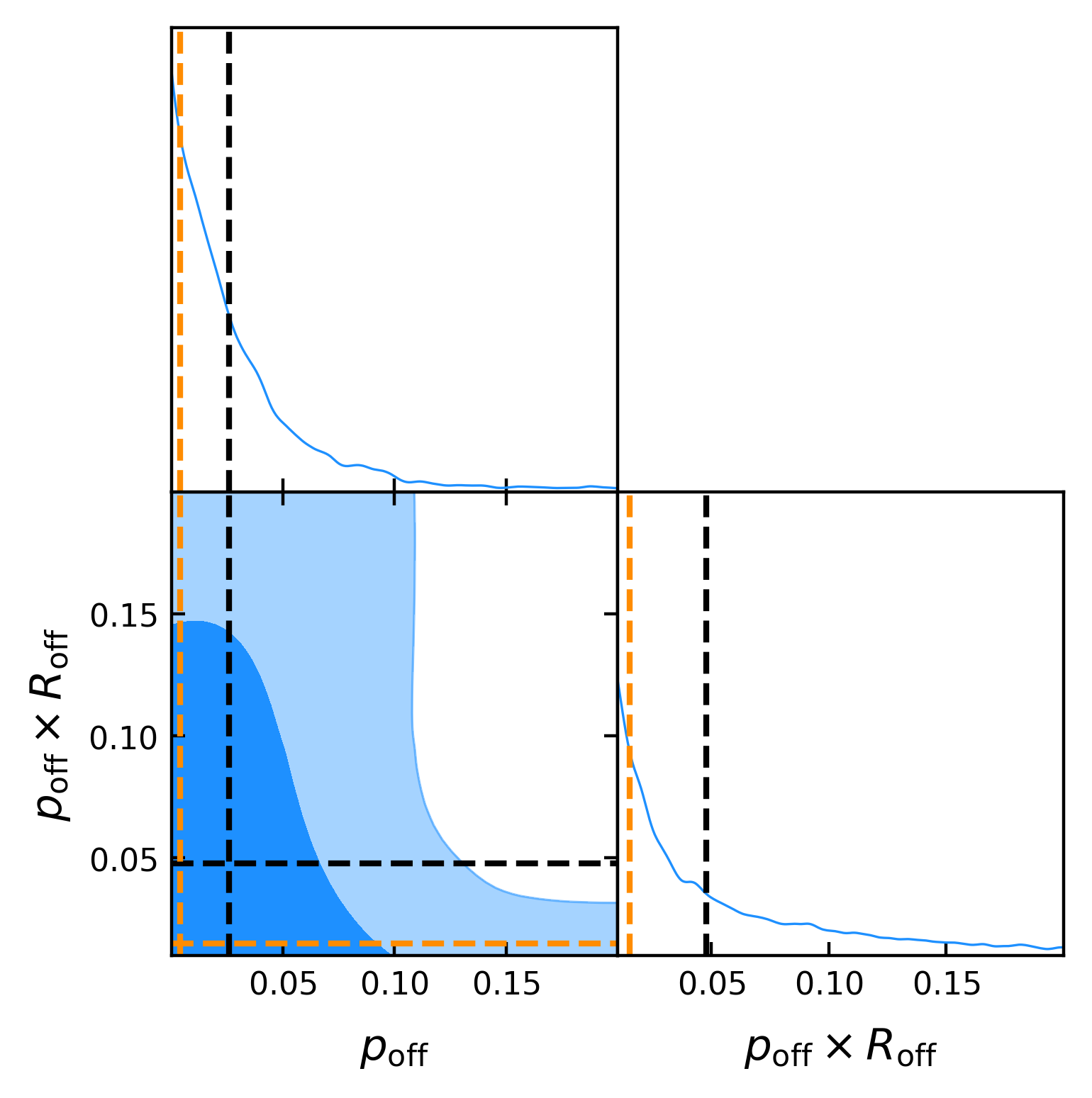}
\caption{Off-centring posterior distributions for the HOD+cosmo+off model. Orange lines mark the MAP estimates, while black lines indicate median values. We show $p_{\mathrm{off}}\times R_{\mathrm{off}}$ instead of just $R_{\mathrm{off}}$ to highlight the distribution of offset distances including both displaced and `true' central galaxies.}
\label{fig:res_4}
\end{figure}

We explore the impact of progressively increasing model complexity in Fig. \ref{fig:res_3}. Here, we adopt the HOD+cosmo model as the baseline configuration since it provides the best joint fit out of the models previously discussed. We find that freeing the off-centring parameters (HOD+cosmo+off curves) makes the fit slightly worse the fit relative to the baseline model, resulting in higher AICc and BIC values. Adding free parameters to a model should either maintain or improve its performance. The small increase in $\chi^2_{\mathrm{tot}}$ likely reflects the data's preference for $p_{\mathrm{off}}=0$ (see Fig. \ref{fig:res_4}), which is a prior boundary. 

Introducing feedback into the previous model (HOD+cosmo+off+fbk curves) yields the best fit to $\Delta\Sigma$. Notably, $\Delta\Sigma$ is much less under-predicted on small scales. Relative to the previous model, the AICc and BIC are reduced. This reduction, however, is not enough to outperform the baseline model. Here, the off-centring effect is no longer consistent with zero, with MAP estimates of $p_{\mathrm{off}}=0.078=7.8\%$ and $R_{\mathrm{off}}=3.277$. The corresponding estimates for feedback parameters are $\mathrm{log}\,M_{\mathrm{c}}=12.450$ and $\theta_{\mathrm{ej}}=1.318$. Due to the moderately low values of $\mathrm{log}\,M_{\mathrm{c}}$ and $\theta_{\mathrm{ej}}$, feedback is actually boosting the small-scale GGL signal compared to the previous fit. Recall from Fig. \ref{fig:fbk} that lower values of feedback parameters enhance the gas density profile on small scales. In this case, the particular parameter configuration leads to a contraction of CLM shells on small scales, with $r_{\mathrm{f}}/r_{\mathrm{i}}<1$. This highlights the flexibility of the BC model, as feedback is typically modelled as a suppression of $P_{\mathrm{mm}}$ rather than an enhancement. However, this flexibility comes at a cost---the BC model can produce unphysical scenarios, as illustrated in Section \ref{subsec:hod_planck}. 

Allowing the satellite concentration normalisation to vary (HOD+cosmo+off+fbk+$R_{\mathrm{c}}$ curves) does not improve the previous model. In fact, $\chi^2_{\mathrm{tot}}$ slightly increases, possibly due to scatter in the MAP estimates. Consequently, the AICc and BIC increase. The MAP estimate for $R_{\mathrm{c}}$ is 0.930---close to the fiducial value of unity. Given this value of $R_{\mathrm{c}}$, one might expect similar parameter estimates between the previous model and the current one. However, there are notable differences in MAP estimates. For instance, $p_{\mathrm{off}}$ shifts from 0.078 to 0.171, and $R_{\mathrm{off}}$ from 3.277 to 2.031. Additionally, $\mathrm{log}\,M_{\mathrm{c}}$ shifts from 12.450 to 14.255, and $\theta_{\mathrm{ej}}$ from 1.318 to 0.474. These changes may be explained by parameter degeneracies, to which the MAP is particularly sensitive. For example, low (high) values of $p_{\mathrm{off}}$ are somewhat degenerate with high (low) values of $R_{\mathrm{off}}$\footnote{These parameters are degenerate to some extent, as each distinctly affects the shapes of the GGL and GC predictions (see panels (a) and (b) of Fig. \ref{fig:rainbow_2}).}. A similar relationship exists between $\mathrm{log}\,M_{\mathrm{c}}$ and $\theta_{\mathrm{ej}}$. Indeed, the shifts in MAP estimates follow the directions of these degeneracies. In addition to parameter degeneracies, there are degeneracies in how $\chi^2_{\mathrm{tot}}$ can be minimised. The previous model achieves the lowest $\chi^2_{\Delta\Sigma}$ and a lower $\chi^2_{\overline{n}_{\mathrm{g}}}$, whereas the current model yields the lowest $\chi^2_{w_{\mathrm{p}}}$. This model is the most complex, with 12 free parameters. We show the resulting posterior distributions in Fig. \ref{fig:res_6}.

\begin{figure*}
\includegraphics[width = 0.85\textwidth]{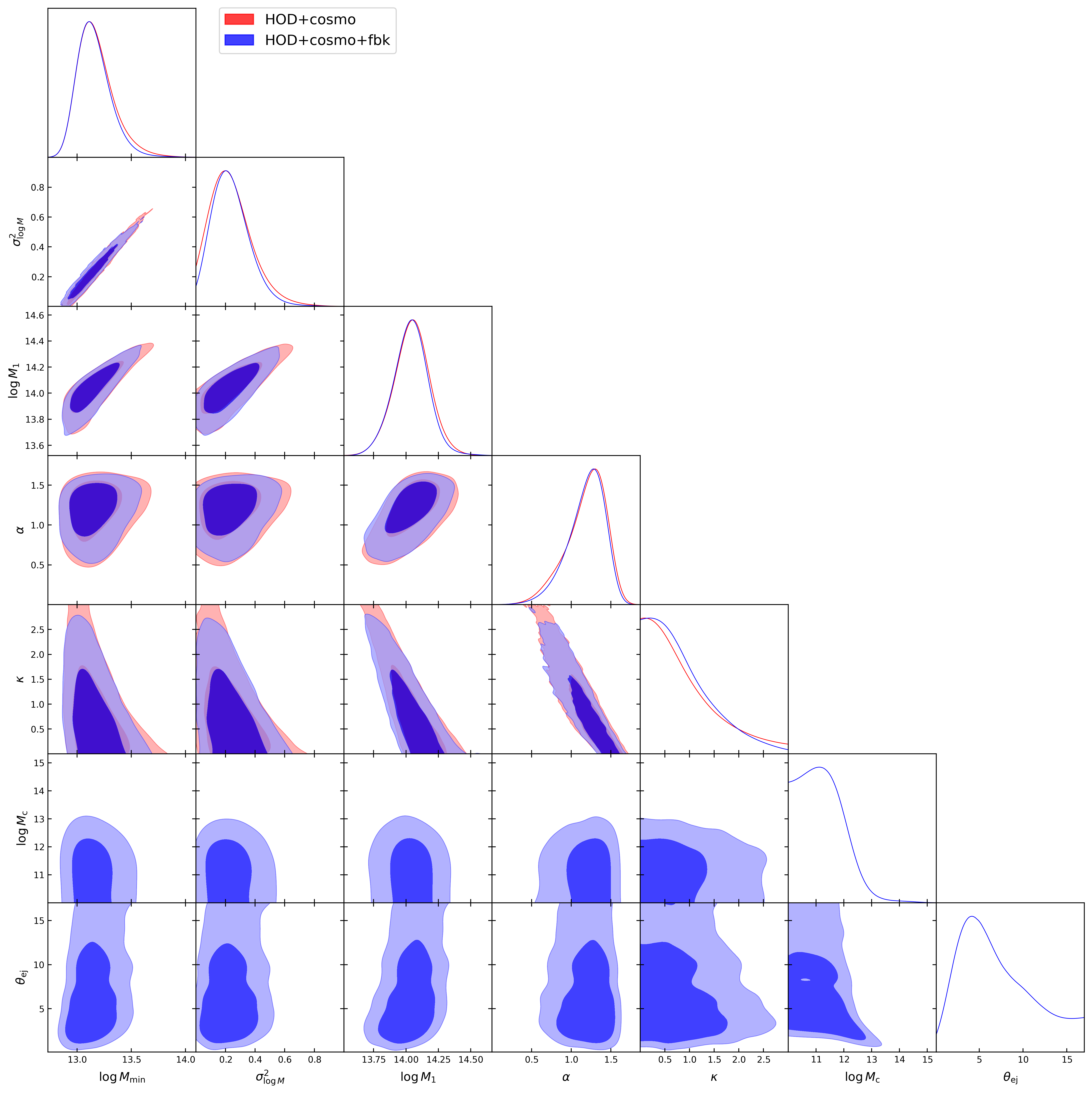}
\caption{Posterior distributions for the HOD+cosmo and HOD+cosmo+fbk models.}
\label{fig:res_5}
\end{figure*}

Since cosmological and feedback parameters appeared to have the largest impact on the joint fit, we constructed a model that varies these alongside the HOD parameters (HOD+cosmo+fbk curves). We find that this model is competitive with the baseline configuration (HOD+cosmo); while the BIC has slightly increased, the AICc has decreased. The MAP estimates for HOD parameters and their posterior distributions (see Fig. \ref{fig:res_5}) are similar between the two models. This similarity is not surprising, given that the baseline model already provides a good fit to $\Delta\Sigma$ on intermediate--large scales, and to $w_{\mathrm{p}}$ across most scales. Our feedback model specifically modifies the small-scale GGL signal. So, feedback parameters can fine-tune the amplitude of $\Delta\Sigma$ on small scales without impacting the fit at larger scales or the fit to $w_{\mathrm{p}}$. The MAP estimates for feedback parameters are $\mathrm{log}\,M_{\mathrm{c}}=11.172$ and $\theta_{\mathrm{ej}}=2.938$. 

Note that all `cosmo' models place $\Omega_{\mathrm{m},0}$ at 0.286---its lowest possible value---and yield similar MAP estimates for $\mathrm{ln}\,(10^{10}A_{\mathrm{s}})$. Additionally, all cosmo models predict nearly identical GGL signals on scales larger than $\sim10\,h^{-1}\,\mathrm{Mpc}$. It is also worth mentioning that $\mathrm{log}\,M_{\mathrm{min}}\sim 13$ and $\mathrm{log}\,M_1\sim 14$ are robust predictions across all models, providing strong evidence that the minimum halo mass required to host a CMASS-like galaxy is $\sim10^{13}\,h^{-1}\,M_{\odot}$.

In addition to the models discussed above, we analysed variants that incorporated halo mass incompleteness. For a detailed description of the incompleteness function and its parameters, we refer readers to \cite{More_2015}. We found that the incompleteness parameters were degenerate with HOD parameters, and that $\mathrm{log}\,M_{\mathrm{min}}$ estimates were unusually low. Since the incompleteness effect did not improve the joint fit, we omitted related models from our discussion.

Comparing all of the different configurations, we have two preferred models: HOD+cosmo and HOD+cosmo+fbk. The HOD+cosmo model yields the lowest BIC score and the second-lowest AICc score, while the HOD+cosmo+fbk model yields the lowest AICc score and the second-lowest BIC score. Regardless, our observables are best fit by a model with flexible cosmology, favouring a slightly lower value of $S_8$ relative to Planck, after corrceting for $m$-bias. 
\begin{table*}
\centering
\begin{tabular}{lccccccr}
\hline
Model & $\chi^2_{\Delta\Sigma}$ & $\chi^2_{w_{\mathrm{p}}}$ & $\chi^2_{\mathrm{tot}}$ & $p$ & AICc & BIC & $k$ \\ \hline
HOD & 45.5 & 22.7 & 69.5 & 0.011 & 80.9 & 89.1 & 5 \\
HOD+$R_{\mathrm{c}}$ & 43.3 & 22.4 & 66.5 & 0.016 & 80.4 & 90.0 & 6 \\
HOD+off & 44.6 & 21.5 & 67.3 & 0.010 & 83.9 & 94.6 & 7 \\
HOD+fbk & 36.5 & 22.6 & 59.9 & 0.045 & 76.6 & 87.3 & 7 \\
HOD+$R_{\mathrm{c}}$+off+fbk & 30.3 & 22.7 & 53.1 & 0.080 & 78.8 & 92.3 & 10 \\
HOD+cosmo & 30.5 & 15.8 & 46.6 & 0.326 & 63.3 & 74.0 & 7 \\
HOD+cosmo+fbk & 23.1 & 16.2 & 39.3 & 0.544 & 61.8 & 74.6 & 9 \\
HOD+cosmo+off & 30.9 & 16.2 & 47.2 & 0.236 & 69.7 & 82.4 & 9 \\
HOD+cosmo+off+fbk & 21.7 & 16.6 & 38.6 & 0.486 & 67.6 & 81.7 & 11 \\
HOD+cosmo+off+fbk+$R_{\mathrm{c}}$ & 22.3 & 15.6 & 38.7 & 0.439 & 71.1 & 85.6 & 12
\end{tabular}
\caption{Statistical values from best-fit models. This table only includes results from fits to both $\Delta\Sigma$ and $w_{\mathrm{p}}$. The number of free parameters is represented by $k$, and the total data vector length across all observables (including $\overline{n}_{\mathrm{g}}$) is 50.} 
\label{table:stats}
\end{table*}
\begin{table*}
\centering
\begin{tabular}{lccccr}
\hline
 & \multicolumn{5}{c}{\textbf{Model}} \\
Parameter & HOD & HOD+$R_{\mathrm{c}}$ & HOD+off & HOD+fbk & HOD+$R_{\mathrm{c}}$+off+fbk \\ \hline
$\mathrm{log}\,M_{\mathrm{min}}$ & 12.983 & 13.004 & 12.997 & 13.016 & 13.227 \\
$\sigma^2_{\mathrm{log}\,M}$ & 0.001 & 0.002 & 0.005 & 0.027 & 0.238 \\
$\mathrm{log}\,M_1$ & 14.114 & 14.152 & 14.106 & 14.105 & 14.133 \\
$\alpha$ & 1.416 & 1.291 & 1.442 & 1.345 & 1.518 \\
$\kappa$ & 0.107 & 0.092 & 0.032 & 0.467 & 0.087 \\
$R_{\mathrm{c}}$ & -- & 0.377 & -- & -- & 2.486 \\
$p_{\mathrm{off}}$ & -- & -- & 0.063 & -- & 0.404 \\
$R_{\mathrm{off}}$ & -- & -- & 4.517 & -- & 2.471 \\
$\mathrm{log}\,M_{\mathrm{c}}$ & -- & -- & -- & 11.432 & 14.028 \\
$\theta_{\mathrm{ej}}$ & -- & -- & -- & 16.918 & 0.380 \\ \hline
 & \multicolumn{5}{c}{\textbf{Model}} \\
Parameter & HOD+cosmo & HOD+cosmo+fbk & HOD+cosmo+off & HOD+cosmo+off+fbk & HOD+cosmo+off+fbk+$R_{\mathrm{c}}$ \\ \hline
$\mathrm{log}\,M_{\mathrm{min}}$ & 13.217 & 13.137 & 13.172 & 13.064 & 13.061 \\
$\sigma^2_{\mathrm{log}\,M}$ & 0.304 & 0.249 & 0.270 & 0.196 & 0.194 \\
$\mathrm{log}\,M_1$ & 14.132 & 14.076 & 14.106 & 13.992 & 13.995 \\
$\alpha$ & 1.359 & 1.341 & 1.411 & 1.358 & 1.340 \\
$\kappa$ & 0.292 & 0.282 & 0.216 & 0.435 & 0.290 \\
$R_{\mathrm{c}}$ & -- & -- & -- & -- & 0.930 \\
$p_{\mathrm{off}}$ & -- & -- & 0.004 & 0.078 & 0.171 \\
$R_{\mathrm{off}}$ & -- & -- & 3.764 & 3.277 & 2.031 \\
$\mathrm{log}\,M_{\mathrm{c}}$ & -- & 11.172 & -- & 12.450 & 14.255 \\
$\theta_{\mathrm{ej}}$ & -- & 2.938 & -- & 1.318 & 0.474 \\
$\mathrm{ln}\,(10^{10}A_{\mathrm{s}})$ & 3.057 & 3.054 & 3.039 & 3.031 & 3.066 \\
$\Omega_{\mathrm{m,0}}$ & 0.286 & 0.286 & 0.286 & 0.286 & 0.286 \\
$\sigma_8$ & 0.761 & 0.761 & 0.756 & 0.752 & 0.765 \\
$S_8$ & 0.743 & 0.743 & 0.738 & 0.734 & 0.748
\end{tabular}
\caption{Best-fit parameter values drawn from the MAP. For the fiducial Planck cosmology, $\sigma_8=0.809$ and $S_8=0.824$.}
\label{table:map}
\end{table*}
\begin{table*}
\centering
\begin{tabular}{lccccr}
\hline
 & \multicolumn{5}{c}{\textbf{Model}} \\
Parameter & HOD & HOD+$R_{\mathrm{c}}$ & HOD+off & HOD+fbk & HOD+$R_{\mathrm{c}}$+off+fbk \\ \hline
$\mathrm{log}\,M_{\mathrm{min}}$ & $13.047^{+0.080}_{-0.045}$ & $13.048^{+0.080}_{-0.045}$ & $13.070^{+0.099}_{-0.057}$ & $13.102^{+0.147}_{-0.077}$ & $13.112^{+0.137}_{-0.086}$ \\
$\sigma^2_{\mathrm{log}\,M}$ & $0.053^{+0.069}_{-0.039}$ & $0.050^{+0.072}_{-0.037}$ & $0.068^{+0.081}_{-0.050}$ & $0.091^{+0.113}_{-0.065}$ & $0.114^{+0.118}_{-0.077}$ \\
$\mathrm{log}\,M_1$ & $14.102^{+0.061}_{-0.078}$ & $14.115^{+0.072}_{-0.010}$ & $14.117^{+0.067}_{-0.070}$ & $14.135^{+0.105}_{-0.091}$ & $14.097^{+0.088}_{-0.092}$ \\
$\alpha$ & $1.262^{+0.141}_{-0.233}$ & $1.248^{+0.145}_{-0.199}$ & $1.317^{+0.120}_{-0.212}$ & $1.228^{+0.170}_{-0.288}$ & $1.323^{+0.175}_{-0.260}$ \\
$\kappa$ & $0.729^{+0.785}_{-0.497}$ & $0.643^{+0.924}_{-0.472}$ & $0.525^{+0.700}_{-0.380}$ & $0.774^{+0.857}_{-0.529}$ & $0.542^{+1.036}_{-0.403}$ \\
$R_{\mathrm{c}}$ & -- & $0.886^{+0.864}_{-0.417}$ & -- & -- & $1.757^{+0.856}_{-0.867}$ \\
$p_{\mathrm{off}}$ & -- & -- & $0.056^{+0.044}_{-0.035}$ & -- & $0.116^{+0.221}_{-0.086}$ \\
$R_{\mathrm{off}}$ & -- & -- & $3.084^{+1.264}_{-1.493}$ & -- & $2.587^{+1.311}_{-0.983}$ \\
$\mathrm{log}\,M_{\mathrm{c}}$ & -- & -- & -- & $12.218^{+4.041}_{-1.461}$ & $11.601^{+3.314}_{-1.115}$ \\
$\theta_{\mathrm{ej}}$ & -- & -- & -- & $13.669^{+2.355}_{-3.840}$ & $7.521^{+6.923}_{-6.977}$ \\ \hline
 & \multicolumn{5}{c}{\textbf{Model}} \\
Parameter & HOD+cosmo & HOD+cosmo+fbk & HOD+cosmo+off & HOD+cosmo+off+fbk & HOD+cosmo+off+fbk+$R_{\mathrm{c}}$ \\ \hline
$\mathrm{log}\,M_{\mathrm{min}}$ & $13.144^{+0.182}_{-0.136}$ & $13.134^{+0.164}_{-0.130}$ & $13.150^{+0.202}_{-0.135}$ & $13.125^{+0.150}_{-0.121}$ & $13.129^{+0.148}_{-0.131}$ \\
$\sigma^2_{\mathrm{log}\,M}$ & $0.224^{+0.153}_{-0.118}$ & $0.226^{+0.135}_{-0.109}$ & $0.225^{+0.165}_{-0.117}$ & $0.223^{+0.123}_{-0.106}$ & $0.233^{+0.130}_{-0.108}$ \\
$\mathrm{log}\,M_1$ & $14.050^{+0.126}_{-0.136}$ & $14.041^{+0.119}_{-0.131}$ & $14.061^{+0.134}_{-0.128}$ & $14.027^{+0.110}_{-0.133}$ & $13.983^{+0.135}_{-0.153}$ \\
$\alpha$ & $1.239^{+0.198}_{-0.305}$ & $1.230^{+0.190}_{-0.271}$ & $1.270^{+0.182}_{-0.302}$ & $1.230^{+0.198}_{-0.273}$ & $1.251^{+0.221}_{-0.302}$ \\
$\kappa$ & $0.693^{+0.903}_{-0.480}$ & $0.702^{+0.805}_{-0.477}$ & $0.620^{+0.828}_{-0.444}$ & $0.658^{+0.814}_{-0.469}$ & $0.943^{+0.988}_{-0.678}$ \\
$R_{\mathrm{c}}$ & -- & -- & -- & -- & $1.929^{+0.740}_{-0.889}$ \\
$p_{\mathrm{off}}$ & -- & -- & $0.026^{+0.105}_{-0.019}$ & $0.063^{+0.134}_{-0.048}$ & $0.070^{+0.153}_{-0.051}$ \\
$R_{\mathrm{off}}$ & -- & -- & $1.843^{+2.223}_{-1.789}$ & $1.834^{+1.987}_{-1.699}$ & $1.985^{+1.964}_{-1.711}$ \\
$\mathrm{log}\,M_{\mathrm{c}}$ & -- & $11.178^{+0.825}_{-0.785}$ & -- & $11.263^{+1.246}_{-0.884}$ & $11.457^{+1.299}_{-0.961}$ \\
$\theta_{\mathrm{ej}}$ & -- & $6.039^{+5.541}_{-3.126}$ & -- & $3.038^{+5.111}_{-1.831}$ & $2.905^{+4.124}_{-2.046}$
\end{tabular}
\caption{Posterior constraints from MCMC chains. We report the 50th percentile as the central value, with lower and upper limits given by the differences between the 50th and 16th percentiles, and the 84th and 50th percentiles, respectively. We quote the median instead of the mean since many of the posterior distributions are skewed.}
\label{table:median}
\end{table*}

\subsection{Constraints from the Kinetic Sunyaev-Zeldovich Effect}
\begin{figure*}
\includegraphics[width = 0.75\textwidth]{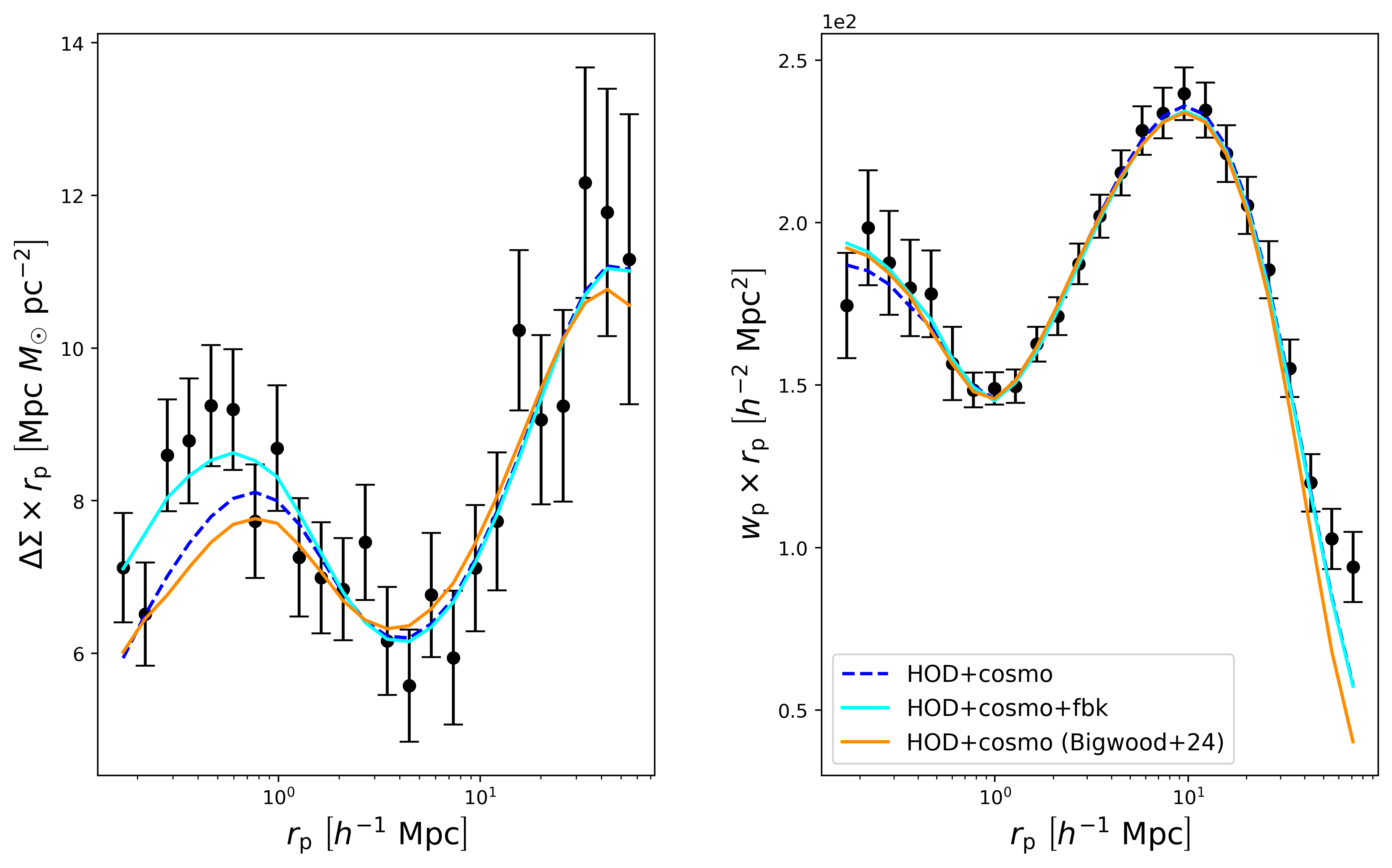}
\caption{Best-fit model adopting the feedback constraints from Table B2, column `WL + kSZ' in \protect\cite{2024MNRAS.534..655B} (solid orange). Our preferred models, HOD+cosmo and HOD+cosmo+fbk, are shown for comparison.}
\label{fig:bigwood}
\end{figure*}

\cite{2024MNRAS.534..655B} constrain feedback using the same BC model adopted in this work. They calibrate feedback by incorporating measurements of the kSZ effect from Atacama Cosmology Telescope\footnote{\url{https://act.princeton.edu}} (ACT; \citealp{2010ApJS..191..423H}) DR5 \citep{2021PhRvD.103f3513S} and fit this along with cosmological parameters to the Dark Energy Survey Year 3 (DES Y3; \citealp{2021ApJS..254...24S}) cosmic shear 2PCFs, $\xi_{\pm}(\theta)$. Since the strength of feedback is degenerate with $S_8$, the kSZ effect helps break degeneracies between feedback and cosmological parameters. We set BC model parameters to their weak-lensing + kSZ constraints and ran an additional MCMC chain varying only HOD and cosmological parameters. Specifically, we used the mean values in Table B2, column `WL + kSZ'. Note that they adopt a slightly different notation, with $\gamma\equiv\gamma_{\mathrm{g}}$, $\delta\equiv\delta_{\mathrm{g}}$, and $\eta\equiv\eta_{\mathrm{star}}$. All other feedback parameters not listed in Table B2 were fixed at the fiducial values in \cite{2024MNRAS.534..655B}. The resulting best-fit model, HOD+cosmo (Bigwood+24), is shown in Fig. \ref{fig:bigwood}. 

From Fig. \ref{fig:bigwood}, we see that the HOD+cosmo (Bigwood+24) model under-predicts the small-scale GGL signal more than the HOD+cosmo model. The cosmological parameters derived from this model are, however, similar to those derived from previous HOD+cosmo models since $S_8$ cannot be raised without over-predicting the large-scale GGL signal. \cite{2024MNRAS.534..655B} report $S_8=0.823^{+0.019}_{-0.020}$, consistent with Planck but with extremely strong negative feedback (in fact, stronger than the predictions of popular hydrodynamical simulations such as \citealp{2018MNRAS.479.5385H} and \citealp{2023MNRAS.524.2539P}). It is important to note, however, that cosmic shear is measured as a function of angular separation, whereas GGL is measured as a function of projected distance. As a result, baryonic effects are mixed across a wide range of angular scales in cosmic shear, while in GGL their influence is more localized (i.e., primarily confined to the 1-halo term, or below $\sim2\,h^{-1}\,\mathrm{Mpc}$). This may explain why \cite{2024MNRAS.534..655B} can strongly suppress $P_{\mathrm{mm}}$ through feedback while still yielding a higher $S_8$. 

Our results highlight the value of GGL as a weak-lensing probe, particularly in the context of the $S_8$ tension. Since the GGL signal is defined in terms of projected distance (given by the spectroscopic redshifts of lenses), baryonic and cosmological effects can be more cleanly disentangled than in cosmic shear, provided that both small and large scales are modelled. This also helps explain why the HOD+cosmo and HOD+cosmo+fbk models yield the same best-fit value of $S_8=0.743$. 

\subsection{Comparison}
To compare our `best' models with results from other studies, we first compute the mean host-halo mass, $\overline{M}_{\mathrm{h}}$:
\begin{equation}
\overline{M_{\mathrm{h}}}\,=\,\frac{\int M\langle N_{\mathrm{c}}(M) \rangle\frac{\mathrm{d}n_{\mathrm{h}}}{\mathrm{d}M}\mathrm{d}M}{\int\langle N_{\mathrm{c}}(M) \rangle\frac{\mathrm{d}n_{\mathrm{h}}}{\mathrm{d}M}\mathrm{d}M}.
\end{equation}
For the HOD+cosmo model, we obtain $\mathrm{log}_{10}\,(\overline{M}_{\mathrm{h}}/h^{-1}\,M_{\odot})=13.351$. Comparable results are presented in \cite{Saito_2016}, who construct mock CMASS catalogues using both a abundance matching (\citealp{2002ApJ...569..101M,2004MNRAS.353..189V}) methodology and an extended version incorporating age matching. For the standard subhalo abundance matching (SHAM) case, they find mean halo masses of $\mathrm{log}_{10}\,(\overline{M}_{\mathrm{h}}/h^{-1}\,M_{\odot})$ = 13.120, 13.340, and 13.660 for central galaxies at redshifts $z$ = 0.445, 0.565, and 0.685, respectively. The corresponding values for the extended SHAM approach are $\mathrm{log}_{10}\,(\overline{M}_{\mathrm{h}}/h^{-1}\,M_{\odot})$ = 13.150, 13.350, and 13.680. Converting our mean halo mass to the \textsc{Rockstar} definition used in \cite{Saito_2016} via \textsc{Colossus}, we get $\mathrm{log}_{10}\,(\overline{M}_{\mathrm{h}}/h^{-1}\,M_{\odot})=13.330$, which closely matches the \cite{Saito_2016} result at $z=0.565$ (recall that our fiducial redshift is $z=0.54$). Since our feedback procedure alters the $M_{200\mathrm{m}}$ of a given halo, we do not compute a mean host-halo mass for the HOD+cosmo+fbk model.

We also calculate the mean satellite galaxy fraction, $\overline{f}_{\mathrm{s}}$:
\begin{equation}
\overline{f}_{\mathrm{s}}\,=\,\frac{\int\langle N_{\mathrm{s}}(M) \rangle\frac{\mathrm{d}n_{\mathrm{h}}}{\mathrm{d}M}\mathrm{d}M}{\int\langle N_{\mathrm{g}}(M)\rangle\frac{\mathrm{d}n_{\mathrm{h}}}{\mathrm{d}M}\mathrm{d}M}\,=\,\frac{\overline{n}_{\mathrm{s}}}{\overline{n}_{\mathrm{g}}}.
\end{equation}
We find $\overline{f}_{\mathrm{s}}=0.085=8.5\%$ for the HOD+cosmo model, and $\overline{f}_{\mathrm{s}}=0.098=9.8\%$ for the HOD+cosmo+fbk model. Note that feedback does not affect the computation of $\overline{f}_{\mathrm{s}}$. In comparison, \cite{Saito_2016} report satellite fractions between 8.0 and $12.5\%$ across the redshift bins previously listed. \cite{White_2011} find satellite fractions between 5.0 and $12.0\%$ (includes $1\sigma$ uncertainties), based on an HOD fit to $w_{\mathrm{p}}$ over approximately $0.4\text{--}23\,h^{-1}\,\mathrm{Mpc}$ and within the redshift range $0.4<z<0.7$. \cite{More_2015} quote satellite fractions of $2.5\text{--}11.4\%$ (includes $1\sigma$ uncertainties) between different stellar mass subsamples in the redshift range $0.47\leq z\leq0.59$. Generally speaking, $\overline{f}_{\mathrm{s}}\sim0.10\pm0.03$ \citep{Lange_2021}. 

Finally, we compute the large-scale effective galaxy bias, $b_{\mathrm{eff}}$:
\begin{equation}
b_{\mathrm{eff}}\,=\,\frac{\int b(M)\langle N_{\mathrm{g}}(M)\rangle\frac{\mathrm{d}n_{\mathrm{h}}}{\mathrm{d}M}\mathrm{d}M}{\int\langle N_{\mathrm{g}}(M)\rangle\frac{\mathrm{d}n_{\mathrm{h}}}{\mathrm{d}M}\mathrm{d}M},
\end{equation}
where $b(M)$ is the linear halo bias at mass $M$. We obtain a large-scale effective galaxy bias of 2.15 for both the HOD+cosmo and HOD+cosmo+fbk models---precisely the same central value reported by \cite{More_2015} for their fiducial stellar mass bin ($11.1\leq\mathrm{log}_{10}(M_*/h^{-2}M_{\odot})\leq12.0$). Since our feedback procedure only modifies $P_{\mathrm{mm}}$ on small scales, $b_{\mathrm{eff}}$ is unaffected (although the scale-dependent galaxy bias would change on small scales). \cite{White_2011} also find a large-scale galaxy bias of $\sim2$---a value commonly quoted in the literature. 

\subsection{Discussion}
Our results suggest that the lensing is low effect is not confined to small scales. The precision of our large-scale $\Delta\Sigma$ measurements now demands more accurate modelling at these scales and enables us to distinguish between different cosmologies. We find that the HOD+cosmo configuration, one of the preferred models with the lowest AICc and BIC scores, does not provide a good fit to $\Delta\Sigma$ on small scales. However, due to its very good fit on intermediate--large scales, HOD+cosmo remains one of the best models, highlighting the importance of large scales in driving our fits. This stands in contrast to the results of \cite{Amon23}, who find that the large-scale data alone lacks the statistical power to distinguish between Planck and lensing cosmologies ($S_8=0.83$ and 0.76, respectively). They likewise report no significant evidence for a mismatch between $\Delta\Sigma$ and $w_{\mathrm{p}}$ on large scales. Note that our $\Delta\Sigma$ error bars are smaller than those in \cite{Amon23} for the various BOSS subsamples. For example, their $1\sigma$ error on $\Delta\Sigma$ is $\sim3.6\,\mathrm{Mpc}\,M_\odot\,\mathrm{pc}^{-2}$ at $r_{\mathrm{p}}\sim53.8\,h^{-1}\,\mathrm{Mpc}$ for CMASS galaxies in the redshift range $0.54<z<0.7$, whereas our corresponding error is $\sim1.9\,\mathrm{Mpc}\,M_\odot\,\mathrm{pc}^{-2}$.

While we find that a slightly lower $S_8$ cosmology is a better fit than the Planck cosmological parameters, we cannot interpret our results as evidence for a low $S_8$. This is mainly because IAs, which impact $\Delta\Sigma$ across a broad range of scales, have not yet been properly investigated\footnote{A complete analysis would additionally require marginalisation over all cosmological parameters.}. If IAs prove to be a significant contaminant, their mitigation could considerably increase the amplitude of the measured GGL signal (or decrease the amplitude of the predicted signal). This could potentially reconcile $\Delta\Sigma$ and $w_{\mathrm{p}}$ without requiring a departure from the fiducial Planck cosmology. Such a scenario is possible, given that a large $A_{\mathrm{IA}}$ of 1 results in a $\sim20\%$ reduction in $\gamma_{\mathrm{t}}$ under the NLA model (see Section \ref{sec:ia}), while the lensing is low effect is at the $\sim7\%$ level.  

Ideally, IAs would be strongly suppressed through tomography since $A_{\mathrm{IA}}$ is degenerate with (positively correlated to) $S_8$. This degeneracy inflates the constraint on $S_8$, reducing our sensitivity to different cosmologies. Ensuring that the lens and source galaxies are well-separated in redshift would reduce the likelihood of physical associations, which produce IAs. This strategy will be possible in the future, once photometric redshifts are available in UNIONS. 

In addition to enabling tomography, individual photometric redshifts yield more optimal lens--source weights (Eq. \ref{eqn:pair}) since physically close pairs are down-weighted. This, in turn, increases the S/N of the GGL measurement for the same source sample. Since our current lens--source weights are not optimal, the statistical uncertainties on $\Delta\Sigma$ should be regarded as upper limits. 

These preliminary results demonstrate the potential of UNIONS data in elucidating the lensing is low problem and showcase the methodology that will allow us to deliver competitive $S_8$ constraints in future work. This study also aims to drive more interest in the lensing is low problem, as we enter the era of high-precision cosmology. 

\section{Conclusions}
\label{sec:conclusions}
In this paper, we revisit the `lensing is low' problem introduced by \citetalias{lilo17} through a joint analysis of the galaxy--galaxy lensing (GGL) and galaxy clustering (GC) signals of BOSS CMASS galaxies. We provide a fresh perspective on the problem by leveraging new GGL measurements from Ultraviolet Near Infrared Optical Northern Survey (UNIONS) data and analysing several extensions of the standard halo occupation distribution (HOD). We adopt the HOD parametrisation of \cite{zheng05}, implemented via the \textsc{Dark Emulator}. This model is characterised by 5 parameters describing the occupation of halos by central and satellite galaxies. We extend this framework to include central galaxy off-centring, feedback, and a modified satellite concentration normalisation. Our feedback method follows the BC models of \cite{2015JCAP...12..049S} and \cite{shirasaki2024massessunyaevzeldovichgalaxyclusters}, and modifies the 1-halo term of the halo--matter cross--power spectrum, $P_{\mathrm{hm}}$. Model parameters are constrained using MCMC sampling, where the GGL and GC signals are fit simultaneously (except for one chain). We also use the observed galaxy abundance as an additional constraint on the fits. Our main results are as follows:

\begin{itemize}
\item 
Adopting the Planck 2018 TT,TE,EE+lowE+lensing+BAO cosmology, the predicted GGL signal from a standard 5-parameter HOD fit to GC alone exceeds the amplitude of our measurements by $\sim7\%$, allowing for $m$-bias. The uncertainties on this are likely at the 3-4\% level, and so we cannot exclude the Planck cosmology at high significance, even before considering IAs. 
\item We can improve the agreement on large scales slightly by lowering the value of $S_8$ relative to Planck. Without the inclusion of and marginalisation over feedback parameters, such models under-predict the small-scale GGL signal. 
\item Feedback in models with free cosmological parameters enhances the small-scale GGL signal (or $P_{\mathrm{mm}}$), in contrast to the typical expectation that feedback suppresses $P_{\mathrm{mm}}$. 
\end{itemize}

Another possible large-scale effect is intrinsic alignments (IAs). We defer IA mitigation to future work where tomography with UNIONS will be possible, to avoid the $A_{\mathrm{IA}}\text{--}S_8$ degeneracy. In this fiducial analysis, setting $A_{\mathrm{IA}}=0$ allowed us to explore the impact of HOD models, feedback, and different cosmologies on the lensing is low problem.

A key astrophysical effect missing from this analysis is assembly bias (AB). However, there is no purely analytic description of AB (unlike feedback), so investigating this effect would require a departure from our \textsc{Dark Emulator} framework. We leave incorporating AB to future work. 

In summary, this work presents new GGL measurements from UNIONS with high precision on large scales. We additionally establish the methodology that will be used to shed light on the lensing is low problem and to deliver robust $S_8$ constraints in forthcoming tomographic UNIONS analyses. The combination of GGL and GC breaks degeneracies between HOD and cosmological parameters. GGL also cleanly separates small- and large-scale effects (as opposed to cosmic shear), helping to disentangle degeneracies between cosmological and feedback parameters. 

\section*{Acknowledgements} MJH and LVW acknowledge support from NSERC through their resepctive Discovery Grants. FHP acknowledges support from CNES. HH is supported by a DFG Heisenberg grant (Hi 1495/5-1), the DFG Collaborative Research Center SFB1491, an ERC Consolidator Grant (No. 770935), and the DLR project 50QE2305. 

We are honoured and grateful for the opportunity to observe the Universe from Maunakea and Haleakala, which both have cultural, historical and natural significance in Hawaii. This work is based on data obtained as part of the Canada-France Imaging Survey, using observations obtained with MegaPrime/MegaCam, a joint project of the Canada-France-Hawaii Telescope (CFHT) and CEA Saclay, on the CFHT, which is operated by the National Research Council of Canada, the Institut National de Science de l’Univers of the Centre National de la Recherche Scientifique of France, and the University of Hawaii. This research is based in part on data collected at Subaru Telescope, which is operated by the National Astronomical Observatory of Japan. Pan-STARRS is a project of the Institute for Astronomy of the University of Hawaii, and is supported by the NASA SSO Near Earth Observation Program under grants 80NSSC18K0971, NNX14AM74G, NNX12AR65G, NNX13AQ47G, NNX08AR22G, 80NSSC21K1572, and by the State of Hawaii.

\section*{Data Availability}
BOSS data is publicly available on the Science Archive Server (SAS): \url{https://data.sdss.org/sas}. Raw and processed UNIONS data is available to members of the Canadian, French, Japanese, and Pan-STARRS communities. All UNIONS data will be publicly available to the international community at the end of the proprietary period. 



\bibliographystyle{mnras}
\bibliography{example} 




\appendix
\section{Cosmological Parameter Dependence of Measurements}
\label{sec:cosmo_dependence}
We perform measurements adopting a fiducial cosmology to convert redshift into distance. However, we vary $\Omega_{\mathrm{m,0}}$ in our modelling, so we must ensure that predictions and measurements correspond to the `same' comoving separations. We must also consider how the amplitudes of our observables change under different cosmologies. To account for these effects, we follow the procedure in \cite{More_2015}. We specifically modify the \textsc{Dark Emulator} predictions for non-fiducial cosmologies. 

First, we rescale $r_{\mathrm{p}}$ in the following manner:  
\begin{equation}
\tilde{r}_{\mathrm{p}}\,=\,r_{\mathrm{p,\,fid}}\left[\frac{\chi_{\mathrm{new}}(\overline{z})}{\chi_{\mathrm{fid}}(\overline{z})}\right],
\end{equation}
where $\overline{z}=0.54$ is the median lens redshift, $r_{\mathrm{p,\,fid}}$ is the $r_{\mathrm{p}}$ value at which a measurement was made, `new' denotes the new cosmology, and `fid' denotes the fiducial cosmology used to make measurements. Now, we can evaluate our observables at $\tilde{r}_{\mathrm{p}}$. However, before comparing predictions and measurements (e.g., during parameter estimation), we must adjust the amplitudes of our predictions: 
\begin{equation}
\Delta\tilde{\Sigma}(r_{\mathrm{p}})\,=\,\Delta\Sigma(\tilde{r}_{\mathrm{p}})\left[\frac{\Sigma_{\mathrm{crit,\,fid}}(\overline{z},\,z_{\mathrm{s}})}{\Sigma_{\mathrm{crit,\,new}}(\overline{z},\,z_{\mathrm{s}})}\right],
\label{eqn:esd_corr}
\end{equation}
\vspace{-1mm}
\begin{equation}
\tilde{w}_{\mathrm{p}}(r_{\mathrm{p}})\,=\,w_{\mathrm{p}}(\tilde{r}_{\mathrm{p}})\left[\frac{E_{\mathrm{new}}(\overline{z})}{E_{\mathrm{fid}}(\overline{z})}\right],
\end{equation}
where $\Delta\Sigma(\tilde{r}_{\mathrm{p}})$ and $w_{\mathrm{p}}(\tilde{r}_{\mathrm{p}})$ represent the \textsc{Dark Emulator} predictions at $\tilde{r}_{\mathrm{p}}$. The choice of $z_{\mathrm{s}}$ is somewhat arbitrary, as stated in \cite{More_2015}. Note that the fraction in Eq. \ref{eqn:esd_corr} differs from the expression in \cite{More_2015}, where the inverse is presented. 

In addition to correcting our predictions of $\Delta\Sigma$ and $w_{\mathrm{p}}$, we modify our prediction of $\overline{n}_{\mathrm{g}}$, as it is subject to a prior: 
\begin{equation}
\tilde{\overline{n}}_{\mathrm{g}}\,=\,\overline{n}_{\mathrm{g,\,new}}\frac{\chi^3_{\mathrm{new}}(z_2)\,-\,\chi^3_{\mathrm{new}}(z_1)}{\chi^3_{\mathrm{fid}}(z_2)\,-\,\chi^3_{\mathrm{fid}}(z_1)},
\end{equation} 
where $\overline{n}_{\mathrm{g,\,new}}$ corresponds to the redshift bin $[z_2,z_1]=[0.7,0.43]$. Note that $\overline{n}_{\mathrm{g}}$ does not need to be corrected when computing $\Delta\Sigma$ or $w_{\mathrm{p}}$.

\section{$\Delta\Sigma$ and $w_{\mathrm{p}}$ Cross-Covariance Test}
\label{sec:cross_cov_test}
We assume that $\Delta\Sigma$ and $w_{\mathrm{p}}$ are uncorrelated during parameter estimation. This assumption is not always valid, and the degree of correlation usually depends on the amount of overlap between the effective areas used to measure $\Delta\Sigma$ and $w_{\mathrm{p}}$. Since the angular overlap between UNIONS and CMASS galaxies is significant, we devised a test to evaluate the assumption of negligible cross-covariance, described below. 

\begin{figure}
\includegraphics[width = \columnwidth]{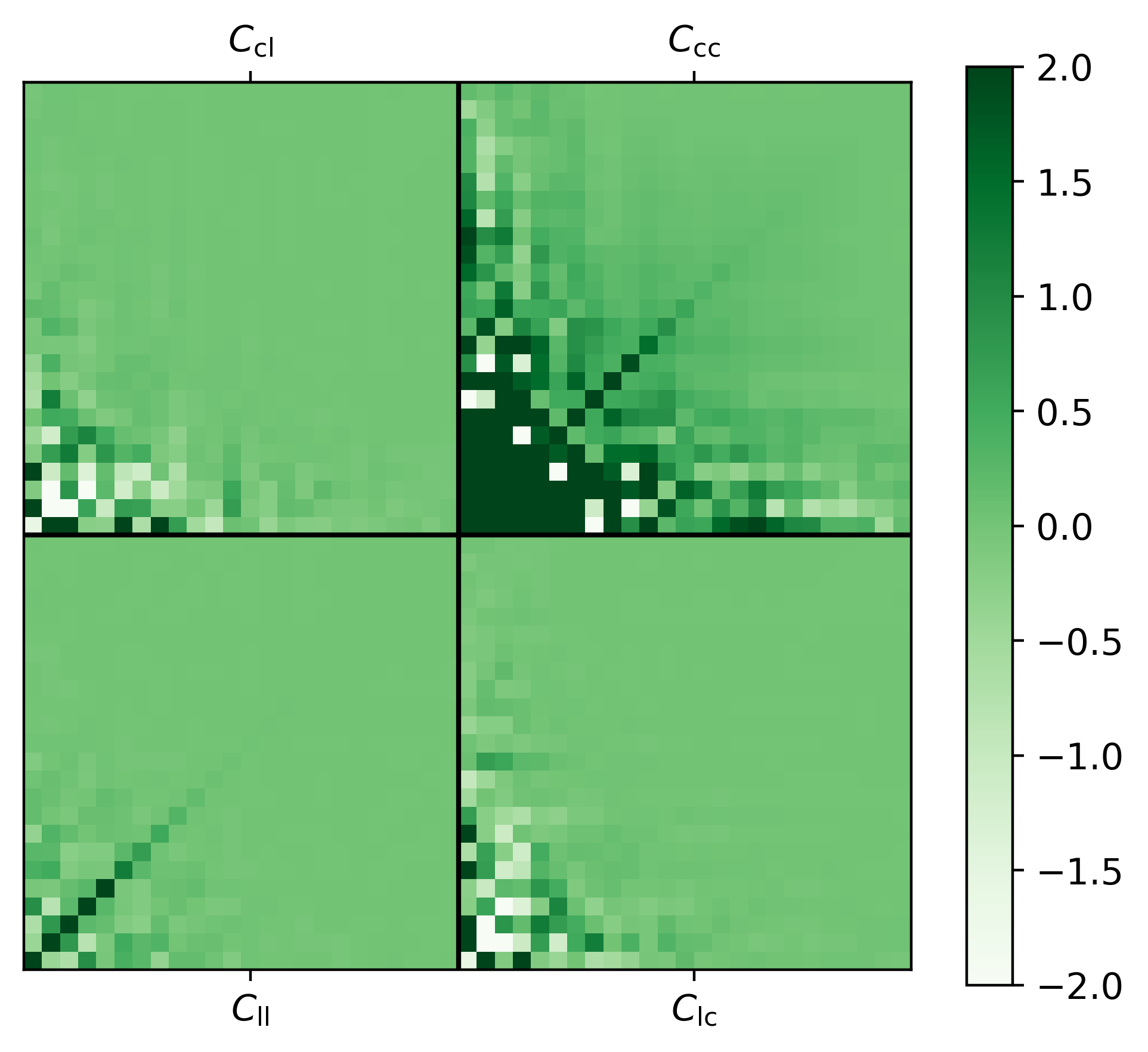}
\caption{Raw joint covariance matrix for $\Delta\Sigma$ and $w_{\mathrm{p}}$. `C' stands for clustering, and `l' for lensing. Each block is an individual covariance matrix. Note that the colour scale is set to the range $[-2,2]$.}
\label{fig:block_cov}
\end{figure}

First, we constructed a block covariance matrix with the $\Delta\Sigma$ and $w_{\mathrm{p}}$ (uncorrected) statistical covariance matrices along the diagonal. We then estimated the cross-covariance blocks once again using jackknife resampling. We computed 150 $w_{\mathrm{p}}$ samples in the UNIONS--CMASS overlap region using the same jackknife patches from our $\Delta\Sigma$ statistical covariance estimation, from which we calculated  $\bm{C}_{\mathrm{JK}}(\Delta\Sigma,\,w_{\mathrm{p,\,overlap}})$ using Eq. \ref{eqn:jk}. After, we approximated the cross-covariance between $\Delta\Sigma$ (by default measured in the survey overlap region) and $w_{\mathrm{p}}$ measured across the entire CMASS footprint by rescaling our cross-covariance matrix that assumes 100$\%$ survey overlap:
\begin{equation}
\bm{C}_{\mathrm{JK}}(\Delta\Sigma,\,w_{\mathrm{p}}) \approx
\bm{C}_{\mathrm{JK}}(\Delta\Sigma,\,w_{\mathrm{p,\,overlap}})\times\frac{A_{\mathrm{overlap}}}{A_{\mathrm{tot}}},
\end{equation}
where $A$ denotes an angular area, and $A_{\mathrm{tot}}$ is the total CMASS footprint area. The expression above assumes that the cross-covariance between $\Delta\Sigma$ and $w_{\mathrm{p}}$ is zero in non-overlapping regions. We ensured that the final block covariance matrix is positive semi-definite. A visualisation of the block covariance matrix is shown in Fig. \ref{fig:block_cov}.  

\begin{figure}
\includegraphics[width = \columnwidth]{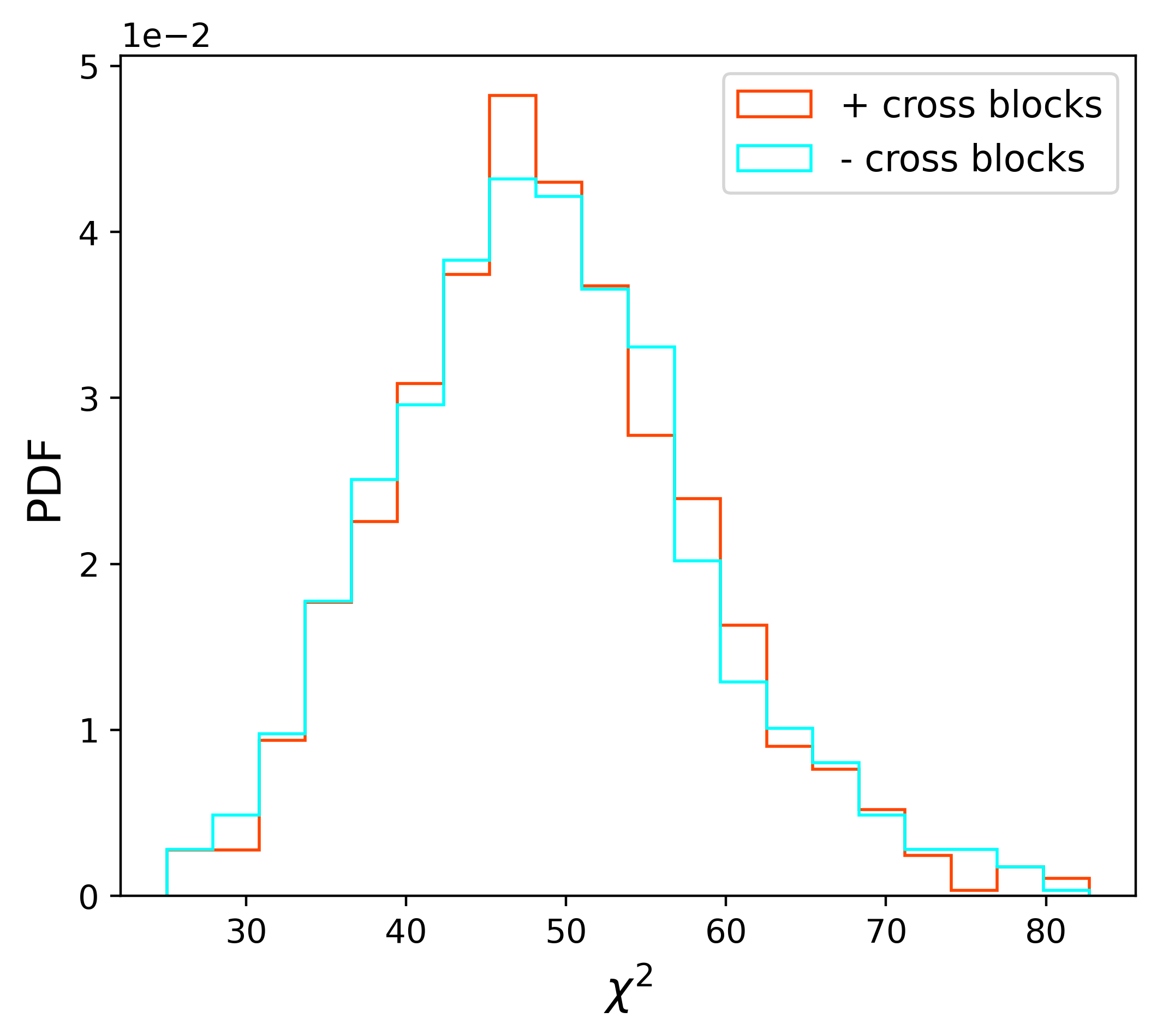}
\caption{$\chi^2$ distributions derived from the joint covariance matrix in Fig. \ref{fig:block_cov}. In one configuration, the cross-covariance blocks are set to zero (- cross blocks). `PDF' on the y-axis stands for `probability density function'.}
\label{fig:chi2}
\end{figure}

Next, we generated fake data vectors from a multivariate Gaussian distribution using the final block covariance matrix. We then computed two sets of $\chi^2$ values using Eq. \ref{eqn:chi2}. The first set used the final block covariance matrix for $\bm{C}^{-1}$, while the second used the same matrix but with the cross-covariance blocks set to zero. The resulting $\chi^2$ distributions are shown in Fig. \ref{fig:chi2}. 
\begin{figure}
\includegraphics[width = \columnwidth]{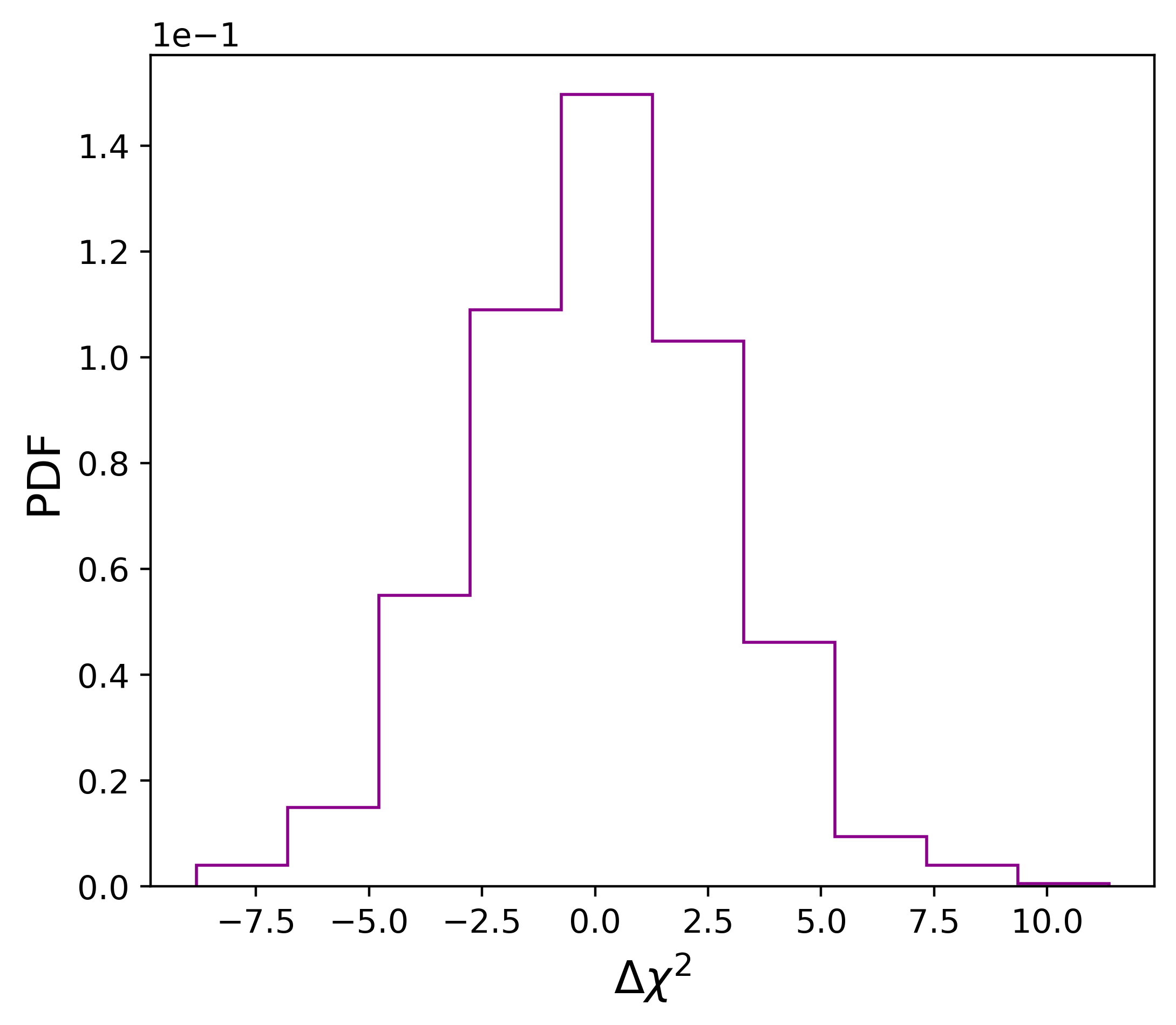}
\caption{$\Delta\chi^2$ distribution from the differences between the datasets in Fig. \ref{fig:chi2}.}
\label{fig:delta_chi2}
\end{figure}

\noindent
The $\chi^2$ distributions in Fig. \ref{fig:chi2} are quite similar, and the $\Delta\chi^2$ distribution in Fig. \ref{fig:delta_chi2} is narrowly peaked (compared to the individual $\chi^2$ distributions) and centred around 0. We thus concluded that ignoring the cross-covariance between our measurements of $\Delta\Sigma$ and $w_{\mathrm{p}}$ does not significantly bias $\chi^2$ values. Note that we did not use the joint covariance matrix during parameter estimation, as the jackknife patches used to estimate the $w_{\mathrm{p}}$ block differ from those used for the other blocks. Properly de-biasing our joint covariance matrix would be a non-trivial task. 

\section{Additional Plots}
\label{additional_plots}
\begin{figure*}
\centering
\includegraphics[width = 1\textwidth]{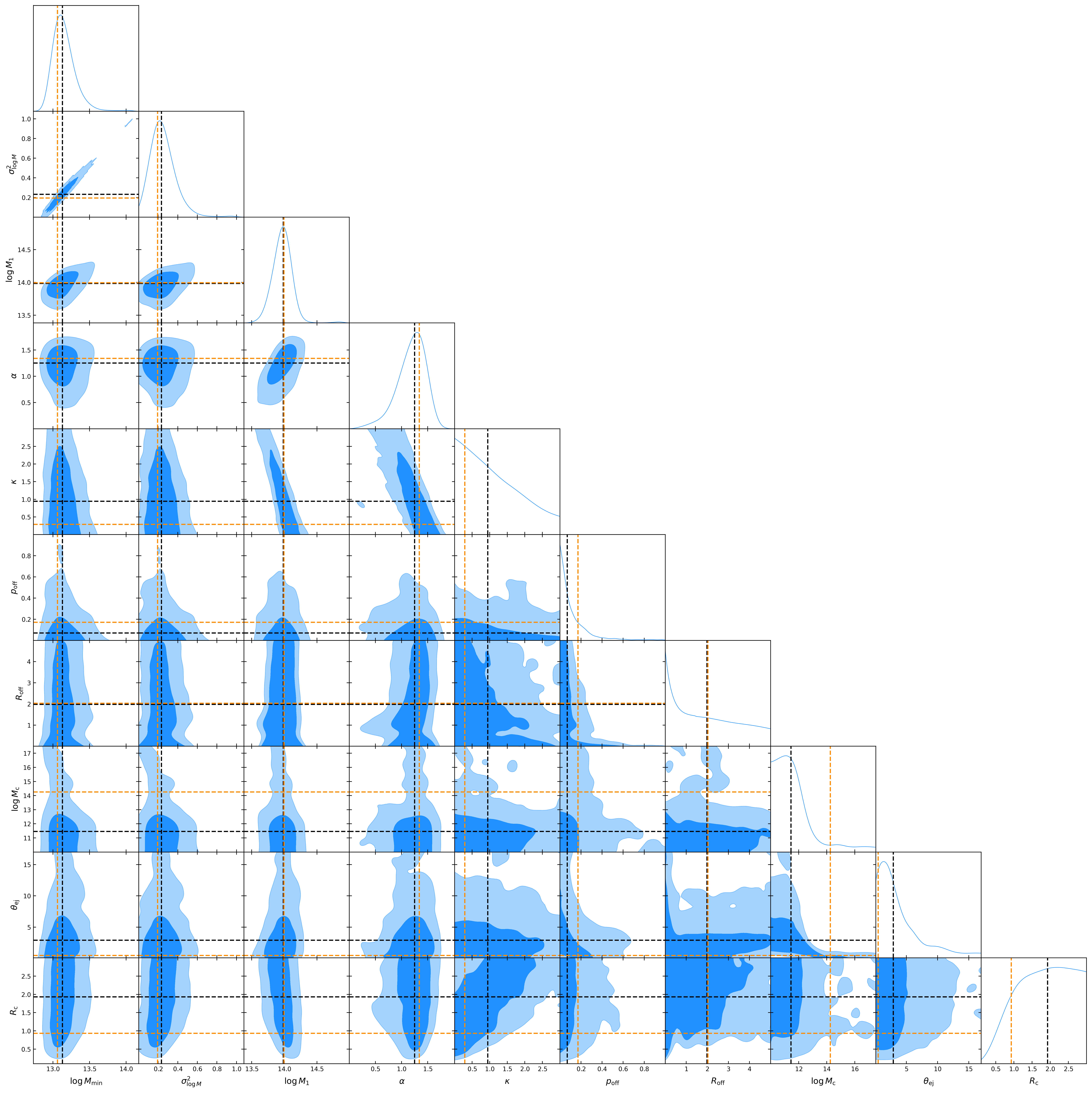}
\caption{Posterior distributions for the HOD+cosmo+off+fbk+$R_{\mathrm{c}}$ model.}
\label{fig:res_6}
\end{figure*}
\begin{table}
\centering
\begin{tabular}{l|r}
\hline
Parameter & Fiducial Value \\ \hline
$\mathrm{log}\,M_{\mathrm{min}}$ & 13.5 \\
$\sigma^2_{\mathrm{log}\,M}$ & 0.5 \\
$\mathrm{log}\,M_1$ & 14.0 \\
$\alpha$ & 1.0 \\
$\kappa$ & 1.0 \\
$R_{\mathrm{c}}$ & 1.0 \\
$p_{\mathrm{off}}$ & 0.2 \\
$R_{\mathrm{off}}$ & 1.0 \\
$\mathrm{ln}\,(10^{10}A_{\mathrm{s}})$ & 3.0 \\
$\Omega_{\mathrm{m,0}}$ & 0.3 \\
\end{tabular}
\caption{Fiducial parameter values used to generate the `rainbow plots' in this Appendix. By default, feedback is not included in the plots.}
\label{table:appendix_table}
\end{table}

\begin{figure*}
\centering
\subfigure[$\mathrm{log}\,M_{\mathrm{min}}$]{\includegraphics[width = 0.49\textwidth]{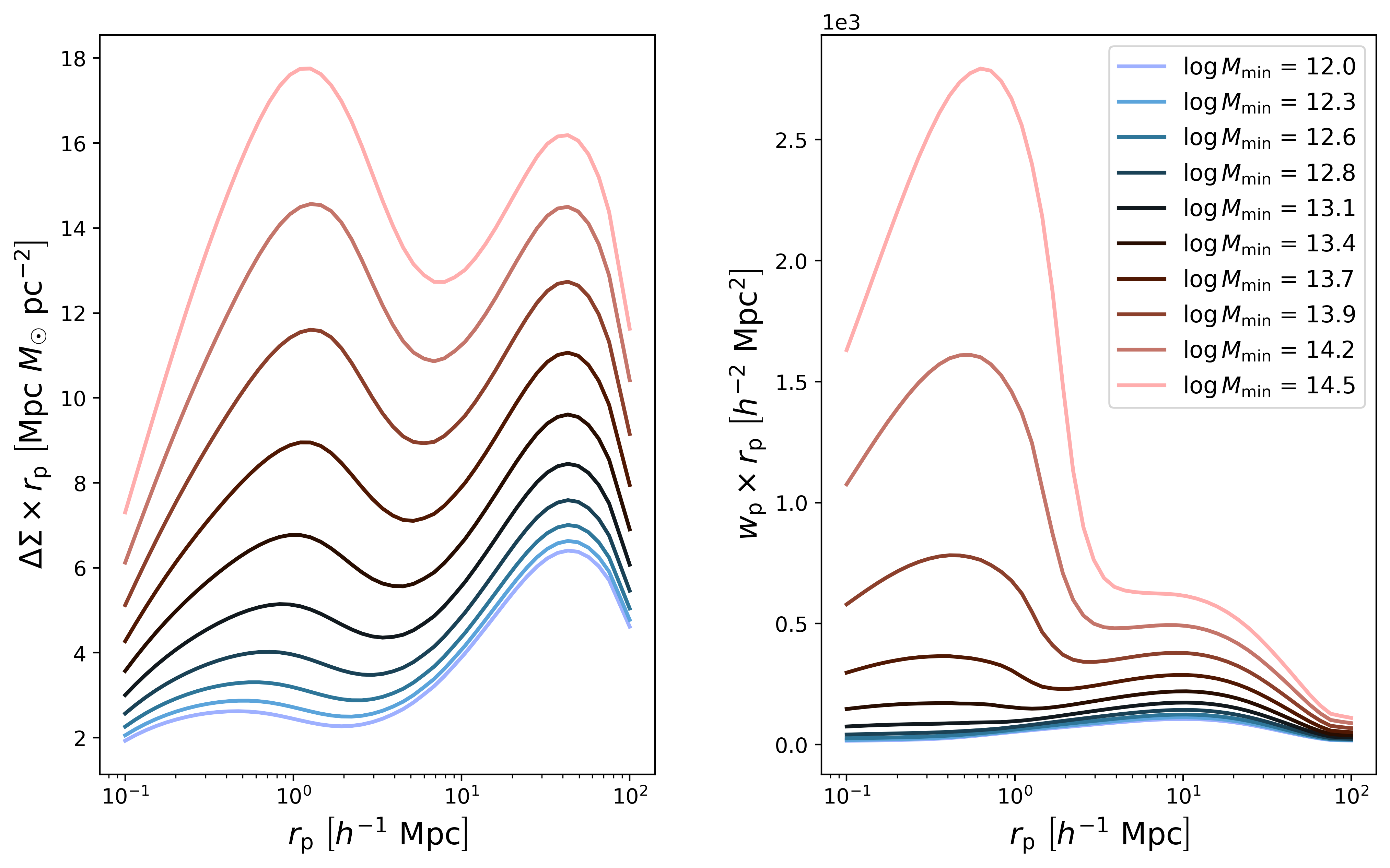}}
\subfigure[$\sigma^2_{\mathrm{log}\,M}$]{\includegraphics[width = 0.49\textwidth]{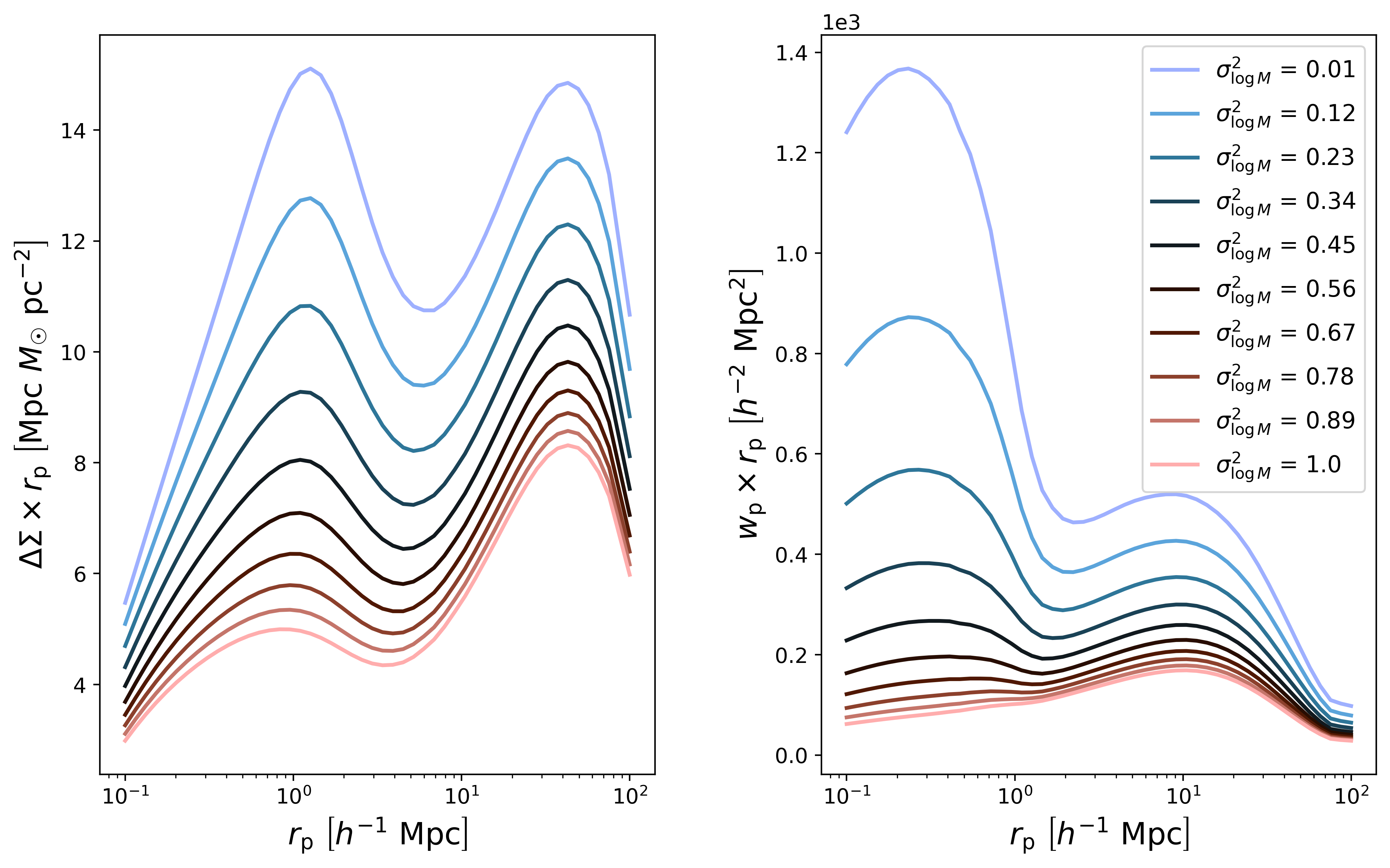}}
\subfigure[$\mathrm{log}\,M_1$]{\includegraphics[width = 0.49\textwidth]{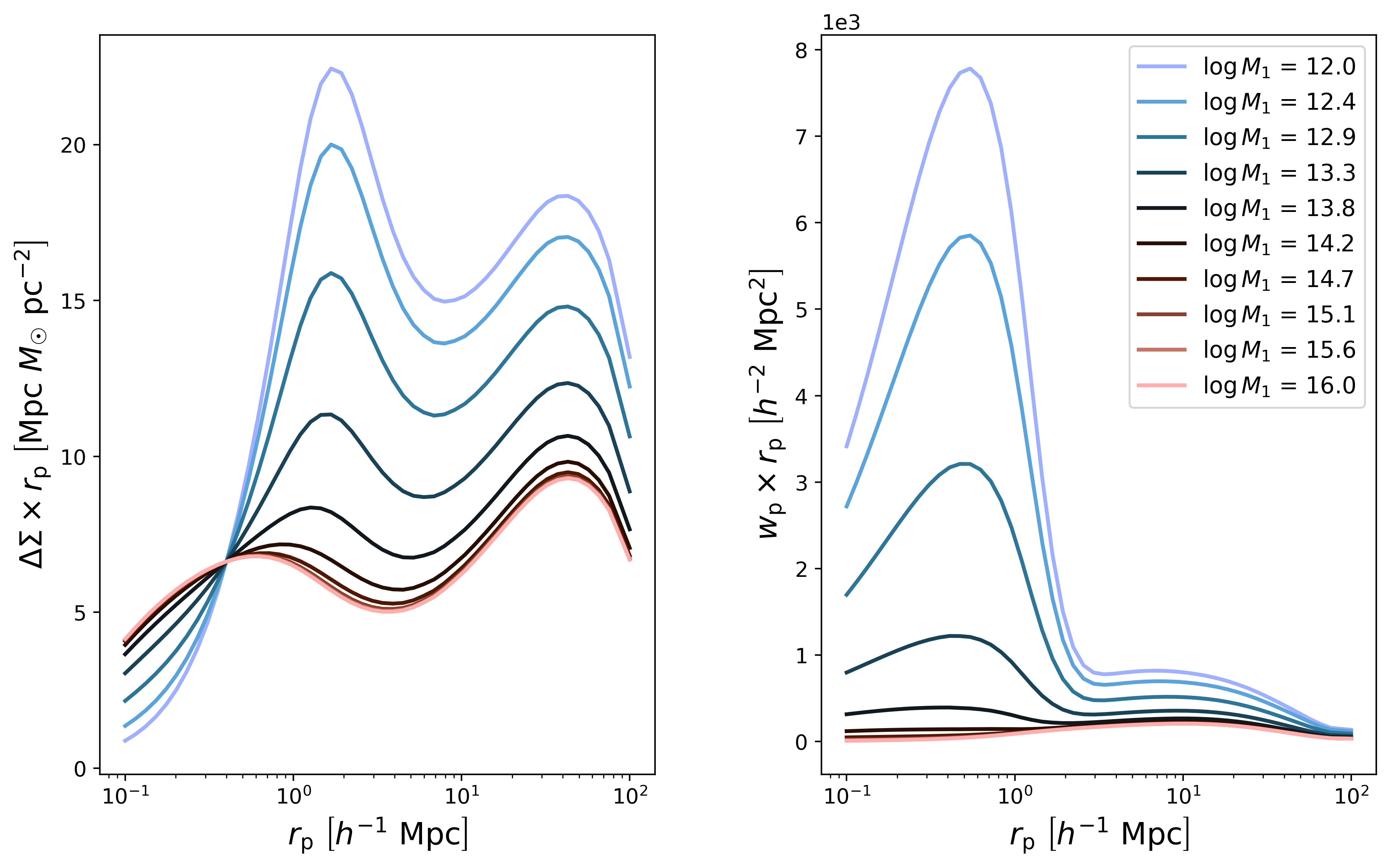}}
\subfigure[$\alpha$]{\includegraphics[width = 0.49\textwidth]{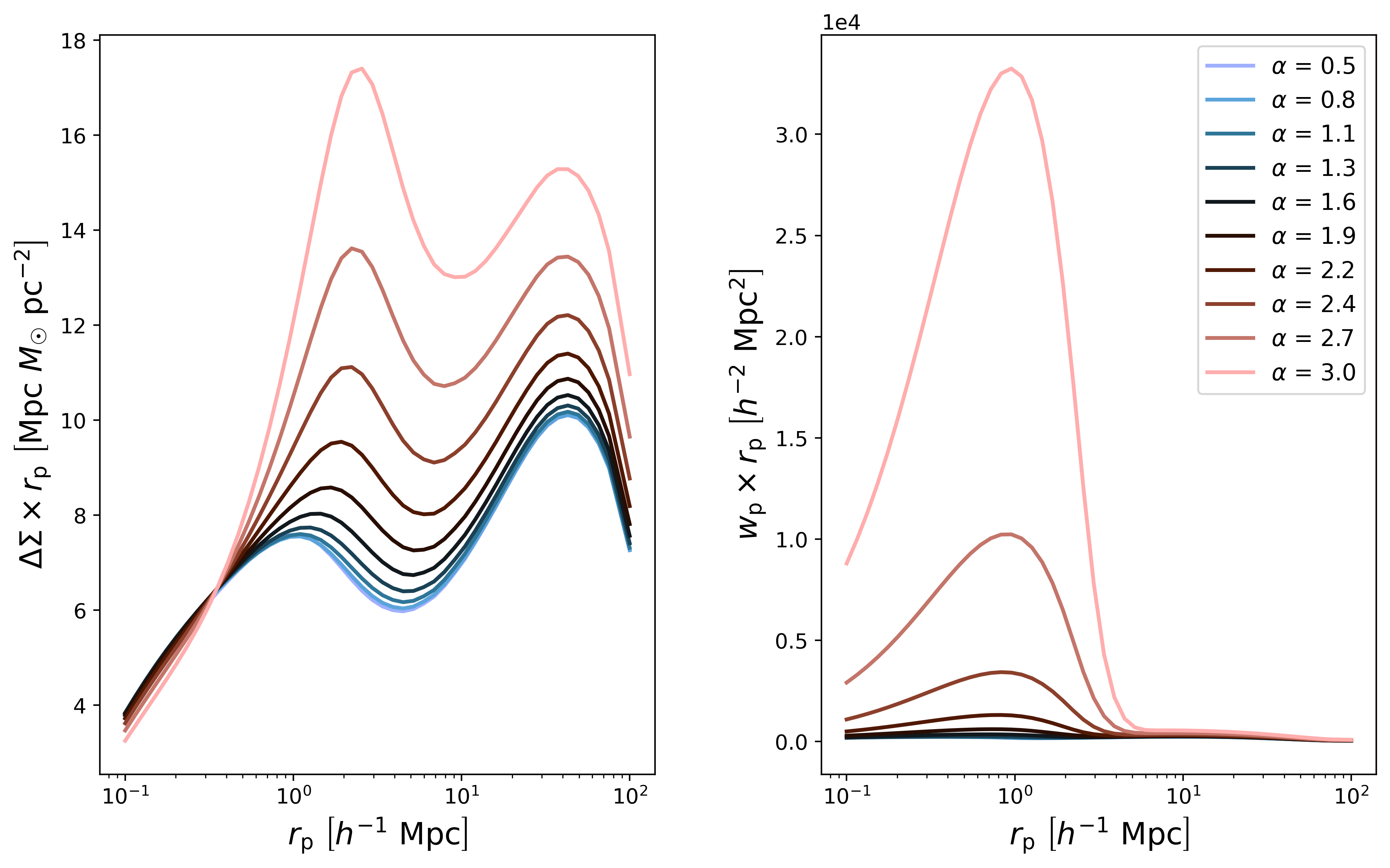}}
\subfigure[$\kappa$]{\includegraphics[width = 0.49\textwidth]{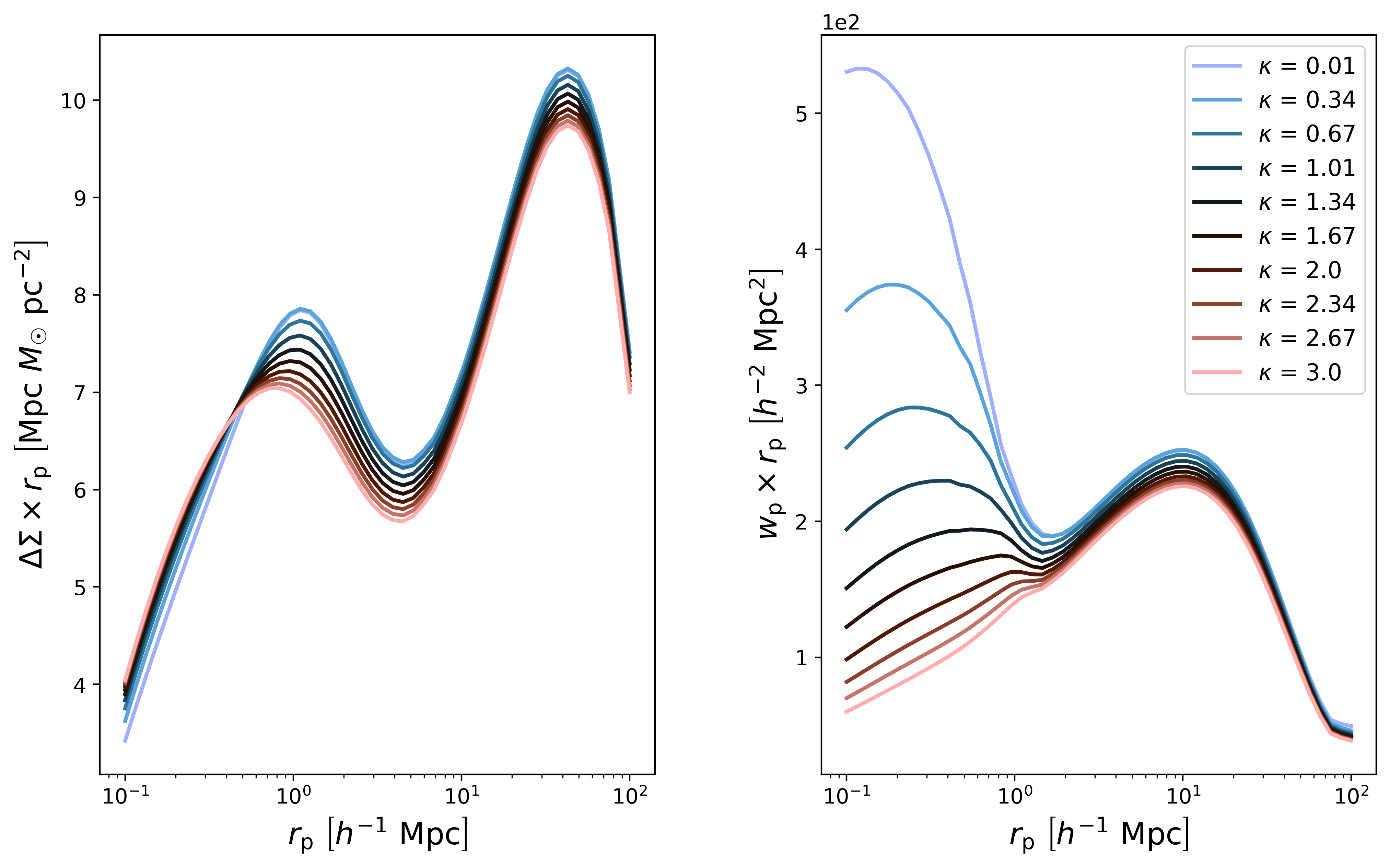}}
\subfigure[$R_{\mathrm{c}}$]{\includegraphics[width = 0.49\textwidth]{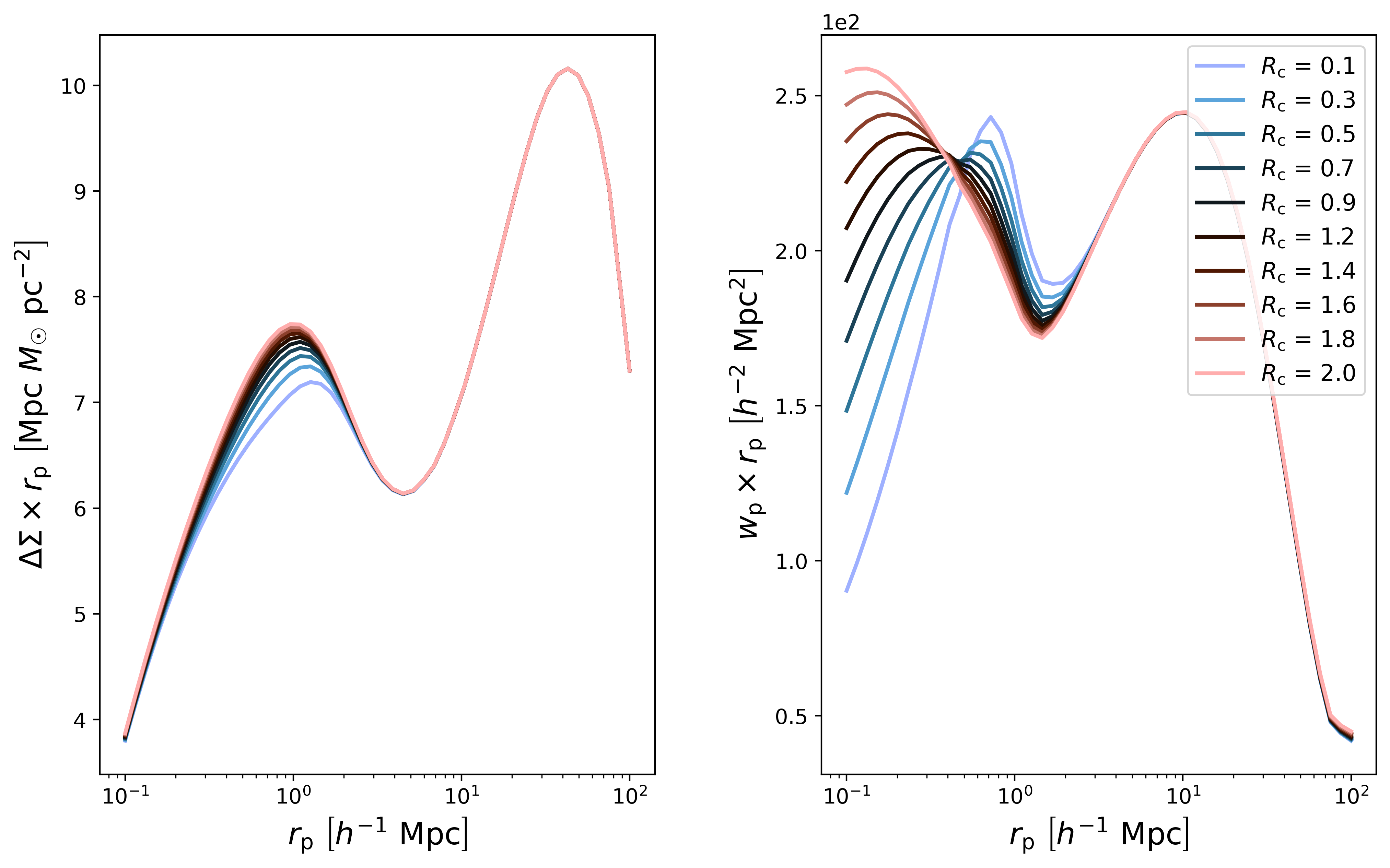}}
\caption{Parameter-dependence of $\Delta\Sigma$ and $w_{\mathrm{p}}$. When fixed, parameters are set to the fiducial values in Table \ref{table:appendix_table}. Feedback is excluded except in panel (e) of Fig. \ref{fig:rainbow_2}.}
\label{fig:rainbow_1}
\end{figure*}
\begin{figure*}
\centering
\subfigure[$p_{\mathrm{off}}$]{\includegraphics[width = 0.49\textwidth]{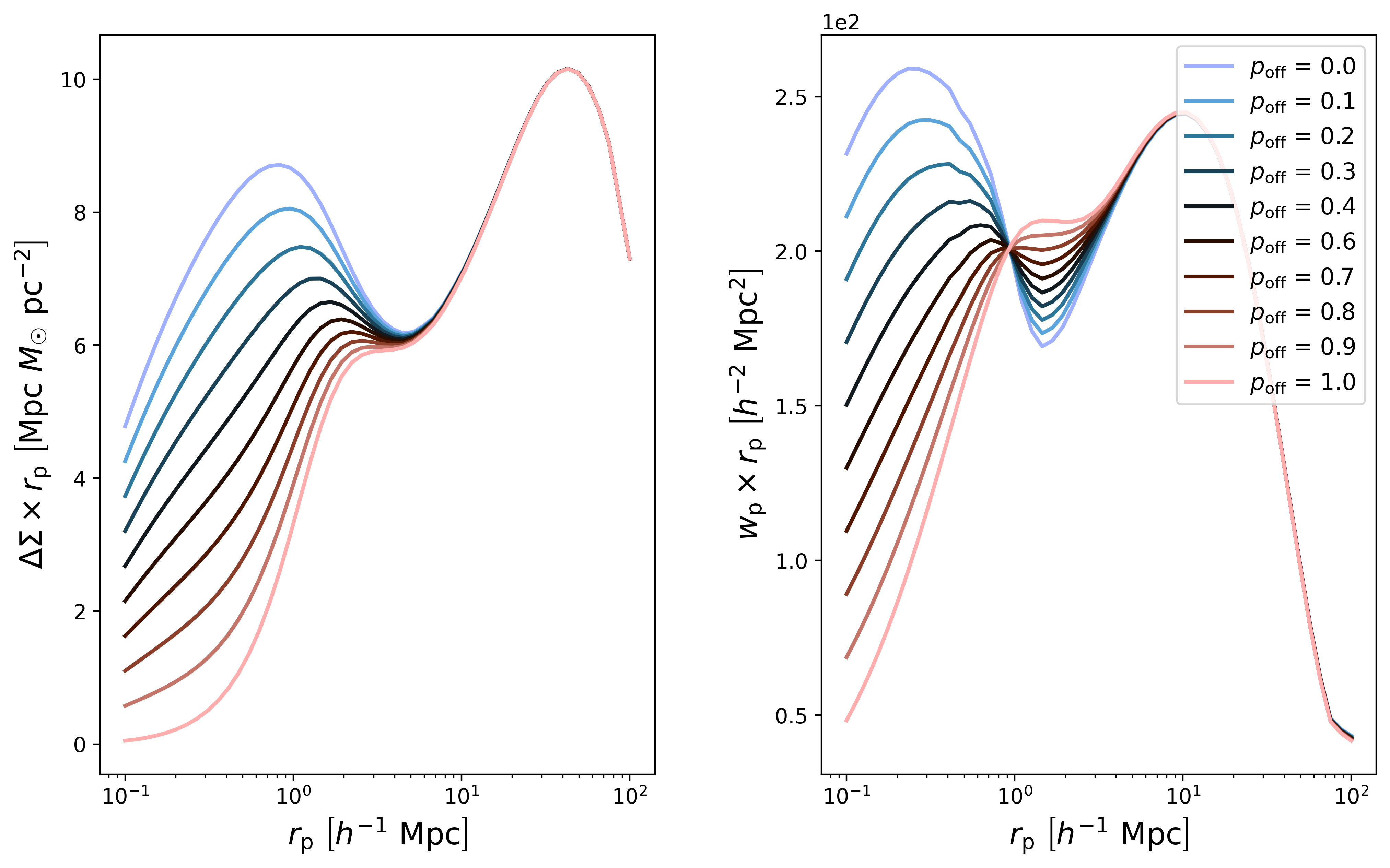}}
\subfigure[$R_{\mathrm{off}}$]{\includegraphics[width = 0.49\textwidth]{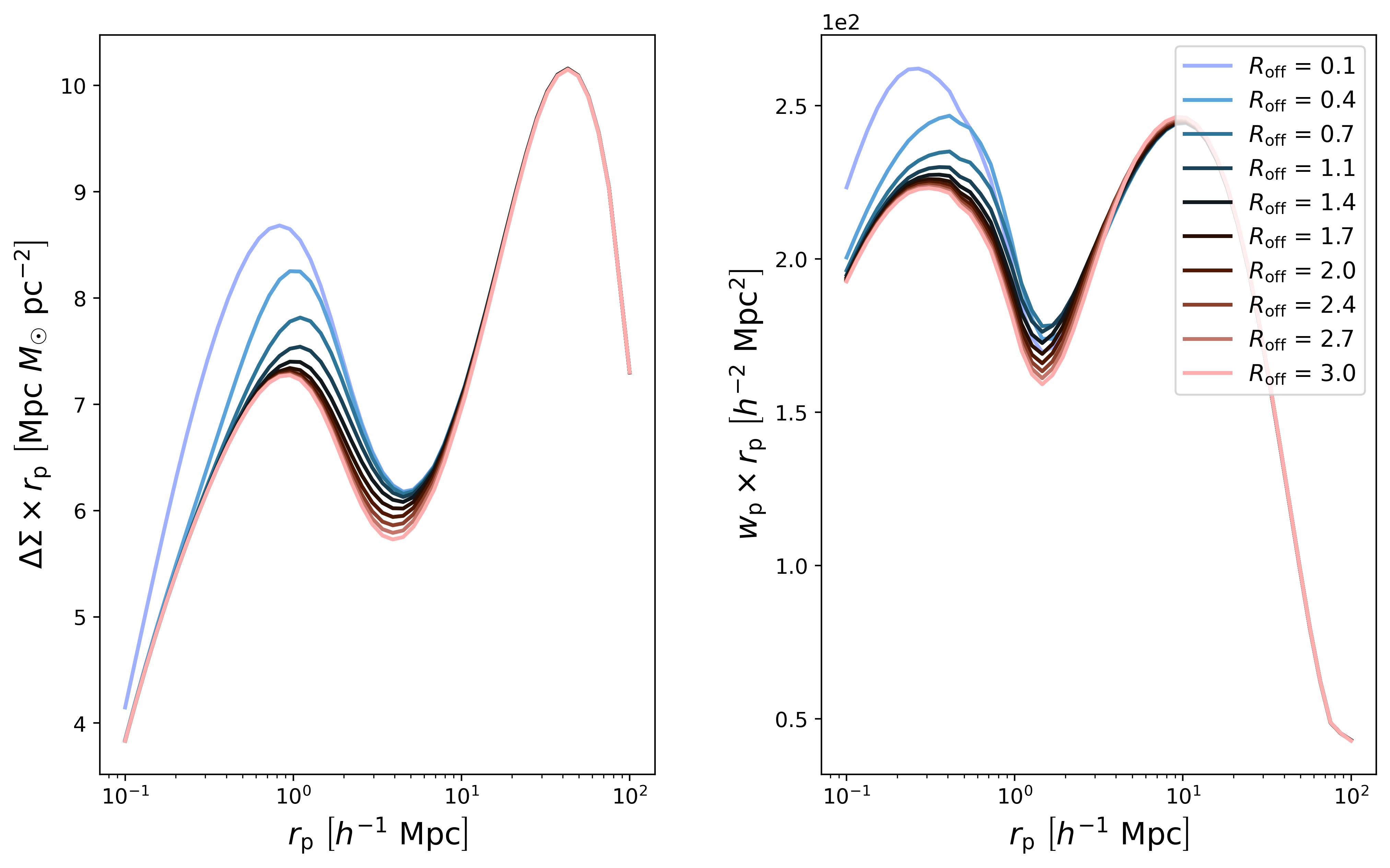}}
\subfigure[$\mathrm{ln}\,(10^{10}A_{\mathrm{s}})$]{\includegraphics[width = 0.49\textwidth]{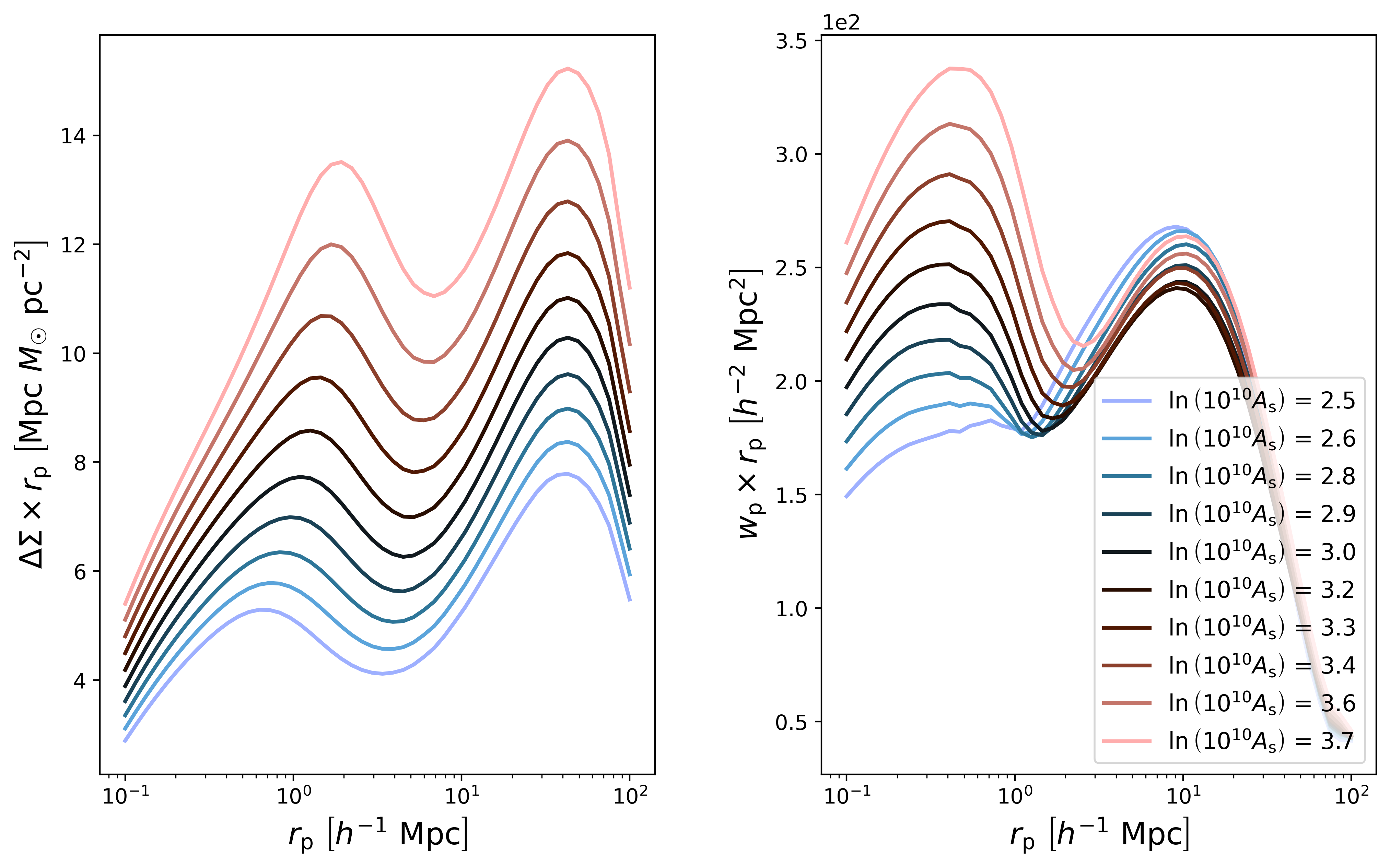}}
\subfigure[$\Omega_{\mathrm{m},0}$]{\includegraphics[width = 0.49\textwidth]{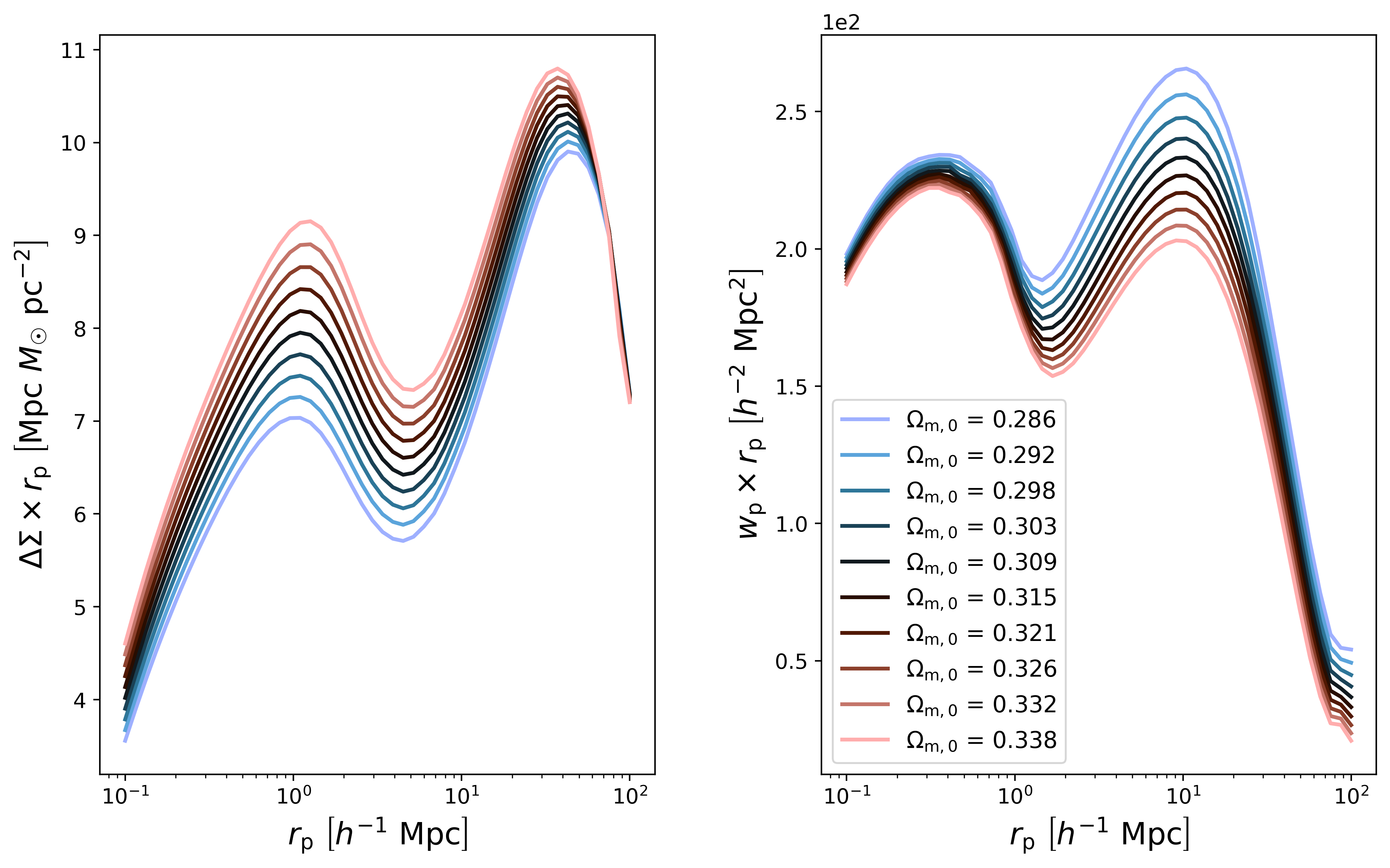}}
\subfigure[$\mathrm{log}\,M_{\mathrm{c}}$ and $\theta_{\mathrm{ej}}$]{\includegraphics[width = 0.49\textwidth]{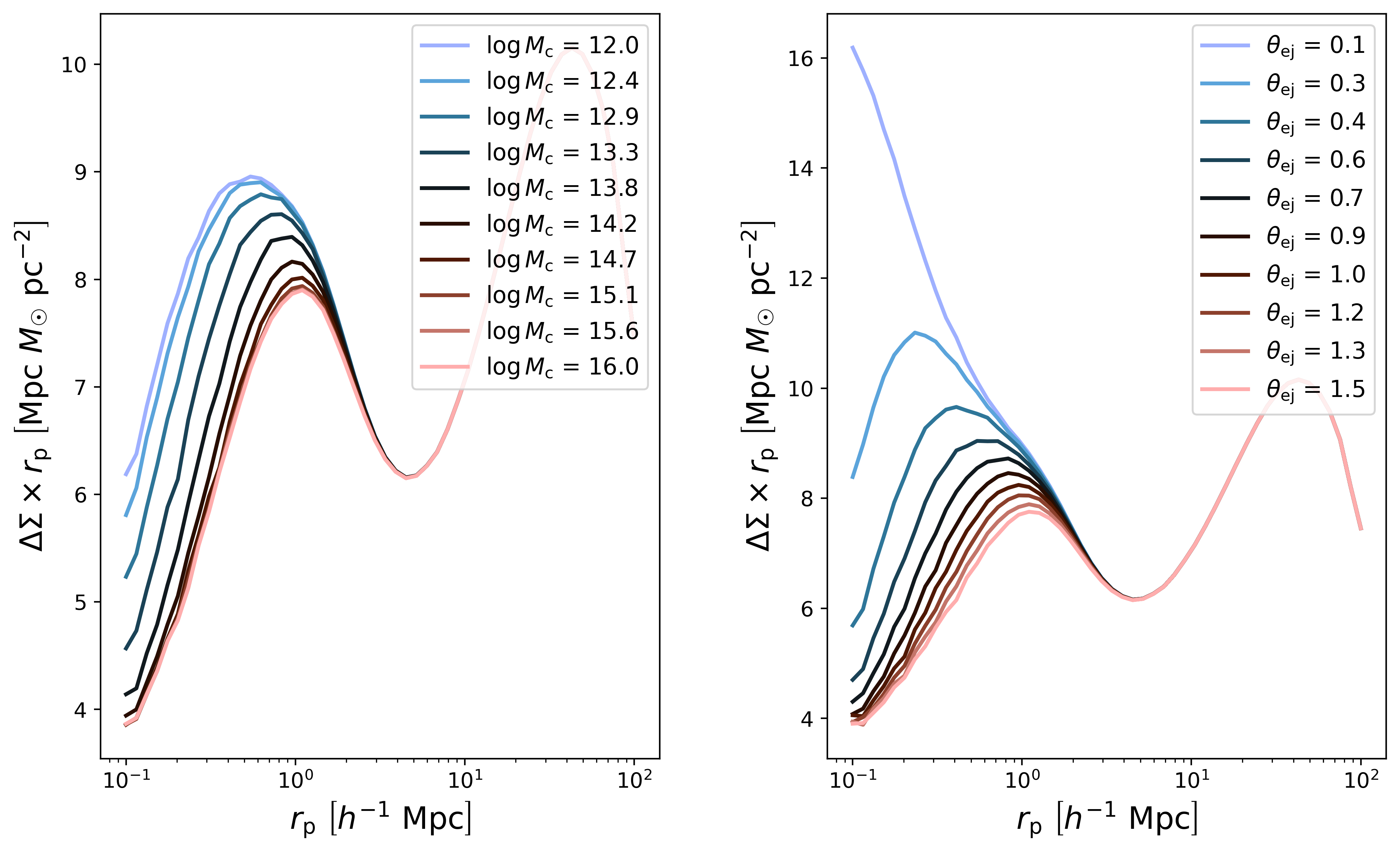}}
\caption{Same as Fig. \ref{fig:rainbow_1}, but for different parameters.}
\label{fig:rainbow_2}
\end{figure*}


\bsp	
\label{lastpage}
\end{document}